\documentclass{aastex631}

\usepackage{xspace} 
\usepackage{xcolor}
\newcommand{\TESS}{TESS\xspace}
\newcommand{\Kepler}{\textit{Kepler}\xspace}

\received{Accepted for publication in The Astronomical Journal, Sept.  2024}
\begin{document}

\title{ DIAmante \TESS AutoRegressive Planet Search (DTARPS): III. Understanding the DTARPS-S Candidate Transiting Planet Catalogs}

\author{Elizabeth J. Melton}
\affiliation{Department of Astronomy \& Astrophysics, Pennsylvania State University, University Park, PA 16802, USA}
\affiliation{Center for Exoplanets and Habitable Worlds, 525 Davey Laboratory, The Pennsylvania State University, University Park, PA, 16802, USA.}
\affiliation{Department of Physics and Optical Engineering, Rose-Hulman Institute of Technology, 5500 Wabash Avenue, Terre Haute, IN 47803, USA}

\author{Eric D. Feigelson}
\affiliation{Department of Astronomy \& Astrophysics, Pennsylvania State University, University Park, PA 16802, USA}
\affiliation{Center for Exoplanets and Habitable Worlds, 525 Davey Laboratory, The Pennsylvania State University, University Park, PA, 16802, USA.}
\affiliation{Center for Astrostatistics, 525 Davey Laboratory, The Pennsylvania State University, University Park, PA, 16802, USA.}

\author{Marco Montalto}
\affiliation{INAF - Osservatorio Astrofisico di Catania, Via S. Sofia 78, I-95123 Catania, Italy}

\author{Gabriel A. Caceres}
\affiliation{EY-Parthenon, 1540 Broadway, New York NY, 10036, USA}

\author{Andrew W. Rosenswie}
\affiliation{Department of Astronomy \& Astrophysics, Pennsylvania State 
University, University Park, PA 16802, USA}
\affiliation{Institut f\"{u}r Physik und Astronomie, Universit\"{a}t 
Potsdam, D-14476 Potsdam, Germany}
\affiliation{Leibniz-Institut f\"ur Astrophysik Potsdam (AIP), An der 
Sternwarte 16, D-14482 Potsdam, Germany}

\author{Cullen S. Abelson}
\affiliation{Department of Astronomy \& Astrophysics, Pennsylvania State University, University Park, PA 16802, USA}
\affiliation{Department of Physics and Astronomy, University of Pittsburgh, 100 Allen Hall, 
3941 O'Hara St.,
Pittsburgh, PA 15260, USA}

\begin{abstract}
The DIAmante \TESS AutoRegressive Planet Search (DTARPS) project, using novel statistical methods,  has identified several hundred candidates for transiting planetary systems obtained from 0.9 million Full Frame Image light curves obtained in the \TESS Year 1 southern hemisphere survey (Melton et al. 2024a and 2024b).  Ten lines of evidence including limited reconnaissance spectroscopy indicate that approximately half are true planets rather than False Positives. Here various population properties of these candidates are examined.  Half of the DTARPS-S candidates are hot Neptunes, populating the 'Neptune desert' found in \Kepler planet samples.   The DTARPS-S samples also identify dozens of Ultra Short Period planets with orbital periods down to 5 hours, high priority systems for atmospheric transmission spectroscopy, and planets orbiting low-mass M stars.  DTARPS-S methodology is sufficiently well-characterized at each step that preliminary planet occurrence rates can be estimated.  Except for the increase in hot Neptunes,  DTARPS-S planet occurrence rates are consistent with \Kepler rates.  Overall, DTARPS-S provides one of the most reliable and useful catalogs of \TESS exoplanet candidates that can be tapped to improve our understanding of various exoplanetary populations and astrophysical processes.  
\end{abstract}

\section{Introduction}

\subsection{The ARPS and DTARPS-S Transit Detection Project}

NASA's Transiting Exoplanet Sky Survey (TESS) mission was predicted to discover several thousand planets during its 2-year prime mission from light curves derived from Full Frame Images \citep[FFIs,][]{Barclay18}.  But it has proved challenging to achieve this goal due to the difficulties of reliably detecting smaller planets and reliably removing astronomical False Positives. Exoplanet discovery  from transit surveys depends critically on the statistical methods used to extract brief periodic dips in brightness in the presence of complex aperiodic photometric variations from the star and instrument.  \TESS Objects of Interest (TOIs) are obtained by a variety of procedures to remove uninteresting photometric trends, detect periodicities, and discriminate transits from contaminating sources \citep{Jenkins20, Guerrero21, Kunimoto22, Tey23}.  

One problem is that most detrending procedures leave behind short-memory stochastic autocorrelation; the residuals are not white noise.  This impedes small planet detection by the widely used Box-Least Squares (BLS) algorithm that assumes Gaussian white noise is present \citep{Kovacs02}. 
Time series modeling methods commonly used in signal processing and econometrics can remove both long-memory trends and short-memory autocorrelation.  ARIMA (autoregressive integrated moving average)  modeling,, also known as Box-Jenkins analysis, has been extensively developed since the 1970s with strong mathematical foundations \citep{Hamilton94, Box15, Feigelson18, Chatfield19}.  ARIMA can be effective in removing  a awide range of temporal behaviors associated with stellar and instrumental variations, including whitening of residuals, leaving transit signals mostly intact.  \citet{Caceres19a} developed a multistage AutoRegressive Planet Search (ARPS) transit detection pipeline that starts with ARIMA detrending, developing a new periodogram based on the Transit Comb Filter (TCF) to replace BLS that can not be used in this situation. TCF periodograms have lower noise levels and better sensitivity to small planets than BLS when autocorrelation is present \citep{Gondhalekar23}.  Features from the ARPS time series analysis are fed into a Random Forest classifier to identify probable exoplanetary transits.  This allowed \citet{Caceres19b} to detect sub-Earth-sized transits in 4-year \Kepler light curves.  

We are now applying an improved ARPS pipeline to analyze $\sim 0.9$ million Full Frame Image light curves from the Year 1 Prime Mission \TESS survey of the southern ecliptic hemisphere.  The DIAmante \TESS AutoRegressive Planet Search (DTARPS) effort is described in our companion papers.  \citet[][henceforth Paper I]{Melton22a} presents the light curves extracted and pre-processed by the DIAmante project \citep{Montalto20}, followed by an improved statistical analysis based on the AutoRegressive Planet Search methodology described by \citet{Caceres19a}. The result is the DTARPS-S Analysis List of 7,743 stars with high sensitivity (recall rate) of confirmed planets, but dominated by False Alarms and False Positives.  \citet[][Paper II]{Melton22b} reduces this list with a multifaceted vetting procedure to improve the purity of the candidate lists, removing False Alarms and reducing the number of False Positives such as blended EBs.  The vetting emerges with 462 objects in the DTARPS-S Candidates catalog with the highest quality plus 310 objects in the DTARPS-S Galactic Plane list that may be subject to more contamination.  

The evaluation of transit catalogs $-$ $Kepler$, TOI, DTARPS-S, and others $-$ depends on their completeness (True Positive Rate) and contamination (False Positive Rate), as well as accurate measurements of period and radii.  The DTARPS-S exoplanet detection procedure is mathematically quite different from those used for the $Kepler$ and TOI samples although the vetting procedures are similar.  We find that the DTARPS-S Analysis List prior to vetting has low levels of erroneous recovery of known False Positives (0.43\% False Positive Rate) and has excellent completeness (92.5\% True Positive Rate) with respect to a training set of injected planets and EBs.  It also is remarkably free from erroneous recovery of astronomically known False Positives and has excellent recall rates for astronomically Confirmed Planets at radii $R \gtrsim 10$~R$_\oplus$, decreasing to $<20$\% recovery for $R \lesssim 3$~R$_\oplus$ (Figures 10, 11, 13 and 15 in Paper I). Final DTARPS-S orbital periods and planetary radius are accurate without systematic biases (Figures 6 and 7 in Paper II).  

The present study (Paper III) is concerned with astronomical interpretations of the DTARPS-S catalogs. We start with a review of the generation of DTARPS-S candidate lists from Papers I and II (\S\ref{sec:PaperI_II}). Section \ref{sec:cand_pure} assesses the purity of the DTARPS-S Candidates catalogs and conclude that at least half of DTARPS-S Candidates are likely to be valid orbiting planets. The dramatic difference between DTARPS-S and \Kepler findings is the large number of DTARPS-S candidates found in the `Neptune desert' (\S\ref{sec:Neptune_desert}).  The study proceeds with identifying DTARPS-S members of astronomically interesting subpopulations: Ultra-Short Period planets (\S\ref{sec:usp}), planets suitable for atmospheric composition study (\S\ref{sec:atmos}), and planets orbiting M dwarfs (\S\ref{sec:m_dwarf}). A careful analysis of planetary occurrence rates derived from the DTARPS-S Candidates catalog is made in \S\ref{sec:occ_rate} with methodology described in Appendix \ref{sec:occ_method}. The paper ends (\S\ref{sec:final_remarks}) with a summary and explanation of DTARPS-S Year 1 results.

 \section{The DTARPS-S Transit Candidates \label{sec:PaperI_II}}

\subsection{DTARPS-S Catalogs and Lists}

The DTARPS-S project is motivated to see whether novel treatments of light curve variations can improve the detections of exoplanets in \TESS survey data (\S1 in Paper I).  The AutoRegressive Planet Search (ARPS) procedure developed by \citet{Caceres19a} has been found to have high sensitivity to smaller planets in 4-year $Kepler$ data \citep{Caceres19b} and is thus a promising avenue for the discovery of transiting systems in \TESS data. 

ARPS is rooted in ARIMA-type fitting that has long been demonstrated to be effective in modeling time series with a wide range of stochastic autocorrelated behaviors \citep{Box15}.  The differencing operation in ARIMA that serves to detrend non-stochastic variations transforms a planet's box-shaped transit to a double-spike pattern. ARPS therefore developed a new periodicity search tool, the Transit Comb Filter (TCF), in place of the traditional Box-Least Squares (BLS) regression approach, for constructing periodograms sensitive to planetary transits \citep{Caceres19a}.  The TCF periodogram is substantially more sensitive to weak transits than BLS when autocorrelation is present in the light curve, and can outperform BLS even for Gaussian white noise  \citep{Gondhalekar23}. 

DTARPS-S analysis starts with 976,814 dwarf and subgiant stars with spectral types F5 to M falling in the footprint of \TESS sectors $1 - 13$ surveyed during Year 1 with identifications in the \TESS Input Catalog \citep{Montalto20}. FGK stars were restricted to $V < 13$ magnitude while M stars were restricted to $V < 16$ magnitude and distance $d < 600$ pc. After some additional preprocessing, 823,099 \TESS light curves were successfully processed through ARIMA modeling and TCF periodograms of the ARIMA residuals (Paper~I).  The missing 15\% of input light curves either had poor ARIMA fits or were missing some `features' required by the machine learning classifier.  

A modern Random Forest classifier \citep{Ishwaran22} is then constructed to select likely planetary transits and to reduce False Alarms and False Positives, such as EBs blended into the large \TESS pixels. After removal of a validation set, the classifier was trained $towards$ 1,048 \TESS light curves with injected planets based on the $Kepler$ distribution of transit periods, depths and durations.  The classifier was trained $against$ 12,475 random light curves and 9,095 light curves injected with False Positive EBs signals as described by \citep{Montalto20}.  After many experiments, the best performing classifier was based on 37 features: 4 host star properties; 7 DIAmante light curve properties; 3 differenced light curve properties; 9 ARIMA residual properties; 3 TCP periodogram properties; and 11 properties of the best-periodogram peak and its folded light curve (Table 1 in Paper I). 

Once a threshold on the Random Forest probability is set, the classifier achieves 92.8\% True Positive Rate, 0.43\% False Positive Rate, and 92\% on various scalar classifier performance metrics (Table 2 in Paper I). The classifier is extremely effective at removing injected False Positives (Figure ~12 in Paper~I).  A heat map of recall rates shows $>80$\% recall rate for radii above 7 (12)~R$_\oplus$) at periods around 2 (10)~days, and $<20$\% for radii below 3~R$_\oplus$ (Figure 17 in Paper~I).  The specificity (also called True Negativity Rate) is universally near 100\% (Figure 22 in Paper~I). We call the list of 7,743 light curves that satisfy this classifier threshold the {\it DTARPS-S Analysis List} (DAL, Table~3 in Paper I). 

However, a False Positive rate of 0.43\% is insufficient to obtain a scientifically reliable sample, as the DAL drawn from the full sample will have at least $0.0043 $ (FPR) times $ 0.9 \times 10^6$ (number of light curves) $\sim 3900$  incorrect identifications.  A multifaceted vetting procedure was therefore pursued in Paper II to improve the DAL reliability.  The vetting started with an image-based analysis for stellar crowding in the \TESS pixel and wobbling of the target centroid during and outside of transits as described by \citet{Montalto20}.  This removed three-quarters of the DAL objects. More objects were eliminated by identification of: False Alarms (no convincing periodicity); False Positives (mostly plausible blended EBs); photometric binaries; and ephemeris matches (leakage of light from EBs). The DTARPS-S Candidates were sorted into two dispositions during vetting. Level 1 candidates clearly passed all of the tests during vetting and had no flags raised. Level 2 candidates either passed a vetting test marginally or raised a potential problem flag after vetting. The result is the {\it DTARPS-S Candidates Catalog} of 462 southern hemisphere stars (Table 1, Figure 5 and associated Figure Set in Paper II)\footnote{
The effectiveness of our vetting procedure in Paper II can also be evaluated by comparing the injected planet recall rates before and after vetting in the planetary Radius-Period cells in Figure~\ref{fig:occ_marg} below. }. 

As the centroid-crowding criteria eliminated virtually all candidates from the Galactic Plane, we chose a subsample of DAL stars that were subject to all vetting steps except the image analysis.  The result of this analysis is the {\it DTARPS-S Galactic Plane list} of 310 southern hemisphere stars (Table 2, Figures 9 and 10 and associated Figure Set in Paper II). This list is not spatially complete.   

\subsection{DTARPS-S Candidates Catalog Overview}

\begin{figure}[t]
    \centering
    \includegraphics[width=0.7\textwidth]{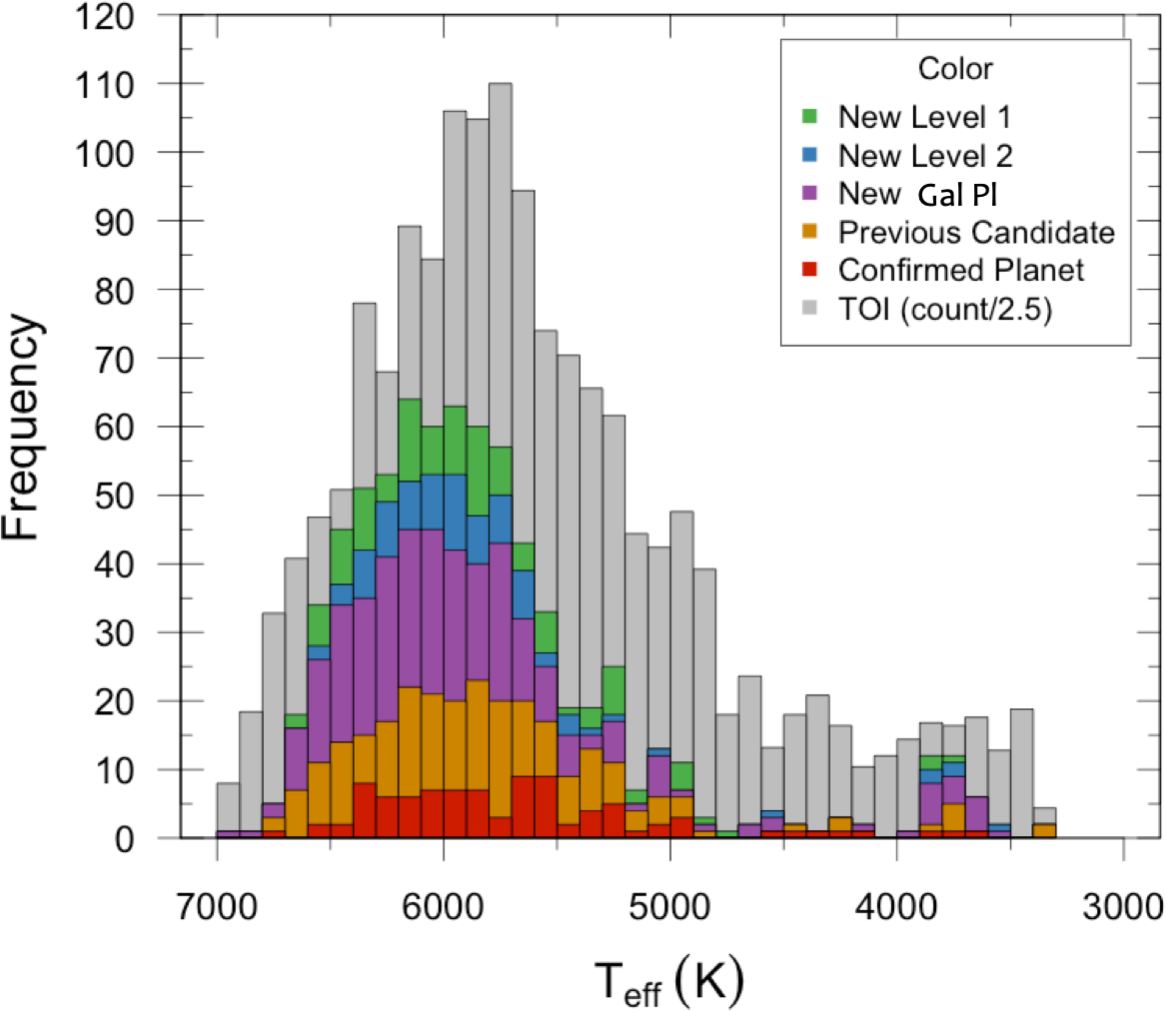}
    \caption{Histogram of the effective temperature of the stellar hosts for newly identified DTARPS-S candidates compared to the TOI hosts from the TOI list (3,000 K $<$ $T_{eff}$ $<$ 7,000 K).  The DTARPS-S objects are split by disposition. The TOI counts divided by 2.5 for convenient visualization.}
    \label{fig:st_temp_hist}
\end{figure}

Figure \ref{fig:st_temp_hist} shows the distribution of stellar temperature for the host stars of DTARPS-S candidates with comparison to the \TESS Objects of Interest.  The Level 1 and Level 2 subsamples together constitute the DTARPS-S Candidates catalog, and dispositions are  obtained from the NASA Exoplanet Archive \citep{NEA-CP} \footnote{In this and later figures, the NASA Exoplanet Archive and \TESS Objects of Interest list were accessed on March 15, 2022.}. The DTARPS-S Candidates' stellar hosts have somewhat hotter temperatures than TOI stellar hosts and shows a secondary peak for M stars, but these characteristics reflect the distribution of temperatures in the input DIAmante light curve sample \citep[][Figure 1]{Montalto20}.  Nearly all of the DTARPS-S candidates orbit F or G dwarf stars while only 7\% (5\%) of the DTARPS-S Candidates have K (M) dwarf hosts.

\begin{figure}[t]
    \centering
    \includegraphics[height=0.35\textheight]{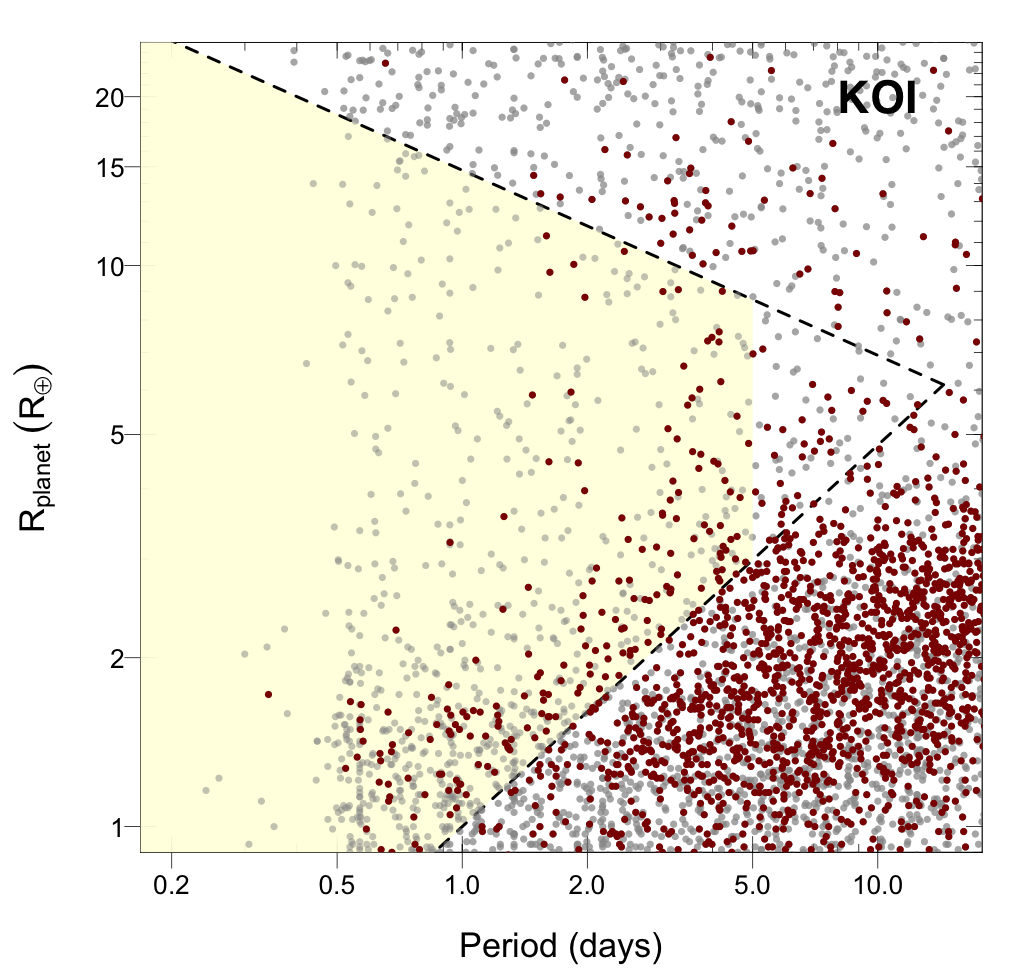} \hspace*{15pt}
    \includegraphics[height=0.35\textheight]{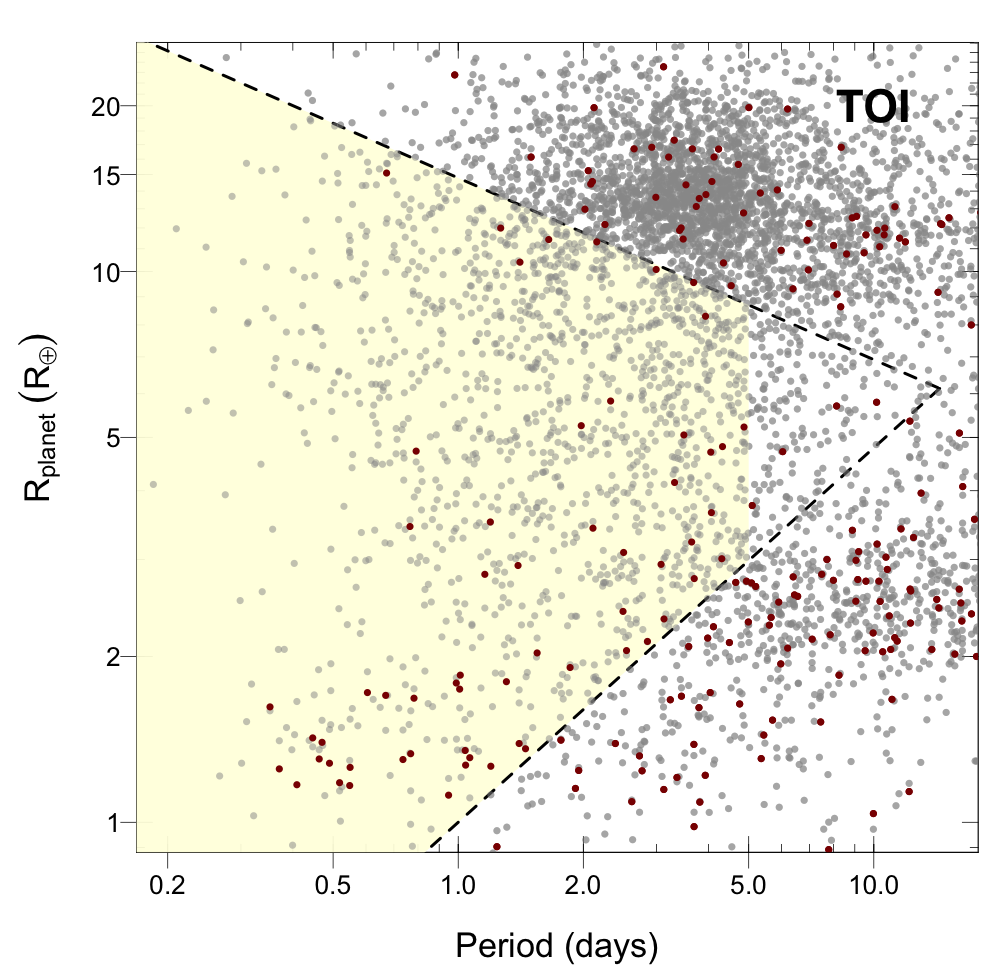}\\ \vspace*{15pt}
    \includegraphics[height=0.55\textheight]{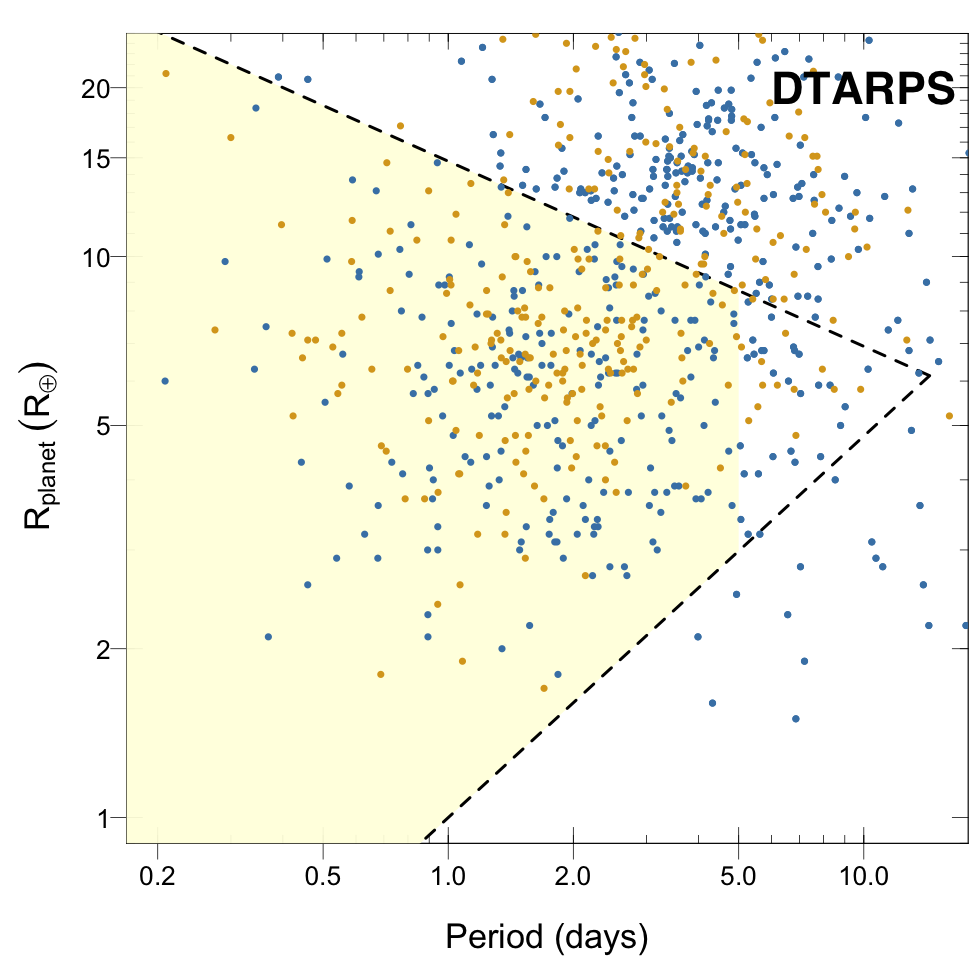}
    \caption{Period-radius diagrams for three wide-field surveys.  Top left: Kepler Objects of Interest in gray symbols with confirmed planets in red \citep{NEA-Kepler}. Top right: \TESS Objects of Interest in gray symbols with confirmed planets in red. These samples, cover the entire celestial sphere.  Bottom: DTARPS-S candidate catalog (blue) and Galactic Plane list (gold) from Melton et al. (2022b). The Kepler Neptune desert,  highlighted in yellow, is bounded by the dashed lines defined by \citet{Mazeh16}.}
    \label{fig:intro_PR}
\end{figure}

The radius-period distribution of the DTARPS-S Candidates is shown in Figure \ref{fig:intro_PR} with comparison to confirmed transiting planets from the NASA Exoplanet Archive. DTARPS-S Candidate catalog objects have inferred orbital periods in the range $0.2-13$ days and planet radii in the range $2-25$~R$_\oplus$.  As predicted by the DTARPS-S heat map of recall rates for planetary injections (Figure 17 in Paper I), DTARPS-S identifies very few  candidates with radii $P <2$ R$_{\oplus}$ or with periods $P > 10$ days. The DTARPS-S candidates are mostly hot planets with sizes ranging from inflated Jupiters to sub-Neptunes\footnote{
We adopt the following nomenclature: planets with $R<$ 1.8 $R_{\oplus}$ are super-Earths, $1.8 < R < 4$ R$_{\oplus}$ are sub-Neptunes,  $4 < R < 8$ R$_{\oplus}$ Neptunes, and $R > 8$ R$_{\oplus}$ are Jovians or Jupiters.}.  Forty-three percent have radii greater than 10 $R_{\oplus}$ and can be considered to be `hot Jupiters'.  Most of the DTARPS-S Candidates with radii below 10 $R_{\oplus}$ fall in Neptune desert region shown as the yellow region in Figure~\ref{fig:intro_PR}.  This important result is discussed in \S\ref{sec:Neptune_desert}.  

The truncation of Candidate properties  at long periods and small radii is intrinsic to the DTARPS-S analysis of \TESS light curves, as revealed by the heat map of recall rates of injected planets in the DTARPS-S Analysis List (Figure 13 in Paper~I).  Recovery rates are poor for planets smaller than $\sim$2~R$_\oplus$ due to the small effective signal-to-noise-ratio for shallow transits in short time series and absent for periods longer than 13~days due to the classifier response to 13.7~day orbital cycles.  

\begin{figure}
    \centering
    \includegraphics[width=0.7\textwidth]{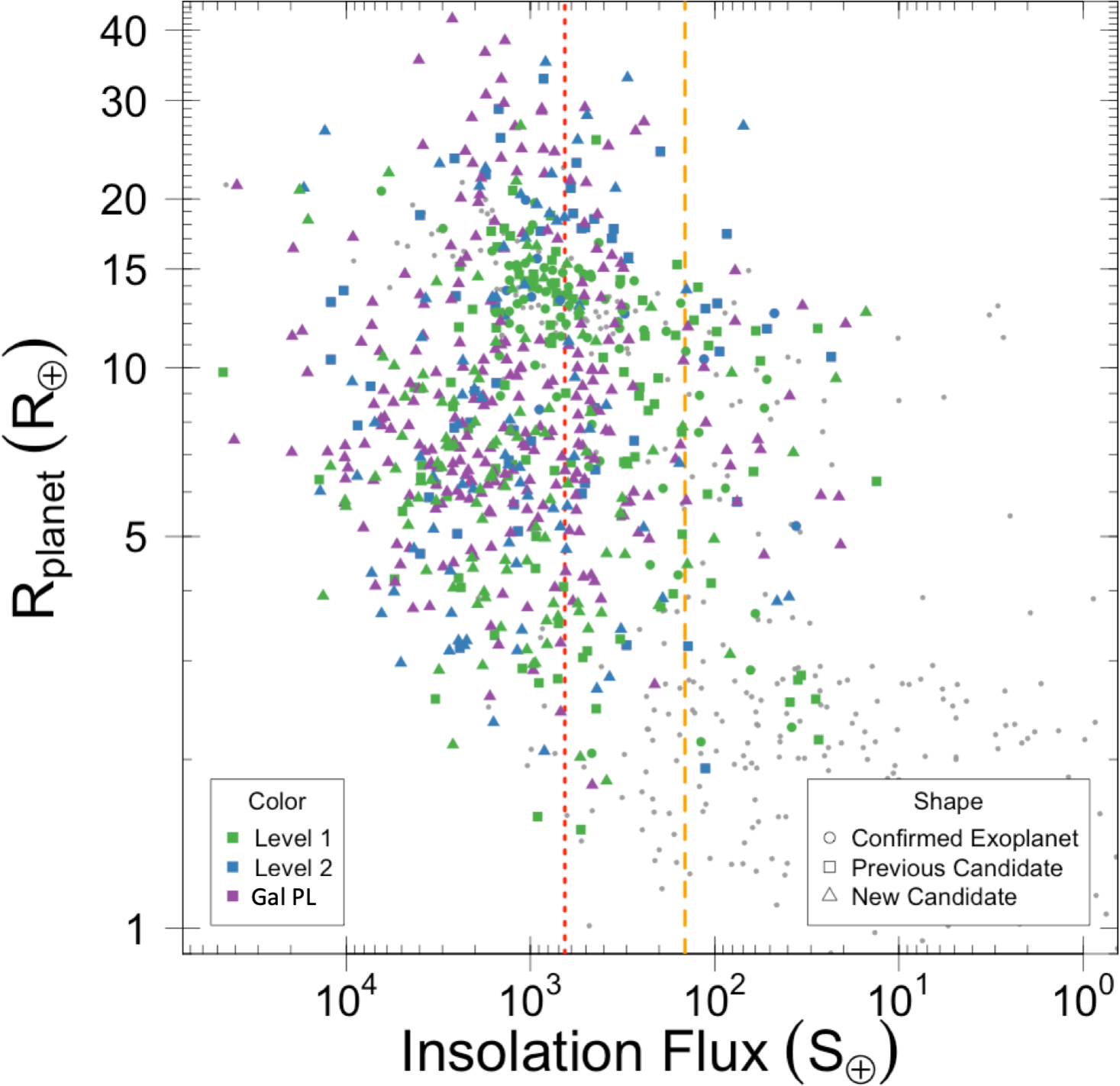}
    \caption{Insolation flux relative to Earth for the DTARPS-S candidates.  The light gray points represent Confirmed Planets from the NASA Exoplanet Archive. The disposition of the DTARPS-S Candidates is indicated by the color of each point and the symbol represents whether the DTARPS-S Candidate is a Confirmed Planet (circle), previously identified planet candidate (square), or a newly identified candidate (triangle). The boundary for Jovian planet inflation (145 $S_{\oplus}$) is marked with an orange dashed line, and the boundary for photoevaporative mass loss in Neptunes (650 $S_{\oplus}$) is marked with the red dotted line.}
    \label{fig:insol_flux}
\end{figure}

DTARPS-S candidates with short periods experience high levels of insolation flux from the host star, 
\begin{equation}
    \frac{S}{S_{\oplus}} = \left(\frac{R_{\star}}{R_{\sun}}\right)^2\, \left(\frac{T_{eff}}{T_{\sun}}\right)^4\, \left(\frac{1 AU}{a}\right)
\end{equation}
where $S_\oplus$ is the solar insolation on Earth, $R_{\star}$ is the stellar host radius, $T_{eff}$ is the stellar host effective temperature, and $a$ is the semi-major axis of the planet candidate, assuming a circular orbit. Figure \ref{fig:insol_flux} shows that nearly all DTARPS-S candidates have insolation fluxes above the 145 $S_{\oplus}$ lower limit calculated to cause radii inflation \citep{Demory11}.  Some  also lie above the 650 $S_{\oplus}$ limit estimated for photoevaporative atmospheric mass loss in Neptune mass planets \citep{Lundkvist16}.  However this limit cannot be directly applied here because atmospheric loss is mass-dependent  \citep{Lopez13, King18} and masses are not available without radial velocity follow-up. 

\section{Quality of the DTARPS-S Candidates Catalog \label{sec:cand_pure}}

The most effective validation tool of transiting planet catalogs is follow-up spectroscopy of target stars and nearby stars that may be blended into the large \TESS point spread function.   EBs will have kilometer-per-second radial velocity variations in contrast to meter-per-second variations for true planets.  Section~\ref{sec:recon} summarizes preliminary reconnaissance spectroscopy of a small subsample of targets from the DTARPS-S Candidate catalog and DTARPS-S Galactic Plane list from Paper~II.  Section~\ref{sec:tricera} then applies the \texttt{TRICERATOPS} statistical validation tool for transiting catalogs. Section~\ref{sec:fp_specificity} discusses EBs contamination.  Section~\ref{sec:summary_validation} assimilates these findings with those in Papers I and II and elsewhere in this paper to give an overall evaluation of the quality of the DTARPS-S 
Candidate catalog.

\subsection{Purity Estimates from Reconnaissance Spectroscopy and External Surveys \label{sec:recon}}

In an effort to give a preliminary evaluation of the reliability of DTARPS-S Candidates, 49 of the 777 Candidates were observed spectroscopically with the Tillinghast Reflector Echelle Spectrograph (TRES)  on the 1.5-meter telescope at the Fred Lawrence Whipple Observatory on Mt. Hopkins, Arizona in early-2022.  A fiber-fed, cross-dispersed, optical spectrograph, TRES is actively performing reconnaissance radial velocity (RV) spectroscopy as part of the \TESS Follow-up Observing Program;  over 1,000 TOIs have been observed with TRES \citep{Bieryla21}.  DTARPS-S results are summarized briefly here; details will be provided by the observers on the ExoFOP-TESS Web site\footnote{{https://exofop.ipac.caltech.edu/tess/}}.  

Targets were chosen to be bright with $V < 12.5$ and accessible from a northern hemisphere telescope in early-2022.  The observed sample of DTARPS-S Candidates are unbiased with respect to DTARPS-S disposition, orbital period, planet radius, and stellar temperature.  Some lie in the Neptune desert and others do not.  

From RV measurements on two or more occasions, 12 of the 49 Candidates exhibited low-amplitude RVs variations in-phase with the photometric transit signal found with DTARPS.  These motions are consistent with a planetary companion with a mass in the range $0.5 \lesssim M \lesssim 7$~M$_J$. Seven objects showed high-amplitude RV variations consistent with contaminating EBs.  The remaining objects' have an ambiguous status from these initial reconnaissance spectroscopy measurements. Sixteen objects have RV measurements in-phase with the photometric transit but with semi-amplitude too small to distinguish to distinguish between a planetary companion and noise and therefore are likely good candidates to observe with a more sensitive spectrograph. Fourteen targets had RV values that were either highly scattered or out of phase with the photometric transit and are unlikely to be good candidates for further follow-up. 

We can estimate the prevalence of true planets in the DTARPS-S Candidates candidates by combining the results from the reconnaissance spectroscopy with the DTARPS-S Candidates with follow-up characterization from external surveys. This estimation is subject to the biases of the variety of external surveys and should be viewed cautiously. Combining the 92 Confirmed Planets from the NASA Exoplanet Archive and 42 previously identified False Positives with the 49 reconnaissance spectroscopy preliminary insights, 66\% of DTARPS-S Candidates with some follow-up information have had their exoplanetary nature either confirmed or not disputed. If this trend held for the rest of the DTARPS-S candidate list, the remaining 595 DTARPS-S Candidates without any form of follow-up yet should contain at least 390 exoplanets. 

We conclude that at least half, and possibly two-thirds or more, of DTARPS-S Candidates are likely to be valid orbiting planets.  However, these estimates from RV and external surveys are not based on randomized subsamples, and the reliability of the conclusion is unknown.  

\subsection{Purity Estimate Using \texttt{TRICERATOPS} \label{sec:tricera}}

\texttt{TRICERATOPS} is a probabilistic vetting and validation tool designed to assist in evaluation of TOI candidate exoplanets  \citep{triceratops21}.  It examines evidence of a False Positive signal leaking into the target pixels from nearby stars, reporting a False Positive probability (FPP) and nearby False Positive probability (NFPP). Here we use \texttt{TRICERATOPS} to get an independent estimate of planet prevalence in the DTARPS-S Candidate catalog and DTARPS-S Galactic Plane list.  \texttt{TRICERATOPS} is a statistical procedure and is used here as an independent estimation of candidate reliability rather than an astronomical validation like spectroscopic follow-up.

For each of the 772 DTARPS-S candidates, \texttt{TRICERATOPS} was used to extract \TESS FFI light curves assuming the target star occupied the center pixel at its celestial position.  The extracted light curve was gridded into 100 evenly spaced bins, and the flux error from the DTARPS-S light curve applied to the \texttt{TRICERATOPS} light curve. Rather than using the suggested boundaries for \TESS FFI data from \citet{Giacalone21}, we explored boundaries based on the FPP and NFPP values for the Confirmed Planets and known False Positives in the DTARPS-S Candidate catalog. 

The Galactic Plane list stars are more strongly affected by the NFPP boundary than the Candidate catalog stars, indicating that they likely suffer more from blended contaminants.  This is expected given their failure of the crowding-centroid tests in DTARPS-S \citep[Paper II \S2.1,][]{Montalto20}.  DTARPS-S Candidates in the Neptune desert tend to have higher FPP and NFPP values than other Candidate stars, suggesting that light leakage contamination rate in the Neptune desert is $\sim$10-30\% higher than for other Candidate stars.  This is consistent with our finding higher fraction of known FPs in the Neptune desert.  

The results from \texttt{TRICERATOPS} agree with an overall estimate that DTARPS-S Candidate catalog suffers 40-50\% contamination of non-planetary objects or a precision of 50-60\%.

\subsection{Purity Estimate Using the {Pr{\v{s}}a} et al. (2022) Eclipsing Binaries Catalog} \label{sec:fp_specificity} 

Paper II (Appendix C.4) compares the DTARPS-S Catalog candidates with the eclipsing binary (EB) catalog presented by \citet{Prsa22} in an attempt to independently evaluate the likely contamination of DTARPS-S Candidates by EBs. Of 1,056 \citet{Prsa22} EBs in the full DIAmante list, 188 appear in the DTARPS-S Analysis List (Paper I) and 67 in the DTARPS-S Candidate catalog and Galactic Plane list (Paper II). Individual cases are noted in the Appendix of Paper II.  But some of the overlaps are spectroscopically Confirmed Planets, suggesting that the Pr{\v{s}}a et al. EB catalog suffers contamination from transiting exoplanets. For this reason, we did not use Pr{\v{s}}a et al.  to modify the disposition labels of the DTARPS-S Candidates  (Paper~II, Appendix C).

Of the 67 catalogued EBs among our candidates, 17 lie inside the hot Neptune desert.  This is a much lower fraction than expected from random selection given that nearly half of the DTARPS-S candidates reside there (\S\ref{sec:Neptune_desert}). The EB specificity (that is, the correct identification of True Negatives by a classifier) of the DTARPS-S classifier alone is 82\%, and the vetted specificity of the DTARPS-S Candidate catalog and Galactic Plane list 94\%. The true specificity to EBs may be somewhat higher due to errors in the Pr{\v{s}}a et al. catalog. 

The negative training set for the DTARPS-S classifier discussed in Paper~I (\S4.2) contains random light curves and injected unequal mass EBs. It does not cover the full morphology of EBs, nor was the distribution of injected periods and radii drawn from a physical sample as in the case of the injected planets.  Despite these limitations, the Random Forest classifier was able to remove 78\% of a diverse population of previously identified False Positives in the DIAmante data set (Paper~I, \S9.3). After the vetting analysis in Paper~II, the DTARPS-S methodology had removed approximately 92\% of the previously identified False Positives. In future applications of ARPS, improvements will be made the classifier including injection of a wider range of False Positive signals in the negative training set.

\subsection{Summary of DTARPS-S Catalog Validation Measures}
\label{sec:summary_validation}

A variety of measures are used to assess the quality of a statistical classification of a transiting exoplanet. Paper~I uses three scalar classification metrics (Matthew’s Correlation Coefficient, Youden’s J Index, and adjusted F-score) as well as bivariate plots (True Positive Rate $vs.$ False Positive Rate ROC curves and Precision $vs.$ Recall curves) to select a final classifier\footnote{
These classifier assessment measures are discussed by \citet{Powers11}, \citet{Chicco20}, and \citet{Tharwat21}.  A convenient source of definitions is the Wikipedia page on classifier confusion matrices, \url{https://en.wikipedia.org/wiki/Confusion_matrix}.
}.  High specificity (True Negatives classified as Negatives) supports other evidence that DTARPS-S is not prone to high rates of erroneous False Positive recovery. Here we add a simple measure called `prevalence', often used in epidemiology: prevalence is the number of True Positive cases in the data divided by the total sample size.  Transit catalogs will reliably represent the true planetary population when both the prevalence that approach 1.00 (i.e. no True Negative cases are present) and the precision approaches 1.00 (i.e. no False Positives are present). 

Prevalence and precision in transit catalogs can be confidently measured only when the full sample has been characterized with radial velocity measurements, a task that lies beyond the scope of our photometry-based analysis.  But Papers I and II, together with the present study, give various indirect indications that the DTARPS-S Candidates catalog has only a modest fraction of False Alarms, blended EBs, or other contaminants.  A low fraction is not unexpected as the Random Forest classifier was carefully trained against thousands of injected EBs \citep[Paper I,][]{Montalto20} and the vetting procedures strictly removed False Alarms and evidence for EBs (Paper II, \S2). 

We summarize here these indications of the high quality of the DTARPS-S catalogs.  
\begin{enumerate}

\item The 7,743-object DTARPS-S Analysis List emerging from Random Forest misclassified only 2 False Positives in the training set of 9,093 True Negatives among the injected EBs (Figure~12, Paper I). The False Positive Rate of the classifier is 0.043\%.  The successful removal of EBs by the classifier is also suggested by sparsity of high-depth objects in the DTARPS-S Analysis List, and the relatively few contaminants in the \citet{Prsa22} EBs catalog (\S3.3). 

\item The False Positive recall rate in the validation set of the DTARPS-S Analysis List prior to vetting is $18$\% (\S6, Paper I). The vetting process removed 78 False Positives leaving 35 known False Positives in the 462-object DTARPS-S Candidate catalog.  These include 17 EBs systems, 5 low-mass EB systems (EBLM catalog), 2 blended EBs systems, 2  single-line spectroscopic EBs systems, and 1  double-line spectroscopic EBs system (Appendices A and B in Paper II). 

\item After vetting procedures are applied, the 772 objects in the DTARPS-S Candidates catalog and Galactic Plane list includes only 43 known False Positives based on NASA Exoplanet Archive dispositions (\S3 and Appendices A and B, Paper II). 

\item Comparison with the \citet{Prsa22} catalog of probable EBs upheld the high specificity (94\%) of the DTARPS-S methodology found with the previously identified False Positives in the full DIAmante list (\S\ref{sec:fp_specificity}).

\item Reconnaissance spectroscopy of a limited subsample of DTARPS-S Candidates suggests that less than half, and possibly fewer than a third, of the catalog is contaminated by EBs (\S\ref{sec:recon}). 

\item Planetary occurrence rates estimated from the DTARPS-S Candidates assuming they are all valid planets does not exceed estimated from rates obtained from $Kepler$ Confirmed Planets for areas other than the Neptune desert (\S\ref{sec:occ_comp} below).  

\item Of the 283 previously identified False Positives in the Neptune desert in the DIAmante data set of $\sim 0.9$ million stars, only 26 are DTARPS-S Candidates (\S\ref{sec:nep_desert}). 

\item Application of the \texttt{TRICERATOPS} statistical vetting and validation tool suggests the precision of the DTARPS-S Candidate catalog is around $50-60$\% (\S\ref{sec:tricera}).  

\item An astrophysical indication of catalog validity for larger radii hot Jupiters appears in Figure~\ref{fig:insol_flux}.  DTARPS-S planet candidates with radii $> 15$~R$_\oplus$ that are likely inflated by stellar heating are largely restricted to insolation fluxes in excess of $\sim 300$~S$_\oplus$.  This suggests that the inflated Jupiters are real and not False Positive cases involving binary stellar companions.

\item Recent studies of young and intermediate age open clusters indicate high fraction of hot Neptune occurrence rates \citep{Fernandes23, Christiansen23}.  Most DTARPS-S Candidates lie close to the Galactic Plane suggesting that they also are relatively young (\S\ref{sec:plan_age}).
\end{enumerate}  

From these findings, we estimate that at least half, and possibly most, of our DTARPS-S Candidate catalog are true transiting exoplanets.  We can roughly compare this to the TOI list produced by the \TESS Science Office.  The dispositions listed by the \TESS Follow-up Observing Program Working Group in the NASA Exoplanet Archive \citep{ExoFop-TOI} are based on nonuniform, incomplete follow-up from a variety of groups and sources. As above (\S \ref{sec:recon}), we use the dispositions for only those objects that have had follow-up performed to gain the best insight possible for the survey. In the full DIAmante star sample, TOIs \citep{ExoFop-TOI} have the following dispositions: 628 Confirmed Planets, 833 False Positives or False Alarms, and 3,996 planetary candidates (including ambiguous planetary candidates) without clarifying disposition.  The precision of the TOI catalog at this time is thus $628 / (628 + 833) = 0.43$. The 10 arguments above give strong indications that the DTARPS-S sample consists of at-least-half true exoplanets. \textit{The TOI sample and the \TESS Year 1 DTARPS-S Catalog produced in Paper~II thus appear to have comparable purity} although this evaluation may be biased by the incompleteness of the spectroscopic confirmations.

We emphasize that the DTARPS-S study is entirely based on photometric properties of the host stars.  Spectroscopic follow-up of the candidates is necessary for confirmation of individual candidates and accurate inference of the \TESS exoplanetary population.

\section{The Neptune Desert Candidate Paradox}
\label{sec:Neptune_desert}

\subsection{DTARPS-S Candidates Compared with \TESS Objects of Interest}
\label{sec:Neptune_TOI}

Figure~\ref{fig:intro_PR} shows that both the DTARPS-S Candidate catalog and the \TESS Objects of Interest have significant fractions of candidates inside the hot Neptune desert as defined by \citet{Mazeh16}.  The DTARPS-S and TOI distributions differ in the proportion of hot Neptune candidates $-$ 50\% for DTARPS-S compared to 25\% for TOI $-$ but the distributions are not very dissimilar. However, when compared independently with the Confirmed Planet list from the NASA Exoplanet Archive that is largely based on the Kepler mission catalog, the DTARPS-S Candidates strongly overpopulate the hot Neptune Desert region (Figure \ref{fig:nep_desert_per_rad}). The DTARPS-S occurrence rates (\S \ref{sec:occ_rate}) agree with the \Kepler rates except in the region of the Neptune desert (especially the central Neptune desert).  In particular, DTARPS-S has an overabundance of candidates with periods between one and two days and radii between 4 and 8 $R_{\oplus}$. 

As only a portion of these candidates have been observed spectroscopically, we can consider the findings of probabilistic vetting studies.  \citet{Magliano23} seek `to perform a homogeneous and statistically controlled validation of the sample of hot Neptunes in the TESS data’.  They perform vetting analysis of 250 TOIs in the central region of the Kepler Neptune Desert region with visual examination of pixel and light-curve level data using the Discovery and Vetting of Exoplanets (DAVE) procedure followed by TRICERATOPS probabilistic estimation of astronomical False Positives.  They rank 75\% TOI central Neptune desert candidates as likely False Positives.  Focusing on 9 candidates with low FP probabilities, 2 are statistically validated upon further analysis. DTARPS-S has 55 candidates in the central desert region defined by \citet{Magliano23} for TOIs.

Their analysis depends on the quality of the construction of the TESS Objects of Interest list that rests on a complex amalgamation of three different analysis and classification systems \citep{Guerrero21}.  Our DTARPS-S analysis is a more straightforward procedure with quantified confusion matrices and completeness heat maps for the classification procedure (Paper I) and multifaceted qualitative procedures for vetting (Paper II).  The DTARPS-S procedure deters identification of eclipsing binaries (either in the target star or blended into the light curve extraction region) in two ways:
\begin{enumerate}
     \item Following the DIAmante project of \citet{Montalto20}, the Random Forest classifier is trained away from EBs by including thousands of simulated eclipsing binaries in the negative training set.  This leads to 100\% `specificity’ (the fraction of known FPs that are correctly classified as FPs) for simulated FPs in the validation set and 90\% specificity for previously identified FPs (Figs.\ 11 and 15 of Paper I).  The final classifier has a True Positive Rate of 92.5\% and a False Positive Rate of 0.43\% from the labeled training set.

     \item The post-classification vetting operation excludes many remaining FPs with four rejection criteria: large inferred planetary radii, differences in even- $vs.$ odd-transits, presence of secondary transits, and presence of out-of-transit curvature in folded light curves (\S2.3, Paper II).  While these are similar to other vetting procedures, we note that the out-of-transit curvature criterion should, in the hot Neptune region, remove all semi-detached EBs as well as many detached EBs with $P \lesssim 2$~d \citep{Prsa11}.  
\end{enumerate}
It is therefore quite possible that the True Positive rate for DTARPS-S candidates is higher than the 25\% success rate found by \citet{Magliano23} for TOIs in the hot Neptune regime.

\subsection{The \Kepler vs. \TESS Discrepancy} \label{sec:nep_desert_Kep_TESS}

An astrophysical explanation for the Neptune desert readily emerged: the extreme ultraviolet (EUV) emission produce by `saturated' magnetic activity during the first $\sim 100$ Myr of solar-type stars will cause rapid loss of inner planetary primordial (H and He) atmospheres \citep[][and others]{Lopez13, Owen13}. \citet{Lundkvist16} calculate that the H/He envelope of Neptune-sized objects would be stripped at insolation fluxes greater than 650 $S_{\oplus}$, leaving remnant super-Earth sized cores.  There is some debate that the atmospheric loss may occur on slower gigayear timescales; stellar EUV emission extends long after after the brief period of saturated dynamos \citep{King21}.  The upper edge of the Neptune desert zone appears to be more resistant to photoevaporation than the lower edge \citep{Ionov18, Vissapragada22}.

However, the observational evidence for the Neptune desert in transit surveys is not completely clear.  It is not entirely empty in the $Kepler$ dataset (Figure~\ref{fig:intro_PR}, top left panel): several dozen confirmed $Kepler$ planets lie near the super-Earth side and a dozen lie in the interior of the desert \citep{Berger18}.  But the planetary population emerging from the \TESS Objects of Interest (TOI), in both targeted and FFI observations, show a considerable population in the desert region (Figure~\ref{fig:intro_PR}, top right panel).  This is supported by the DIAmante study (M20) and our work here: the DTARPS-S Candidates catalog has 208 stars in the desert region defined by \citet{Mazeh16} and the Galactic Plane list has an additional 177 objects.  These are listed in Table~\ref{tab:neptunedesert.members} with detail,ls provided in the Machine Readable Tables of Paper~II.  If various \TESS-based studies are combined, hundreds of potential planets in the Neptune desert regime have been identified (Figures~\ref{fig:intro_PR}, \ref{fig:nep_desert_per_rad}, and \ref{fig:nep_des_comp}). 

Candidates emerging from analysis of \TESS photometric light curves can  suffer contamination from False Positive systems.  But a growing number of hot Neptune TOIs are being spectroscopic confirmed as true exoplanets.  These include TOI 132b \citep{Diaz20}, 332b \citep{Osborn23}, 442b \citep{Dreizler20}, 532b \citep{Kanodia21}, 674b \citep{Murgas21}, 824b \citep{Burt20}, 849b \citep{Armstrong20}, 908b \citep{Hawthorn23}, 969b \citep{LilloBox23}, 1272b \citep{Martioli22}, 1288b \citep{Knudstrup23}, 1347b \citep{Hord24}, 1408c \citep{Korth24}, 1410b \citep{Hord24}, 1696b \citep{Mori22},
1710b \citep{Konig22}, 1853b \citep{Naponiello23}, 2196b \citep{Persson22}, 2266b \citep{Parviainen24}, 2374b \citep{Hacker24}, 3071b \citep{Hacker24}, 3261b \citep{Nabbie24}, 4010b \citep{Kunimoto23}, 3261b \citep{Nabbie24}, 4479b \citep{EsparzaBorges22}, 5126b \citep{Fairnington24}, and 5398c \citep{Mantovan24}. 

 Hot Neptunes discovered outside of the TOI list include GJ 436b \citep{vonBraun12}, HAT-P-11b \citep{Bakos10}, HATS 37Ab \citep{Jordan20}, HATS 38b \citep{Jordan20}, HD 18599b \citep{Vines23},  K2 100b \citep{Barragan19}, Kepler 4b \citep{Borucki10}, Kepler 94b \citep{Marcy14}, Kepler 98b \citep{Marcy14}, LTT 9779b \citep{Jenkins20}, NGTS 4b \citep{West19}, NGTS 5b \citep{Eigmuller19}, NGTS 14Ab \citep{Smith21}, and WASP 156b \citep{Bourrier23}.  Others are emerging from photometric surveys of open clusters implying high occurrence rates in stars with ages $\lesssim 1$~Gyr  \citep{Fernandes23, Christiansen23}.  

 While most of these planets have been found along the edges of the \citet{Mazeh16} hot Neptune Desert, some have been confirmed in the central hot Neptune region. Although the central region of the desert is not well-defined, hot Neptunes lying away from the desert boundaries include: HATS 38b, K2 100b, LTT 9779b, NGTS 4b, TOI 132b, TOI 332b,  TOI 532b, TOI 824b, TOI 849b, TOI 1347b, TOI 1410b, TOI 2196b, TOI 3261b, TOI 1853b, TOI 2196b, TOI 3071, TOI 3261b, TOI 4010b, and TOI 4479b.  

Several groups have suggested that this milder deficit of hot Neptunes, especially near the borders of region, be called an `oasis' \citep{Murgas21}, `savanna' \citep[e.g.][]{Bourrier23, Szabo23, Kalman23}, or `ridge' \citep{CastroGonzalez24} as it is becoming increasingly clear that planets are present in this region even if there are mechanisms driving the planets out.  The statistical analysis discussed above by \citet{Magliano23} suggests that one-quarter are likely true planets  while another analysis of 30 similar TOIs estimates half are valid planets \citep{Mistry22}. 

Radial velocity surveys also indicate an occurrence rate $\sim$2 times greater than obtained from the \Kepler transit survey \citep[][and references therein]{Winn15}. \citet{Guo17} found that radial velocity surveys are biased towards high metallicity targets around which hot Jupiters and hot Neptunes are preferentially found \citep{Dai21}, but this did not fully account for the discrepancy. \citet{Moe21} found that the discrepancy might arise from suppression of larger planets around close binary stars. 

It may appear difficult to explain why the \TESS survey, mostly based on single-sector 27-day light curves, is more successful that the \Kepler survey, with 4-year light curves, in identifying easily detectable planetary signals.  \Kepler is sufficiently sensitive to detect Earths and super-Earths at short periods, so sub-Neptunes and Neptunes have very strong signals that cannot be missed.  We address this issue here focusing on the DTARPS-S catalog findings.

\subsection{DTARPS-S Hot Neptune Candidates \label{sec:nep_desert}}

\begin{figure}[th]
    \centering
    \includegraphics[width=0.49\textwidth]{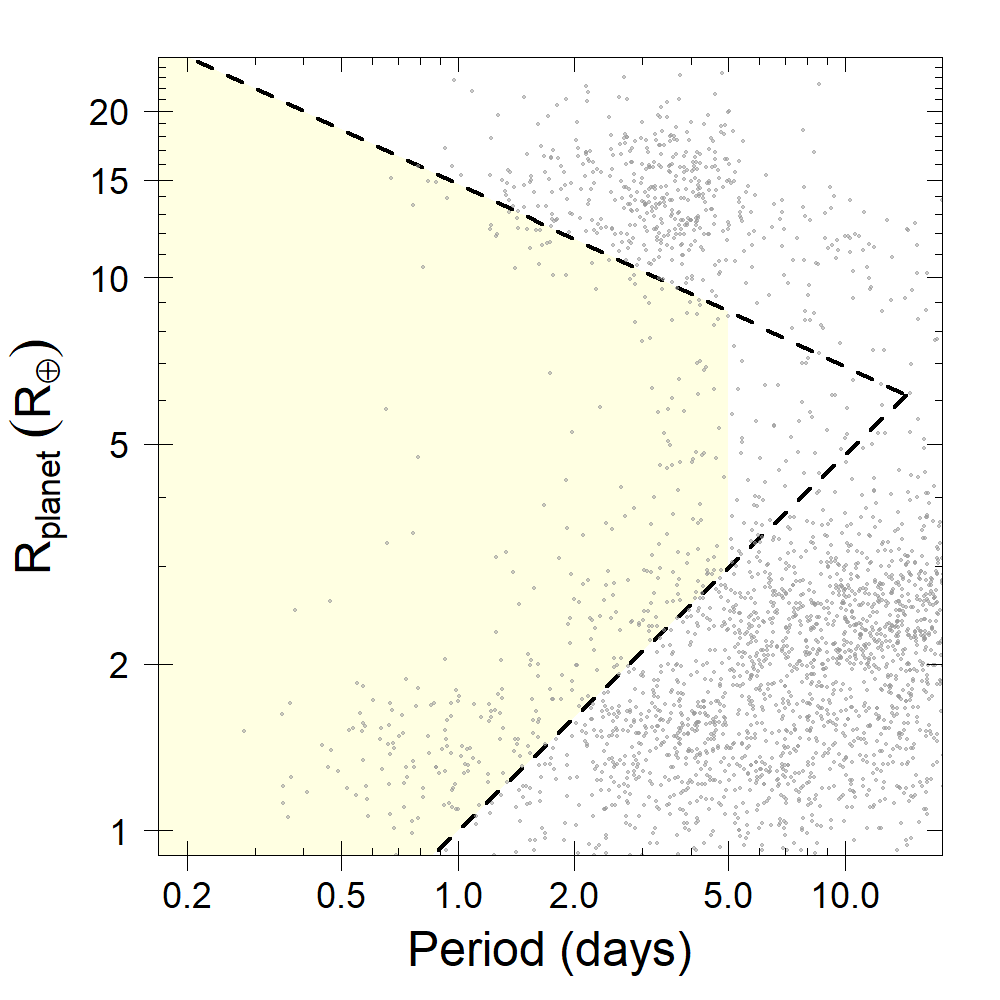}
    \includegraphics[width=0.49\textwidth]{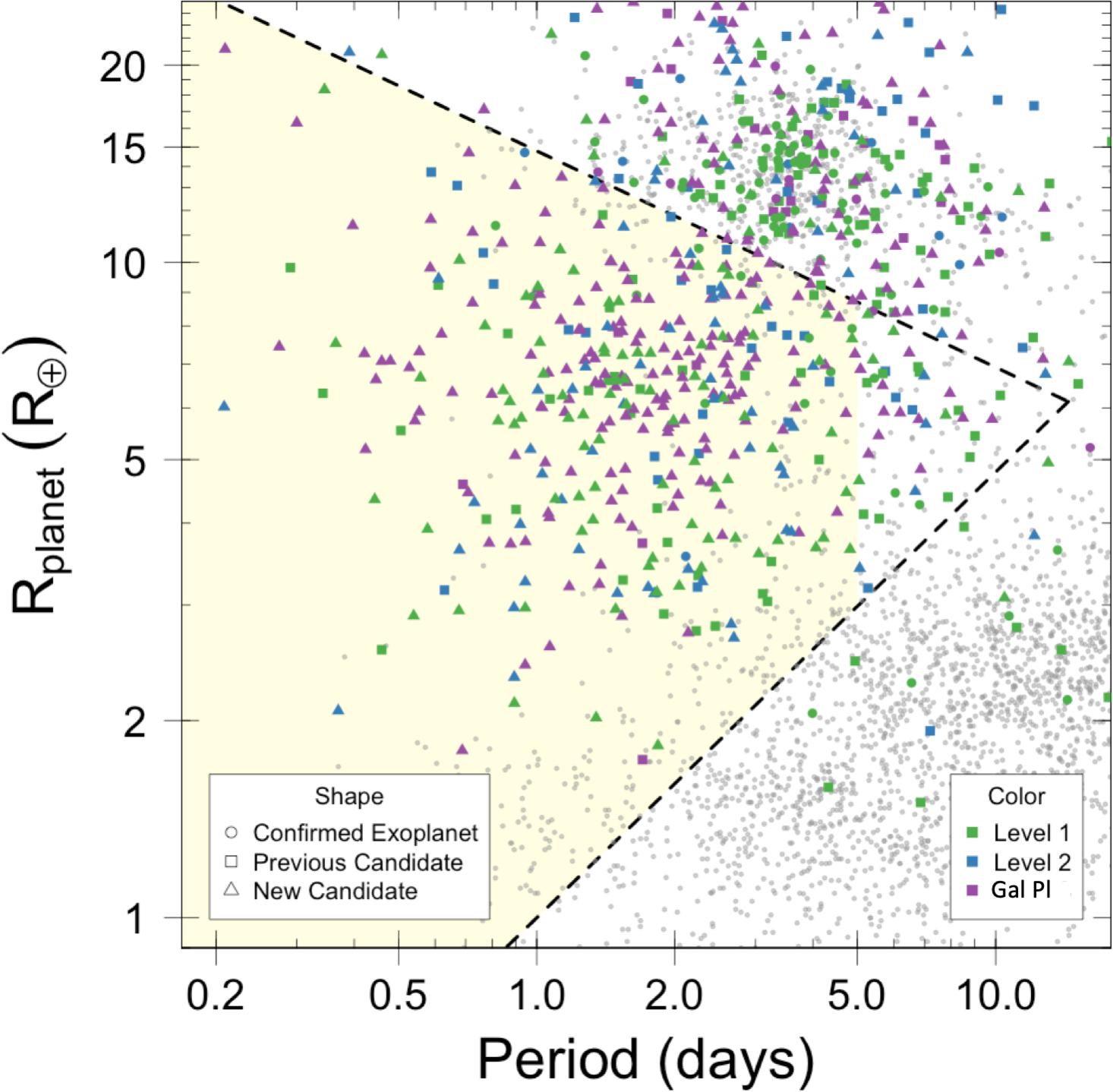}
    \caption{Detailed view of the period-radius diagram.  The left panel shown only the Confirmed Planets from the NASA Exoplanet Archive (light gray points).  The right panel adds the Confirmed Planets,  DTARPS-S Candidates and Galactic Plane objects.  The symbol of each point represents whether it is a Confirmed Planet (circle), previously identified planet candidate (square), or a newly identified candidate (triangle).  The Neptune desert, highlighted in yellow, is bounded by the dashed lines defined by \citet{Mazeh16}.  \label{fig:nep_desert_per_rad}}
\end{figure}

\begin{figure}[b]
    \centering
    \includegraphics[width=0.49\textwidth]{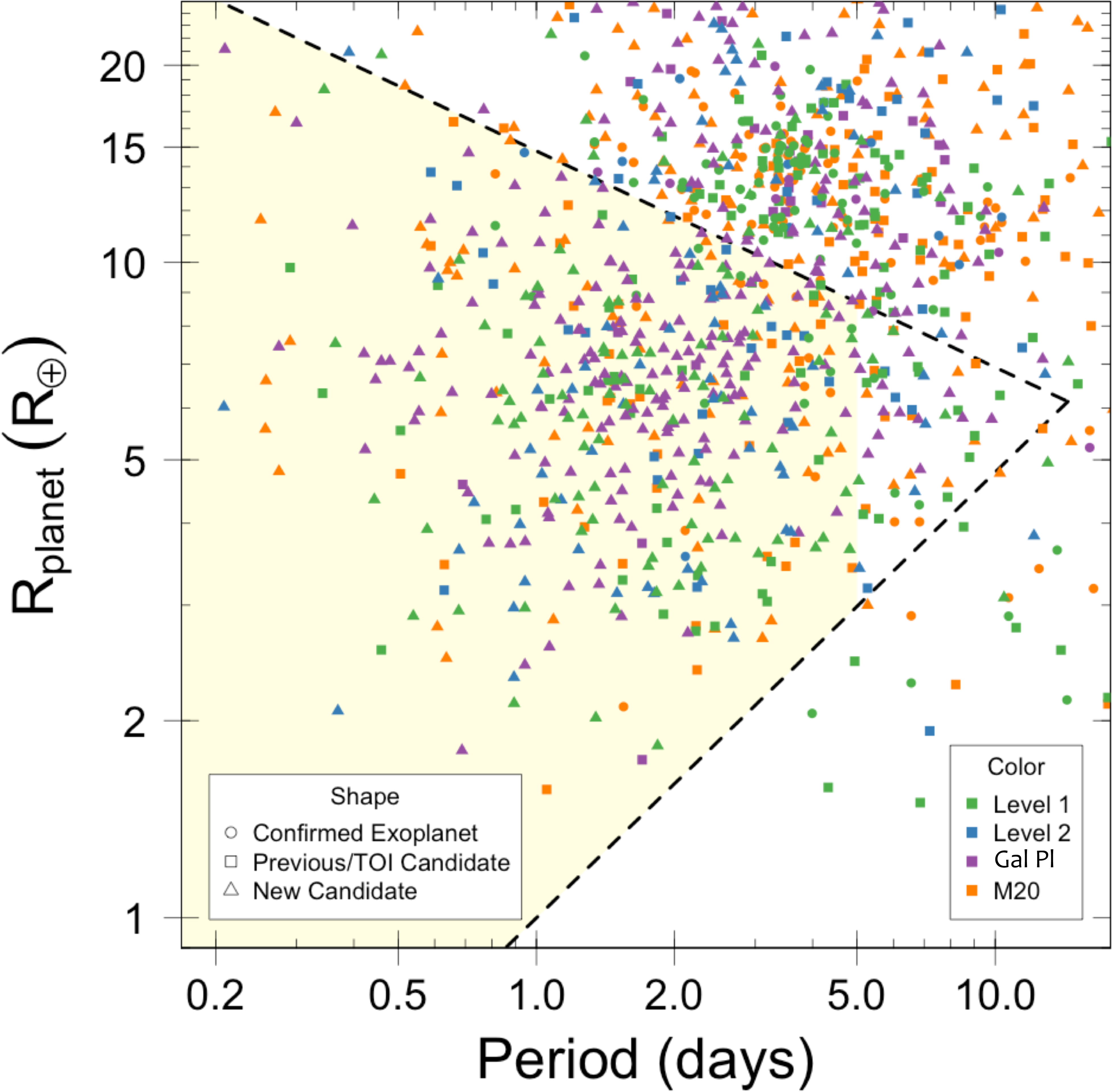}
    \includegraphics[width=0.49\textwidth]{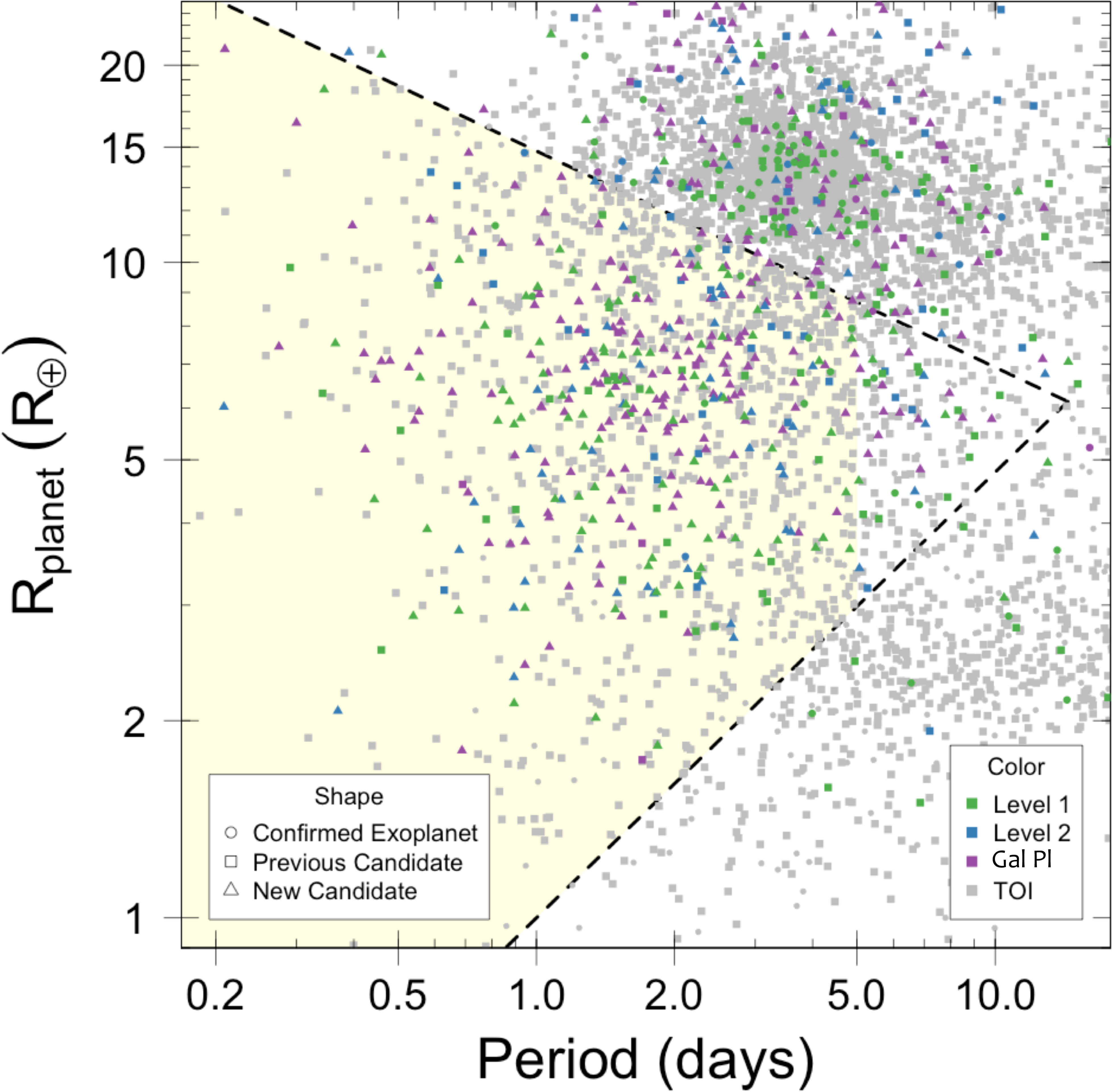}
    \caption{Neptune desert region in period-radius space for the M20 candidates (left) and the TOI list objects (right). The candidates identified in M20 are plotted in orange and the Confirmed Planets and planet candidates from the TOI list are plotted in gray. DTARPS-S candidates are colored by their disposition level. The symbol of each point represents whether it is a Confirmed Planet (circle), previously identified planet candidate (square) or a newly identified candidate (triangle). }
    \label{fig:nep_des_comp}
\end{figure}

Figure \ref{fig:nep_desert_per_rad} shows the DTARPS-S  period-radius plot in the Neptune desert region with more detail than in Figure~\ref{fig:intro_PR}. The boundaries in the planet period-radius diagram shown as dashed lines were derived by \citet{Mazeh16} from $Kepler$ data; their Neptune desert does not extend beyond 5 days.  The distribution show little evidence for a decrease in DTARPS-S candidates as one crosses the upper boundary of the Neptune desert.  Rather, a localized excess of candidates is present at the center of the desert around $6 < R < 8$~R$_\oplus$ and $1 < P < 3$~day.  Half of the DTARPS-S transit candidates, 385 of 772 objects (Table~\ref{tab:neptunedesert.members}), fall within the Neptune desert region defined by \citet{Mazeh16}.   Detailed figures (light curves, periodograms, folded light curves) and tabular information for individual DTARPS-S objects are available in the Figure Sets and Machine Readable Tables in Paper II.  

The DTARPS-S hot Neptune candidates include 12 Confirmed Planets and 62 previously identified planet candidates (Appendices A and B in Paper II).  In the full DIAmante data set of 0.9 million  light curves, there are 283 previously identified False Positives in the Neptune desert; however, the DTARPS-S analysis eliminated all but 24 of them as DTARPS-S Candidates.  This 92\% specificity rate for known False Positives is higher than the 76\% specificity rate for entire DTARPS-S Analysis List prior to vetting (\S10.5, Paper I) indicating that the DTARPS-S vetting process (Paper II) was effective in removing False Positives from the list emerging from the DTARPS-S statistical analysis (Paper I) in the Neptune desert. 

The sample of objects with follow-up observations to differentiate False Positives from true planetary transits is too inhomogeneous to reliably infer contamination estimations from the False Positive specificity rate.  False Positive rates for KOI and TOI objects are higher for candidates in the hot Neptune desert region than in the region outside of the hot Neptune desert, but these False Positive rate calculations are estimates too. Great caution needs to be exercised when making inferences using Confirmed Planets and Previously Identified False Positives due to the inhomogeneous nature of spectroscopic surveys, the different reporting rates, and the relative ease or difficulty of confirming different classes of planets and False Positives.

\begin{deluxetable}{rrrrrrrrrrrr} 
\label{tab:neptunedesert.members}
\tablefontsize{\footnotesize}
\tablecaption{Neptune Desert DTARPS-S Systems}
\startdata
& \\
\multicolumn{12}{c}{\bf DTARPS-S Candidates Catalog (N=208)} \\
   2$^a$\phn{}\phn{} &   5\phn{}\phn{}\phn{} &   6\phn{}\phn{}\phn{} &   7\phn{}\phn{}\phn{} &   8\phn{}\phn{}\phn{} &  11$^e$\phn{}\phn{} &  16$^b$\phn{}\phn{} &  17\phn{}\phn{}\phn{} &  19\phn{}\phn{}\phn{} &  21$^d$\phn{}\phn{} &  22$^f$\phn{}\phn{} &  23\phn{}\phn{}\phn{} \\ 
  24\phn{}\phn{}\phn{} &  25\phn{}\phn{}\phn{} &  26\phn{}\phn{}\phn{} &  29\phn{}\phn{}\phn{} &  30\phn{}\phn{}\phn{} &  32$^b$$^e$\phn{} &  34$^e$\phn{}\phn{} &  36\phn{}\phn{}\phn{} &  38$^b$\phn{}\phn{} &  39$^a$\phn{}\phn{} &  41$^b$\phn{}\phn{} &  42\phn{}\phn{}\phn{} \\ 
  44\phn{}\phn{}\phn{} &  45\phn{}\phn{}\phn{} &  46$^a$\phn{}\phn{} &  48$^e$\phn{}\phn{} &  50\phn{}\phn{}\phn{} &  53$^b$\phn{}\phn{} & 54$^b$\phn{}\phn{} &  57$^a$\phn{}\phn{} &  59\phn{}\phn{}\phn{} &  61$^b$\phn{}\phn{} &  64$^f$\phn{}\phn{} &  65\phn{}\phn{}\phn{} \\
  66$^e$\phn{}\phn{} &  68\phn{}\phn{}\phn{} &  70\phn{}\phn{}\phn{} &  73\phn{}\phn{}\phn{} &  75\phn{}\phn{}\phn{} &  76\phn{}\phn{}\phn{} &  79$^b$\phn{}\phn{} &  80\phn{}\phn{}\phn{} &  82\phn{}\phn{}\phn{} &  85\phn{}\phn{}\phn{} &  87\phn{}\phn{}\phn{} &  88$^b$\phn{}\phn{} \\ 
  90$^c$\phn{}\phn{} &  95\phn{}\phn{}\phn{} &   96\phn{}\phn{}\phn{} &  98\phn{}\phn{}\phn{} &  99\phn{}\phn{}\phn{} & 100\phn{}\phn{}\phn{} & 103$^a$\phn{}\phn{} & 107$^c$\phn{}\phn{} & 108\phn{}\phn{}\phn{} & 109\phn{}\phn{}\phn{} & 110\phn{}\phn{}\phn{} & 112\phn{}\phn{}\phn{} \\
 113\phn{}\phn{}\phn{} & 114\phn{}\phn{}\phn{} & 115$^d$\phn{}\phn{} & 116$^a$\phn{}\phn{} & 117$^d$$^e$\phn{} & 119\phn{}\phn{}\phn{} & 120\phn{}\phn{}\phn{}  & 122\phn{}\phn{}\phn{} & 123\phn{}\phn{}\phn{} & 124\phn{}\phn{}\phn{} &  125\phn{}\phn{}\phn{} & 131\phn{}\phn{}\phn{} \\
 134\phn{}\phn{}\phn{} & 137\phn{}\phn{}\phn{} & 138\phn{}\phn{}\phn{} & 147\phn{}\phn{}\phn{} & 150$^b$\phn{}\phn{} & 152\phn{}\phn{}\phn{} & 153\phn{}\phn{}\phn{} & 155$^e$\phn{}\phn{} &  156$^b$\phn{}\phn{} & 157\phn{}\phn{}\phn{} & 158\phn{}\phn{}\phn{} & 161\phn{}\phn{}\phn{} \\
 164$^a$\phn{}\phn{} & 168$^d$\phn{}\phn{} & 169$^f$\phn{}\phn{} & 173$^b$\phn{}\phn{} & 175$^e$\phn{}\phn{} & 176\phn{}\phn{}\phn{} & 177\phn{}\phn{}\phn{} & 178\phn{}\phn{}\phn{} & 180\phn{}\phn{}\phn{} & 186\phn{}\phn{}\phn{} & 188\phn{}\phn{}\phn{} & 189\phn{}\phn{}\phn{} \\
 191$^b$\phn{}\phn{} & 192$^a$$^e$\phn{} & 194$^b$\phn{}\phn{} & 195$^b$\phn{}\phn{} & 198\phn{}\phn{}\phn{} & 201\phn{}\phn{}\phn{} & 204\phn{}\phn{}\phn{} & 205$^b$$^e$\phn{} & 208$^a$\phn{}\phn{} & 209$^a$\phn{}\phn{} & 211$^b$\phn{}\phn{} & 213$^b$$^e$\phn{} \\
 215$^a$\phn{}\phn{} & 222\phn{}\phn{}\phn{} &  223$^d$\phn{}\phn{} & 225$^b$$^e$\phn{} & 228$^d$$^e$\phn{} & 229$^c$\phn{}\phn{} & 233$^d$\phn{}\phn{} & 235$^b$\phn{}\phn{} & 239\phn{}\phn{}\phn{} & 240\phn{}\phn{}\phn{} & 242\phn{}\phn{}\phn{} & 246\phn{}\phn{}\phn{} \\ 
 247\phn{}\phn{}\phn{} & 249\phn{}\phn{}\phn{} & 253\phn{}\phn{}\phn{} & 255\phn{}\phn{}\phn{} & 258\phn{}\phn{}\phn{} & 259$^b$\phn{}\phn{} & 261\phn{}\phn{}\phn{} & 264$^e$\phn{}\phn{} & 265$^d$\phn{}\phn{} & 271\phn{}\phn{}\phn{} &  272$^d$\phn{}\phn{} & 273\phn{}\phn{}\phn{} \\
 274$^e$\phn{}\phn{} & 276\phn{}\phn{}\phn{} & 281\phn{}\phn{}\phn{} & 284\phn{}\phn{}\phn{} & 289\phn{}\phn{}\phn{} & 290$^b$\phn{}\phn{} & 291$^d$\phn{}\phn{} & 292$^a$\phn{}\phn{} &  298\phn{}\phn{}\phn{} & 303\phn{}\phn{}\phn{} & 304\phn{}\phn{}\phn{} & 305\phn{}\phn{}\phn{} \\
 310\phn{}\phn{}\phn{} & 313\phn{}\phn{}\phn{} & 315$^e$\phn{}\phn{} & 322$^e$\phn{}\phn{} & 323\phn{}\phn{}\phn{} & 326\phn{}\phn{}\phn{} &   328$^b$\phn{}\phn{} & 329\phn{}\phn{}\phn{} & 330$^b$$^e$\phn{} & 331\phn{}\phn{}\phn{} & 332\phn{}\phn{}\phn{} & 333\phn{}\phn{}\phn{} \\
 335$^c$$^e$\phn{} & 337\phn{}\phn{}\phn{} & 338$^c$$^e$\phn{} & 339$^e$\phn{}\phn{} &  343\phn{}\phn{}\phn{} & 344\phn{}\phn{}\phn{} & 345\phn{}\phn{}\phn{} & 349\phn{}\phn{}\phn{} & 351\phn{}\phn{}\phn{} & 354$^e$\phn{}\phn{} & 356\phn{}\phn{}\phn{} & 359\phn{}\phn{}\phn{} \\
 363\phn{}\phn{}\phn{} & 364$^d$\phn{}\phn{} & 367\phn{}\phn{}\phn{} & 370\phn{}\phn{}\phn{} & 371\phn{}\phn{}\phn{} & 374\phn{}\phn{}\phn{} & 378\phn{}\phn{}\phn{} & 379$^e$\phn{}\phn{} & 380\phn{}\phn{}\phn{} & 382\phn{}\phn{}\phn{} & 383\phn{}\phn{}\phn{} & 384\phn{}\phn{}\phn{} \\ 
  385\phn{}\phn{}\phn{} & 386$^b$\phn{}\phn{} & 388$^c$\phn{}\phn{} & 389\phn{}\phn{}\phn{} & 390$^e$\phn{}\phn{} & 391\phn{}\phn{}\phn{} & 392$^b$$^e$\phn{} & 393\phn{}\phn{}\phn{} & 396\phn{}\phn{}\phn{} & 403\phn{}\phn{}\phn{} & 404\phn{}\phn{}\phn{} & 407$^b$$^e$\phn{} \\
  410$^a$\phn{}\phn{} & 412\phn{}\phn{}\phn{} & 413\phn{}\phn{}\phn{} & 417\phn{}\phn{}\phn{} & 419\phn{}\phn{}\phn{} & 420$^b$\phn{}\phn{} & 424\phn{}\phn{}\phn{} & 426$^b$\phn{}\phn{} & 434$^b$\phn{}\phn{} & 435\phn{}\phn{}\phn{} & 439\phn{}\phn{}\phn{} & 440\phn{}\phn{}\phn{} \\ 443\phn{}\phn{}\phn{} & 452\phn{}\phn{}\phn{} & 453\phn{}\phn{}\phn{} & 457\phn{}\phn{}\phn{} \\
\enddata
\end{deluxetable}

\begin{deluxetable}{rrrrrrrr} 
\tablefontsize{\footnotesize}
\setcounter{table}{1}    
\startdata
& \\
\multicolumn{8}{c}{\bf Galactic Plane list (N=177)} \\
    883943\phn{}\phn{} &   1605476$^g$$^h$ &   4616346$^b$\phn{} &   4784880\phn{}\phn{} &   5108864\phn{}\phn{} &   6432352\phn{}\phn{} &   9432774\phn{}\phn{} & 10320635\phn{}\phn{} \\
    11232328\phn{}\phn{} &  13139556\phn{}\phn{} &   13739039\phn{}\phn{} &  19519368$^b$\phn{} &  19937775$^g$\phn{} &  24830294\phn{}\phn{} &  25505404\phn{}\phn{} & 25585464\phn{}\phn{} \\
    25585493\phn{}\phn{} &  26047594\phn{}\phn{} &  34844046\phn{}\phn{} &  36209863$^f$$^g$ & 36916955\phn{}\phn{} &  48605325\phn{}\phn{} &  50375091\phn{}\phn{} &  52297327\phn{}\phn{} \\
  53588284\phn{}\phn{} &  54387451\phn{}\phn{} &  60349516\phn{}\phn{} &  61498024\phn{}\phn{} &  61755686$^g$\phn{} &  62759308\phn{}\phn{} & 64326510\phn{}\phn{} &  64573956$^g$\phn{} \\ 
  65751396$^g$$^h$ &  66306843\phn{}\phn{} &  68010197\phn{}\phn{} &  71572657\phn{}\phn{} &  71728605\phn{}\phn{} &  73043215\phn{}\phn{} &  73191957\phn{}\phn{} &  79143083$^b$\phn{} \\
  79938660\phn{}\phn{} &  80045246\phn{}\phn{} &  80135726\phn{}\phn{} &  80556961\phn{}\phn{} &  80709429$^g$\phn{} &  81591410$^g$\phn{} &  81676788\phn{}\phn{} &  81739674\phn{}\phn{} \\
  81746258\phn{}\phn{} &  89192798\phn{}\phn{} &  93517731\phn{}\phn{} &  94239926\phn{}\phn{} &  94695074$^g$\phn{} &  99249755\phn{}\phn{} &  99771082\phn{}\phn{} &  99935720\phn{}\phn{} \\
  120272891\phn{}\phn{} & 120331990\phn{}\phn{} & 120616194\phn{}\phn{} & 123886171\phn{}\phn{} &  124244886\phn{}\phn{} & 124498746\phn{}\phn{} & 125018207\phn{}\phn{} & 125201129\phn{}\phn{} \\
  125640034$^g$$^h$ & 139775416\phn{}\phn{} & 141462999\phn{}\phn{} & 141831460\phn{}\phn{} & 142123542\phn{}\phn{} & 142363812\phn{}\phn{} &  143350972$^d$\phn{} & 143525808\phn{}\phn{} \\
  143831261\phn{}\phn{} & 146323580\phn{}\phn{} & 148938758\phn{}\phn{} & 153610688\phn{}\phn{} &   167974648\phn{}\phn{} & 168115016\phn{}\phn{} & 168343381\phn{}\phn{} & 168598493\phn{}\phn{} \\
  170112990\phn{}\phn{} & 175320880\phn{}\phn{} & 176242777\phn{}\phn{} & 176380570\phn{}\phn{} &   177068644\phn{}\phn{} & 177405795$^g$$^h$ & 177411679\phn{}\phn{} & 177722855$^d$\phn{} \\
  177895571\phn{}\phn{} & 178120324\phn{}\phn{} &  178265008\phn{}\phn{} & 178580001$^g$\phn{} & 
  179159972$^d$\phn{} & 187567207\phn{}\phn{} & 187919451\phn{}\phn{} & 200516718\phn{}\phn{} \\ 200600277\phn{}\phn{} & 206897666\phn{}\phn{} & 219382473$^g$\phn{} & 220294417$^g$\phn{} & 231159377$^g$$^h$ & 231383819\phn{}\phn{} & 231385335\phn{}\phn{} & 231387465\phn{}\phn{} \\
 232395760$^f$\phn{} & 234047033$^g$\phn{} & 234091431\phn{}\phn{} & 234146625$^g$\phn{} &  234406675\phn{}\phn{} & 234830667\phn{}\phn{} &  235117667$^g$\phn{} & 235183588\phn{}\phn{} \\
 235507238\phn{}\phn{} & 235548135\phn{}\phn{} & 237566605\phn{}\phn{} & 237594977\phn{}\phn{} &  247830747\phn{}\phn{} & 250464109\phn{}\phn{} & 256994765\phn{}\phn{} & 257167116$^g$\phn{} \\
 262414864\phn{}\phn{} & 262605715\phn{}\phn{} & 264630054\phn{}\phn{} & 265045591$^g$$^h$ & 265446888$^g$\phn{} & 265905545$^g$\phn{} & 266009692$^g$\phn{} & 266657385$^g$$^h$ \\
 266741562\phn{}\phn{} & 269085762\phn{}\phn{} &  271098608\phn{}\phn{} &  271374913$^b$\phn{} & 279640579$^g$\phn{} & 280115402$^g$\phn{} & 280615140\phn{}\phn{} & 281907361\phn{}\phn{} \\
 282387432\phn{}\phn{} & 282510346\phn{}\phn{} & 282996918\phn{}\phn{} &  292068999\phn{}\phn{} &  295516295\phn{}\phn{} & 296179330$^g$\phn{} & 308051471\phn{}\phn{} & 316877482$^g$$^h$ \\
 317022315$^b$\phn{} & 318662985\phn{}\phn{} & 318899586\phn{}\phn{} &  319014919\phn{}\phn{} & 322512607$^g$$^h$ & 332390253\phn{}\phn{} &  405116473\phn{}\phn{} & 409258019\phn{}\phn{} \\
 409377933\phn{}\phn{} & 413166353\phn{}\phn{} & 415080840\phn{}\phn{} &  415404805\phn{}\phn{} & 415559926\phn{}\phn{} & 422022095\phn{}\phn{} & 422899417\phn{}\phn{} & 432068814\phn{}\phn{} \\
 434400698\phn{}\phn{} & 437935798\phn{}\phn{} & 438429401\phn{}\phn{} &  441131571\phn{}\phn{} & 443122499\phn{}\phn{} & 443961200\phn{}\phn{} & 714417948\phn{}\phn{}  \\
\enddata
\tablenotetext{a}{Confirmed planet \quad $^b$ Planetary candidate \quad $^c$ Ambiguous planet candidate \quad $^d$ False Positive \quad  $^e$ DIAmante candidate \quad $^f$ Bright host star (HD, CD) \quad $^g$ Ultra Short Period candidate ($P < 1.0$ day) \quad $^h$ Extreme Ultra Short Period candidate ($P < 0.5$ day)}
\end{deluxetable}

\subsection{Possible Sources of Contamination Among DTARPS-S Hot Neptunes}
\label{sec:Neptune_contam}

The abundance of DTARPS-S candidates in the hot Neptune desert could come from a number of different sources; the DTARPS-S methodology could be flawed so that it skews the resulting candidate list to create a significantly different distribution of candidates than is found using more traditional methods of planet candidate identification, there could be a high fraction of astrophysical contaminants in the DTARPS-S catalog that were not accounted for during the vetting process such as rapidly rotating star spots, the fundamental differences between the \TESS stellar host sample and the \Kepler sample may have an underlying difference in the occurrence rates of planets due to stellar age or other effects, there may be other astrophysical explanations on the order of small timescales that caused planets to be rejected on the long baseline \Kepler timescales but not the short baseline \TESS timescales, the vetting accuracy may vary as a function of period and radius, there may be a larger population of hot Neptunes than previously known due to a lack of robust candidate follow-up. 

We discuss many of these possibilities below. However, a robust analysis is hampered by the paucity of spectroscopic follow-up studies of previously identified candidates.  The vetting accuracy of both \TESS and \Kepler candidates are widely accepted to vary as a function of period and radius, but there is no robust analysis of how well the vetting analysis either rejects false positives or retains true planets.  Like DTARPS-S, \Kepler has posted the results of injections into the \textit{Robovetter} pipeline.  These injection recoveries don't account for the later stages of human vetting. Often the best tool available to measure vetting accuracy is by using reported results of follow-up observations performed on candidates.  The reported follow-up results for candidates are dominated by observational biases.  However, they are the best available measure of the likely results of following up a similar exoplanet candidate signal.

 We attempted to analyze the effectiveness of our vetting process in different radius-period spaces.  However, with only $\sim$ 250 confirmed planets and previously identified false positives in our entire DTARPS-S analysis list (see Paper I), there is not a large enough sample to derive false positive rates in different regions of the radius-period diagram.  This issue is addressed more fully in \S \ref{sec:occ_rate} and \S \ref{sec:occ_method} where we present very preliminary occurrence rates from DTARPS-S and compare them with \Kepler occurrence rates. It is extremely likely that the DTARPS-S false positive rate varies as a function of both candidate period and radius, just as the \TESS and \Kepler surveys.

 The true explanation for the proportionally overabundance of hot Neptune DTARPS-S candidates is likely a combination of all of the aforementioned factors. While it is fundamentally unquantifiable without robust observational follow-up, the varying false positive rate of candidates across the period-radius space could possibly be a very strong contributor to the strong proportion of hot Neptune candidates.  

\textbf{Errors in DTARPS-S Methodology}  ~~  
The DTARPS-S methodology has three stages: selecting, extracting and pre-processing $\sim 1$ million stars from the \TESS Year 1 survey (M20); applying the ARPS ARIMA models, TCF periodogram, and Random Forest classifier to identify potential planetary transits with high recall (Paper I); and applying well-established vetting procedures to reduce False Alarms and False Positives and improve precision (Paper II).  While the injection sample for the Random Forest classifier was drawn from the \Kepler planet sample, due to the use of the Adaptive Neighbor SMOTE algorithm \citep{smotefamily}, there were some injected planetary signals in the Neptune desert used for the training set teaching the classifier what signals in this region would look like.  The training sample also had many injected eclipsing binary signals with periods spanning the hot Neptune desert region to train the classifier away from false positives (Paper I). The list of candidates from the classifier were vetted in a similar manner to candides from the \TESS and \Kepler pipelines (Paper II).

While it is possible an overabundance of Neptune desert objects in the DTARPS-S catalog is a byproduct of errors in these analysis steps, evidence does not support this.  First, the TOI list that is based on different methodologies has a considerable population in the Neptune desert region of the period-radius diagram (Figure~\ref{fig:intro_PR}, top right panel). There over 800 TOI planet candidates in the Neptune desert region of which $\simeq 100$ are Confirmed Planets (\S\ref{sec:nep_desert_Kep_TESS} above). Second, the DIAmante planet candidates also populate the Neptune desert region with period-radius distribution similar to the DTARPS-S candidates.  The relationships between the TOI, M20 and DTARPS-S candidate lists are shown in Figure~\ref{fig:nep_des_comp}.  The three \TESS analysis efforts are all compatible with a profusion of hot Neptune candidates,  including some in the central desert region.

But over half of the False Positives and False Alarms reported for the TOI list lie in the Neptune desert.\footnote{These fractions are based on incomplete follow-up reports on the Community TOI Web site and may not represent the true population contamination rate.} The fact that DTARPS-S identifies many candidates in the Neptune desert yet achieves a lower false discovery rate$^4$ (79\% for TOI and 57\% KOI) suggests that the careful selection of candidates by the DTARPS-S Random Forest classifier and the vetting process were able to remove a high fraction of False Positive contaminants from the candidate list.   In particular, the DTARPS-S vetting step removing folded light curves with out-of-transit curvature will remove most short-period eclipsing binaries where tidal distortions and mutual stellar illumination must be present.

If the DTARPS-S methodology was fundamentally flawed (in terms of either training set or vetting analysis), it would be expected that the candidate distribution would be warped or distorted compared to the distribution of candidates seen by other surveys.  This single sector analysis of \TESS data in the DTARPS-S program prevents us from having a significant recall rate of planetary signals with radii less than 4 $R_\oplus$. Looking at the right panel of Figure \ref{fig:nep_des_comp} for candidates only with periods greater than 4 $R_\oplus$, the distribution of DTARPS-S candidates follows the distribution of \TESS candidates in the Neptune desert.  There is a noticeable lack of hot Jupiters in the DTARPS-S candidates compered with \TESS candidates that we believe is due to our strong training set of eclipsing binaries that strongly attenuated the passing of signals with larger depths. The DTARPS-S catalog hot Neptunes are notable for their proportionality in the candidate catalog, not their presence.

\begin{figure}[b]
    \centering
    \includegraphics[width=0.49\textwidth]{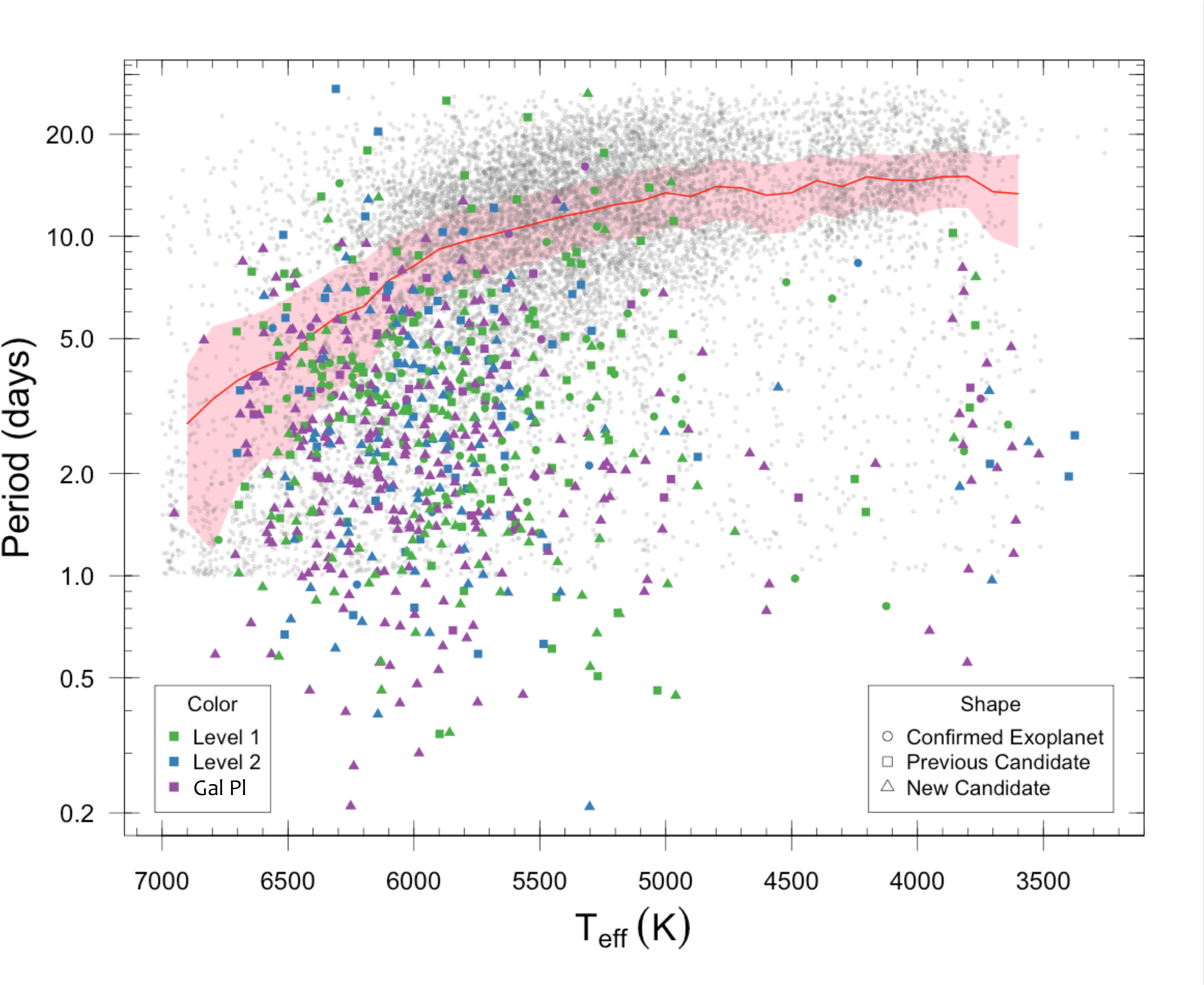}
    \includegraphics[width=0.49\textwidth]{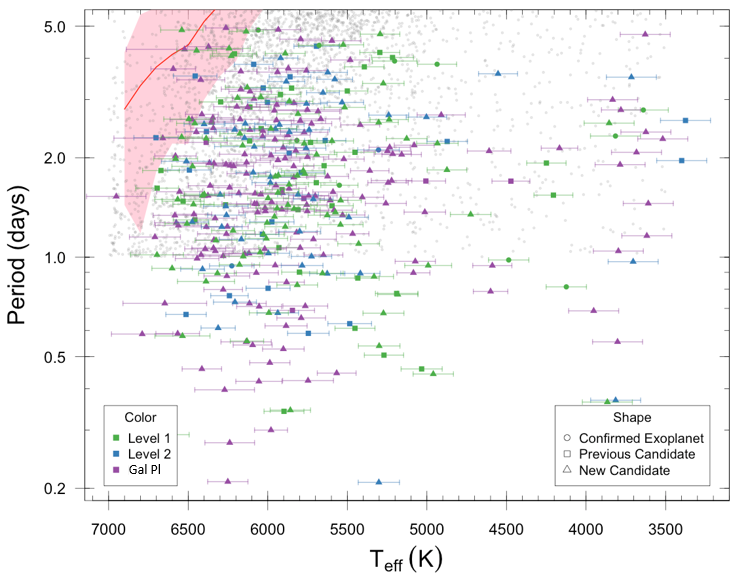}
    \caption{Period of confirmed planets and DTARPS-S candidates as a function of  effective temperature of the host star.  The gray points in the background are the rotation periods for the \Kepler stars and the red shaded area indicates the average stellar rotation period with 1$\sigma$ error bars around the red line measured by \citet{Nielsen13}.  The entire region is shown in the left plot, the right plot shows the candidates with periods $P < 5$ days (most of which are in the Hot Neptune Desert) with error bars on their stellar host temperatures.  The DTARPS-S candidate symbols represent previously confirmed planets (circle), previously identified planet candidates (square), and  new planetary candidates (triangle). }
    \label{fig:st_rot}
\end{figure}

\textbf{Contamination by Rotating Spotted Stars} ~~ All transit detection procedures must differentiate between quasi-periodic variations from rotationally modulated spotted stars and the strictly periodic variations from planetary transits. Contamination from stellar rotation for the Neptune Desert with periods $P < 5$ days  requires rapid rotation characteristic of some F stars and very young solar-type stars.  Pre-main sequence and ZAMS stars will only appear in the DTARPS-S Galactic Plane list and can not be significant contaminants in the high Galactic latitude DTARPS-S Candidates catalog. But F stars are common in the DTARPS-S input data set.  

The average rotation period from \Kepler stars are plotted as a red line in Figure \ref{fig:st_rot} as a function of stellar effective temperature based on rotation measurements by \citet{Nielsen13}.  The red shaded area span the upper and lower 34th percentile values from the median.  Only 36 DTARPS-S Candidates in the Neptune desert overlap the red shaded area in Figure \ref{fig:st_rot}; most have periods shorter than rotating main sequence stars. The DTARPS-S candidates overlapping the red shaded region in Figure \ref{fig:st_rot} may be contaminated by rotational star spots. The overlap suggests that no more than 10\% of DTARPS-S periodicities in the Neptune desert can be attributed to star spots.

\begin{figure}[t]
    \centering
    \includegraphics[width=0.49\textwidth]{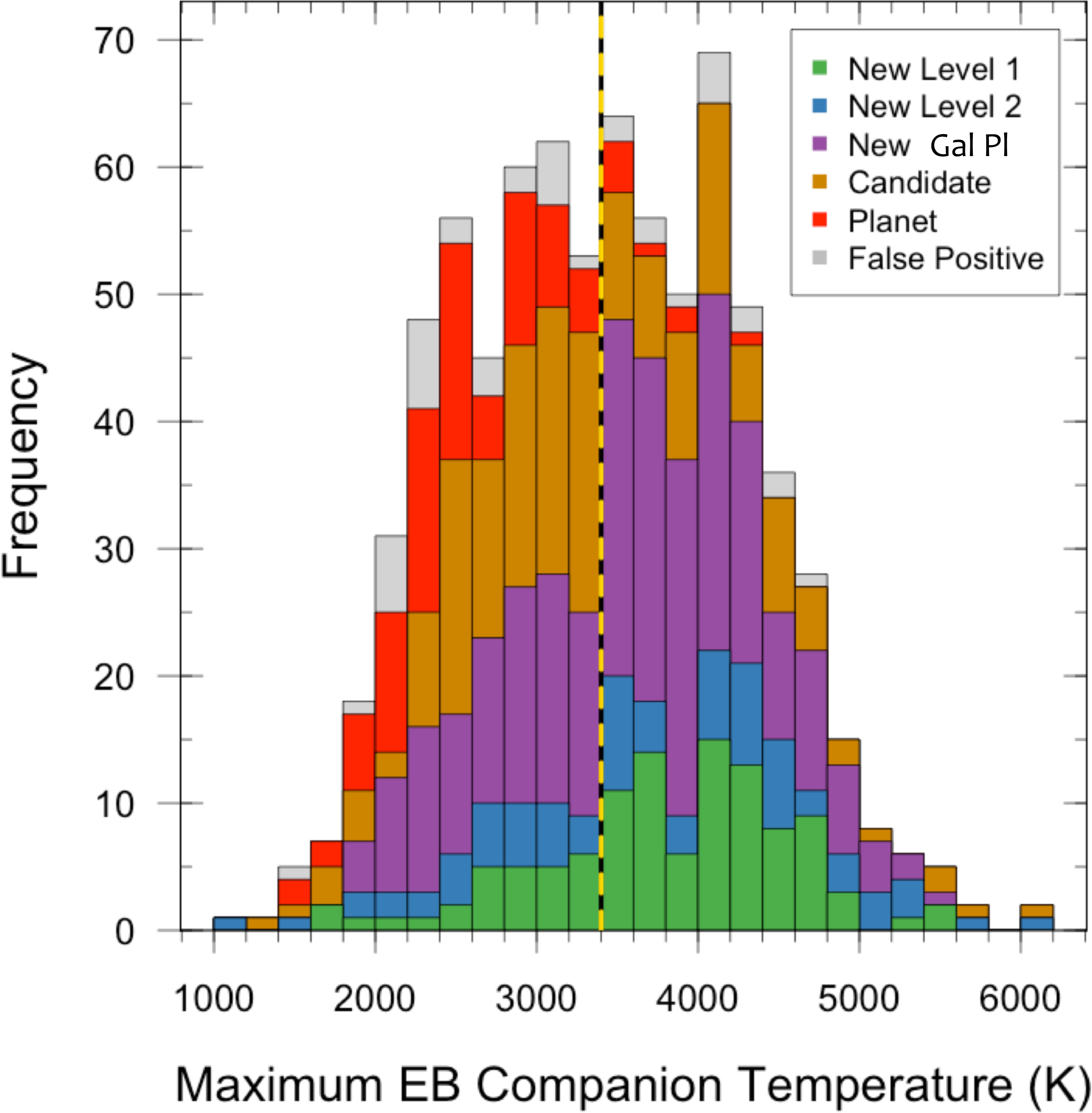}
    \includegraphics[width=0.49\textwidth]{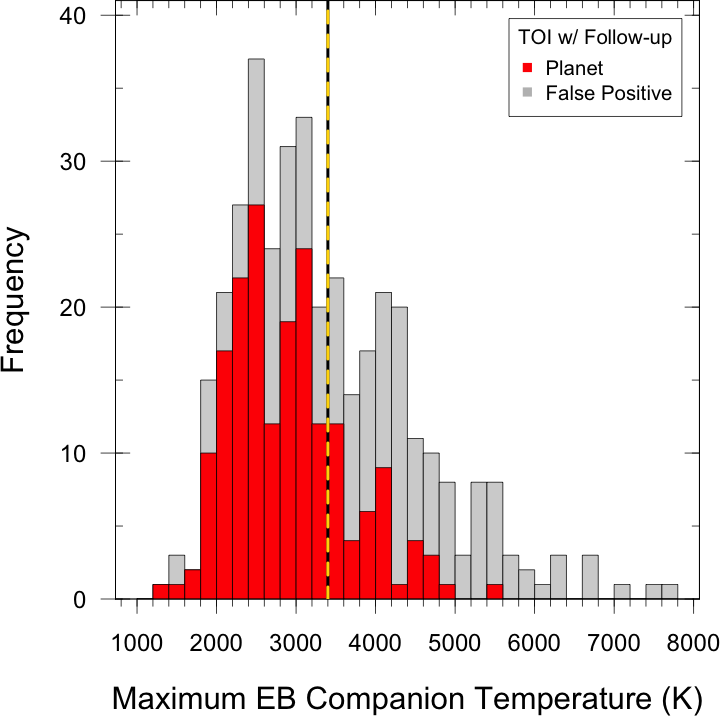}
    \caption{Maximum effective temperature of the smaller star if the DTARPS-S candidate (left) were an unequal mass EB system and if the TOI systems with follow-up (right) were unequal mass EB systems. Each bar of the histogram is colored to show the labels. Gray represents previously identified False Positives, red represents Confirmed Planets from the NASA Exoplanet Archive, and gold represents previously identified planetary candidates. Newly identified are colored by their disposition level: Level 1 (green) and Level 2 (blue) DTARPS-S Candidates or Galactic Plane list (purple). The temperature of a M3 dwarf is marked by the dashed black and yellow line in each histogram. }
    \label{fig:max_temp}
\end{figure}

\textbf{Contamination from EBs with Cool Companions} ~~
The light curves for EBs of stars with two different temperatures and radii are often characterized by a secondary eclipse whose depth is noticeably shallower than the primary eclipse. Assuming both stars are on the main sequence and have an eccentricity of zero, the ratio of the secondary to primary eclipse depth is proportional to $(T_2/T_1)^4$ where $T_2$ is the effective temperature of the cooler star \citep{Armstrong14}.  A vetting step was designed to remove such light curves from the DTARPS-S candidate lists by comparing the depths of alternate transit signals to each other (even/odd transit test; \S2.3 in Paper II).  

However, if the secondary is an M-type star, the eclipse may be sufficiently weak that it would not be identified in the light curve noise. These low mass EB (EBLM) systems are well known contaminants of exoplanet surveys and require radial velocity follow-up to discriminate from planetary systems \citep{Triaud17}.  By setting the depth of the secondary transit to half of the spread of the out-of-transit light curve, we can estimate the maximum temperature of a stellar companion masquerading as a planetary transit (Figure \ref{fig:max_temp}).  

EBLMs typically have F and G primary stars and M secondary stars.  If we set a lower temperature limit of 3,400 K (M3 star), we can estimate an upper limit on the fraction of EBLMs in the DTARPS-S Candidate sample. Figure \ref{fig:max_temp} gives the maximum temperature of a secondary companion that would have gone unnoticed during the DTARPS-S vetting process. A high maximal temperature for a secondary companion indicates either a noisy light curve or a small primary depth for the DTARPS-S candidate. Most of the previously identified False Positives (gray) and Confirmed Planets (red) in the DTARPS-S candidate list, indicating that existing spectroscopic characterization is poor for cool M star companions. 

The right panel of Figure \ref{fig:max_temp} shows a similar analysis performed on TOI objects that had follow-up observations performed (that is, with TFOP disposition of False Positive, False Alarm, Confirmed Planet or Known Planet). It shows about three-quarters of the TOI objects whose theoretical maximum effective temperature for a stellar companion exceeds an M3 dwarf are likely to be False Positives.  Adopting the false discovery rate from the TOIs for the DTARPS-S candidates and assuming therefore that at most three-quarters of the DTARPS-S Candidates whose maximum temperature of a stellar companion is greater than or equal to a M3 dwarf is indeed a EBLM, then we estimate that up to 30$-$40\% of DTARPS-S candidates could be EBLMs.  We note, however, that only five known EBLM systems are found in the 772 DTARPS-S Candidate catalog (Appendices A and B, Paper II).  This suggests that the true contamination rate is substantially lower than this upper limit.

\textbf{Summary} ~~
We tentatively conclude that DTARPS-S methodology and rotating spotted stars are not strong contributors of contamination among DTARPS-S hot Neptunes, while EBLMs could be responsible for a significant fraction.  But even pessimistic estimates suggest that over half of the reported DTARPS-S hot Neptunes are true transiting planets.  Again we stress that the vetting of candidates is known in other surveys to vary widely across period radius space and it is likely also varying in the DTARPS-S catalog. While a potentially significant contributor of contamination, we find no strong evidence of an enhanced false positive rate given the extremely limited information in non-uniform spectroscopic follow-up efforts hampered by observational biases (\S \ref{sec:cand_pure}.

\subsection{Possible Astrophysical Explanations for the \Kepler-\TESS Discrepancy} \label{sec:nep_explain}

\subsubsection{Planetary age effect}
\label{sec:plan_age}

The `Neptune desert’ is remarkably evident in the sample of confirmed Kepler transiting exoplanets (Figure~\ref{fig:intro_PR}, left panel).  The Kepler stellar sample, by virtue of field location above the Galactic Plane, has a wide range of ages; most have ages $2-10$ Gyr with median around 4.5 Gyr \citep{Berger20}. Analysis of Gaia mission kinematics indicates that most \TESS G-type stars (but not K-type stars) have ages $\lesssim 2$~Gyr with a more youthful distribution than \Kepler G-type stars  \citep{Sagear24}.  Importantly, recent studies of younger open clusters near the Galactic Plane indicate hot Neptunes are more prevalent at ages $<1$ Gyr.  \citet{Fernandes23} report a hot Neptune occurrence rate of 93\%$\pm$38\% from TESS observations of 5 clusters with ages $15-450$ Myr and \citet{Christiansen23} report an occurrence rate of 79\%-107\% in the $600-800$ Myr Praesepe cluster from the K2 mission, several times the rate seen in K2 field stars.  As summarized in section \S\ref{sec:nep_desert_Kep_TESS} above, numerous recent studies are also confirming hot Neptunes in field star TOI candidates observed with TESS.  Other TESS-based evidence for more hot Neptune-mass planets orbiting younger stars has been reported by \citet{Vach24}. 

These empirical results imply that the earlier astrophysical calculations of rapid atmospheric photoevaporation of Neptune atmospheres within $\sim 0.1$~Gyr \citep[][and related studies]{Lopez13, Owen13} need to be revised so that atmospheres are still commonly present at ages $\sim 1$~Gyr.

\begin{figure}[b]
    \centering
    \includegraphics[width=0.45\textwidth]{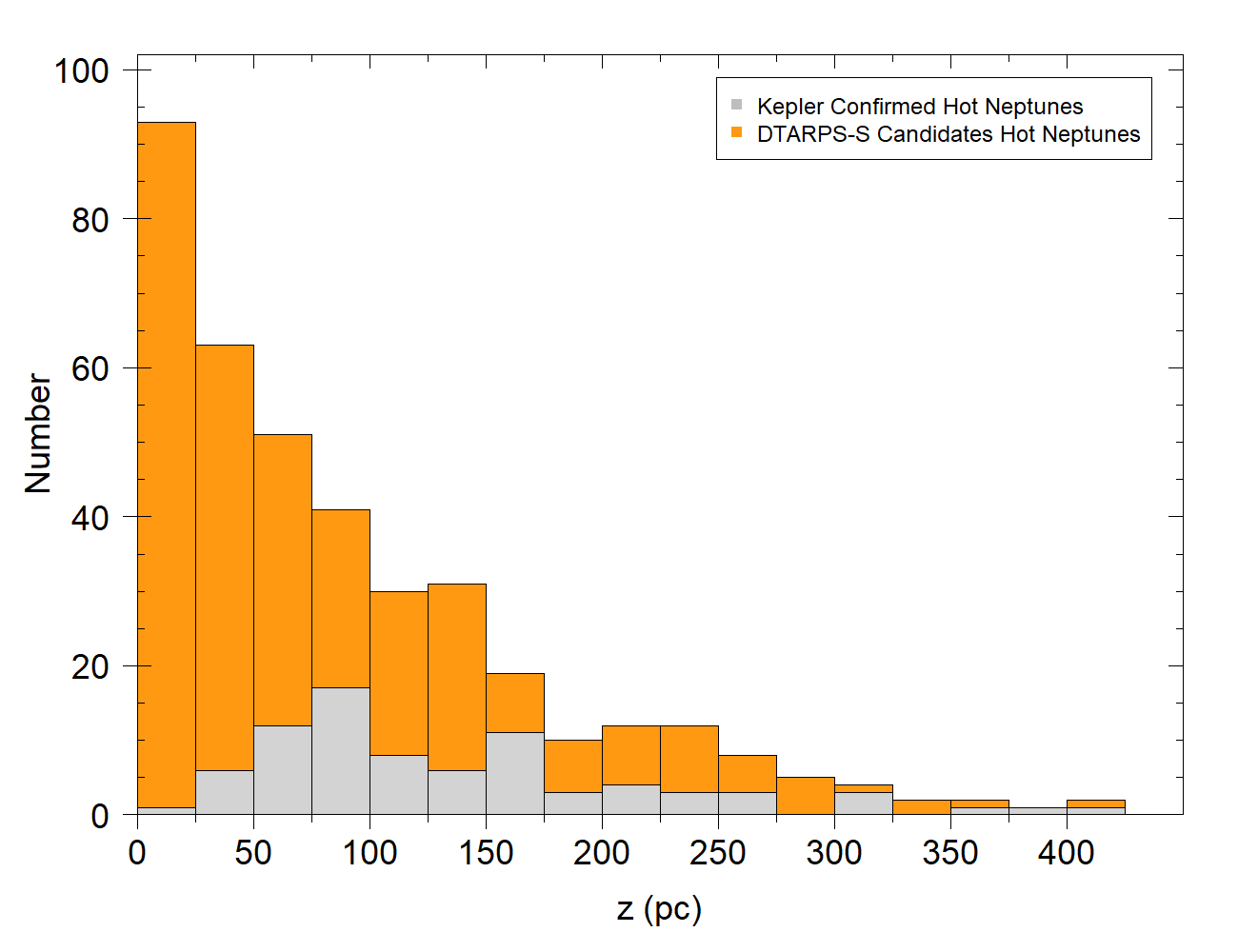}
    \includegraphics[width=0.45\textwidth]{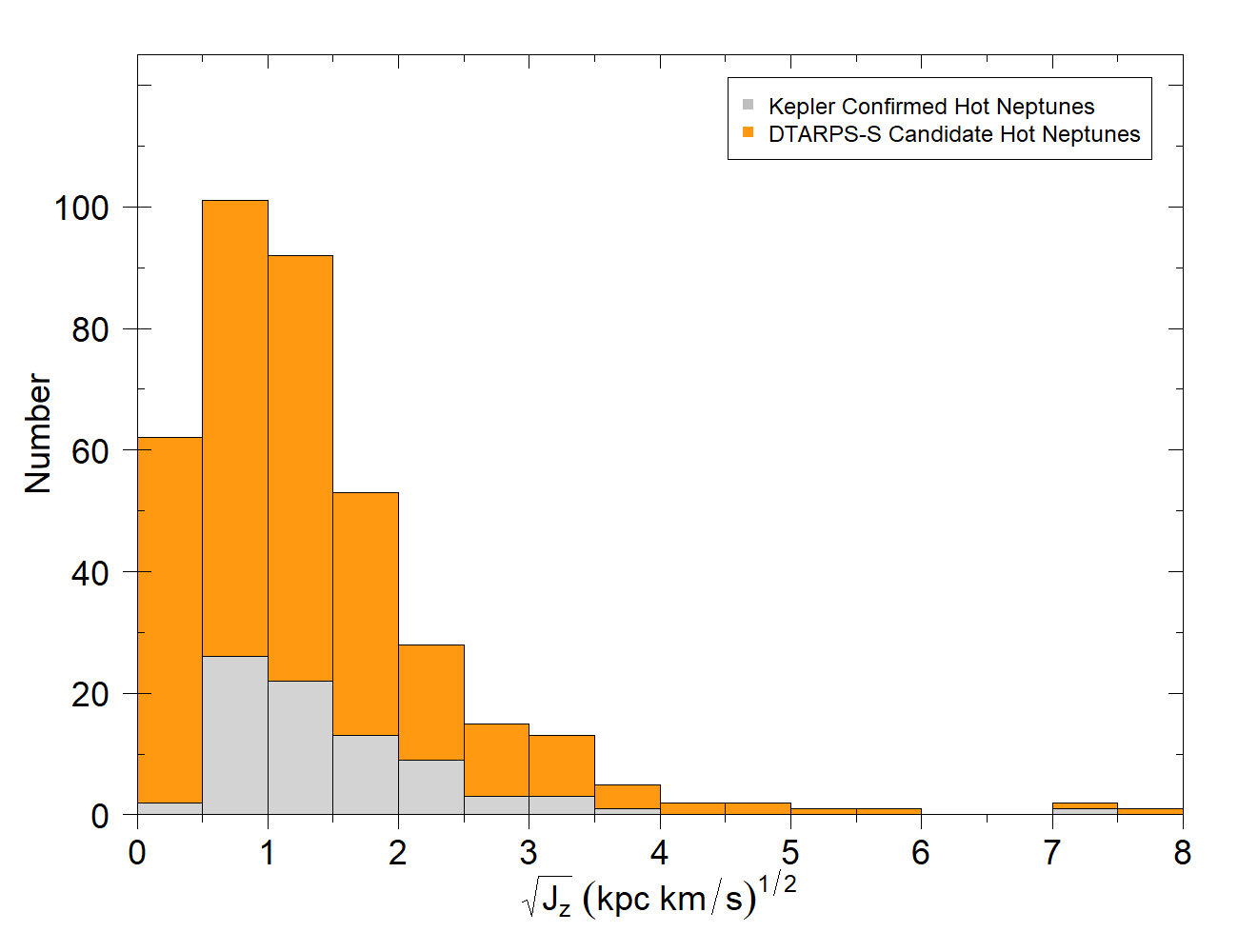}
    \caption{Distributions of absolute distance from the Galactic plane (left) and Galactic vertical action (right) for the DTARPS-S hot Neptune candidates and the Kepler Confirmed hot Neptunes. The DTARPS-S hot Neptune Candidates include hot Neptune candidates from the Galactic Plane catalog. The Confirmed Planets from Kepler are plotted in gray. DTARPS-S candidates are colored by their disposition level. The two distributions are superimposed on top of each other to ease comparison between the two populations.}
    \label{gal_z.fig}
\end{figure}

We can investigate the ages of stars with DTARPS-S Candidate transiting planets in three ways.  First, the celestial locations combined with Gaia parallax distances give estimates of the height $z$ of each star above the Galactic Plane.  Figure~\ref{gal_z.fig} (left panel) shows that 55\% of DTARPS-S hot Neptune candidates have $|z| \leq 80$~pc.  For comparison, Gaia-derived open clusters with ages  $\lesssim 0.5$~Gyr have characteristic Galactic heights $|z| \lesssim 100$ pc while clusters with ages $>2$~Gyr can have $|z| > 500$ pc \citep{Tarricq21}.  At least half of the DTARPS-S hot Neptune candidates may thus be younger than $\sim 0.5$~Gyr.

Second, Galactic vertical action $J_x$ is a measure of the height above the plane that a star may reach, even if it is relatively close to the plane today.  Calculating $J_z$ values using the \texttt{GALPY} package \citep{Bovy15}, we find that nearly all have $\sqrt{J_z} < 3$~kpc~km~s$^{-1}$ (Figure~\ref{gal_z.fig}, right  panel).  For comparison, in an analysis of APOGEE and Gaia data,  \citet{Ting19} find that $\sqrt{J_z} < 3$~kpc~km~s$^{-1}$ is not a strong predictor of stellar ages, although values $>3$~kpc~km~s$^{-1}$ are imply stellar ages $\gtrsim 3$~Gyr.

Third, the sum of DTARPS-S occurrence rates in appropriate cells of Figure \ref{fig:dtarps_occ} below are approximately the same as  values estimated in the open cluster studies of \citet{Fernandes23} and \citet{Christiansen23} where the stellar ages are known to be $0.5-1$~Gyr. The hot Neptune candidates obtained from the DTARPS-S analysis (\S\ref{sec:Neptune_desert}) thus support the open cluster findings of high occurrence rates among younger stars. The main difference is that DTARPS-S produces a large sample of hot Neptunes with poor evaluations of planetary age, while open cluster ages are reliable but give very small hot Neptune samples. 

A simple, though not unique, planet evolution scenario can be formulated to account for the Kepler, DTARPS-S and open cluster results.  The planet formation process produces populations of short (even ultra-short) period planets with a continuous unimodal distribution of masses from hot Jupiters through hot Neptunes to hot rocky planets.  The intermediate-mass population identified by DTARPS-S might be called the `young Neptune forest’.  On timescales around 1~Gyr, slow photoevaporation occurs so the rocky cores are revealed.  After several Gyr, the hot Neptune region of the radius-period diagram is depleted to produce the `old Neptune desert’ seen in the Kepler planet catalog. 

\subsubsection{Planetary properties effects}

\begin{figure}[b]
    \centering
    \includegraphics[width=0.5\textwidth]{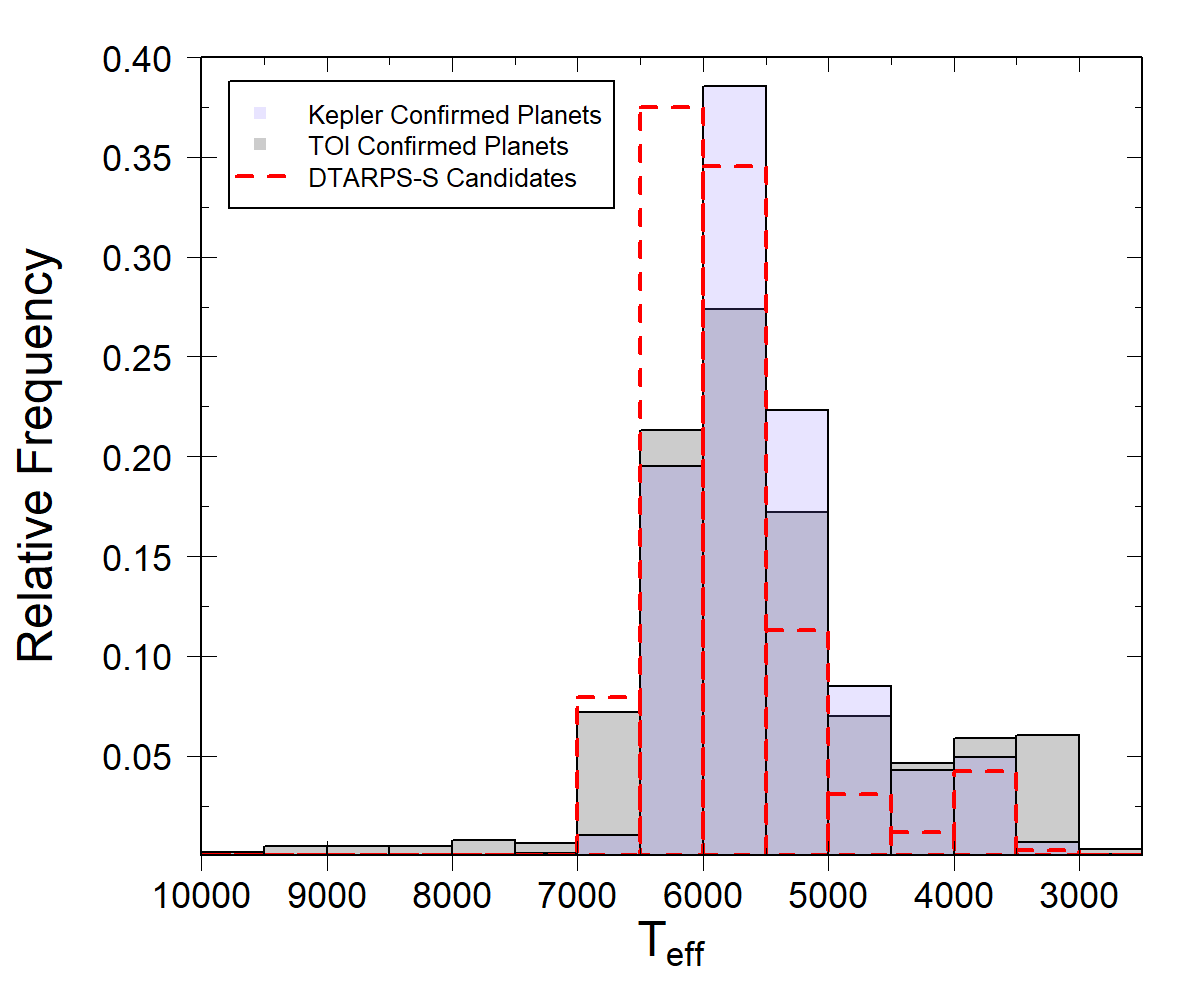}
    \caption{Relative frequency of the effective temperature of the stellar host star for \Kepler confirmed planets (purple) and confirmed planets in the TOI list (gray) normalized by the sample size for each population.  The DTARPS-S Catalog Candidates are marked by the dashed red histogram. The difference between the median effective temperature for the \Kepler planets is only 150 K lower than the median effective temperature of the confirmed planets in the TOI list.  }
    \label{fig:nep_des_temp}
\end{figure}

\citet{Szabo23} found that the boundaries of the hot Neptune desert (or `savanna' as they refer to the region) depended strongly on the planet radius, mass, stellar host effective temperature, and stellar host mass as well as a weak dependence on stellar metallicity.  The distribution of \TESS stars with confirmed planets and DTARPS-S Candidates have slightly higher median effective temperatures than the \Kepler confirmed planets but not enough to explain the strong discrepancy between Kepler and \TESS hot Neptune findings (Figure \ref{fig:nep_des_temp}).  

An alternative approach to understanding the \Kepler-\TESS discrepancy on the hot Neptune population is that these planets are real but have unexpected properties that exclude incorporation into KOI lists but allow their inclusion into TOI lists.  If a larger population of hot Neptunes is truly present, they must appear at intermediate stages of \Kepler analysis (such as TCE lists) as the signal-to-noise of their periodic dips would be high in \Kepler data.  But perhaps they are vetted away at a later stage of analysis and are thus absent from KOI lists and associated follow-up programs.  The Kepler catalogs involve complicated vetting criteria such as those described by \citet{Thompson18}:
\begin{quote}
``The creation of this KOI catalog depends on four different transit fits: (1) the original DV transit fits, (2) the trapezoidal fits performed on the ALT Garcia (2010) detrended light curves, (3) the supplemental DV transit fits, and (4) the MCMC fits … The Robovetter assigns FP TCEs to one or more of the following false-positive categories: 1. Not-Transit-Like (NT): a TCE whose light curve is not consistent with that of a transiting planet or EBs. These TCEs are usually caused by instrumental artifacts or noneclipsing variable stars. … 2. Stellar Eclipse (SS): a TCE that is observed to have a significant secondary event, V-shaped transit profile, or out-of-eclipse variability that indicates that the transit-like rent is very likely caused by an EBs. Self-luminous, hot Jupiter with a visible secondary eclipse are also in this category, but they are still given a disposition of PC.”
\end{quote}

The transit crossing event (TCE) list for \Kepler \citep{NEA-K-TCE} has over 16,000 TCEs identified with periods less than 20 days, 35.0\% of which are initially identified as possible transits in the hot Neptune Desert identified by \citet{Mazeh16}.  After additional vetting of the TCE list, only 23.0\% of the Kepler Objects with periods less than 20 days \citep{NEA-Kepler} lie in the hot Neptune Desert. The \TESS TCE list \citep{MIT-T-TCE} for the Year 1 data (sectors 1 - 13) has a similar number of events with periods less than 20 days (just over 15,000) and yet 29\% of the TCEs are initially identified as transits in the hot Neptune Desert.  Over 33.5\% of the TESS Object of Interest \citep{TOI}, a vetted list based on TCEs, lie in the hot Neptune Desert. 
Thus while \Kepler preferentially vetted away TCEs in the hot Neptune Desert, \TESS preferentially kept TCEs in the hot Neptune Desert.  This underlying discrepancy between \Kepler and \TESS TCEs and TOIs helps explain why DTARPS-S identified many hot Neptune Desert candidates. In order to help spark the needed discussion on the differences between the \Kepler and \TESS candidates, we now speculate on potential astrophysical explanations. 

It is possible to imagine astrophysical properties in the planet or star that might not satisfy these \Kepler vetting criteria but would be found in our DTARPS-S analysis of shorter (typically 27-day) \TESS light curves:
\begin{enumerate}

\item Gaseous hot Neptunes with atmospheres that are unstable on timescales of months-years but not days-weeks. If the atmospheres puff up and contract irregularly in months-years, this might cause irregular variations in depth that trigger rejection by the \Kepler \emph{Robovetter}. These effects might not be present in hot Jupiters due to their higher gravity and/or more effective atmospheric temperature redistribution by global winds. This explanation, however, might only explain a small population of hot Neptunes as their atmospheres are expected to fully escape on timescales shorter than a gigayear. 

\citet{Kalman23} discuss several possible evolutionary paths for young planets formed in the hot Neptune desert to evolve out of the desert area.  These include hyperinflation of low-mass gas giants and planet-growth scenarios.  These might lead to a varying transit depth for hot Neptune candidates on year-long timescales. 

\item Rocky hot sub-Neptunes might have molten surfaces and interiors subject to irregular volcanic eruptions on timescales of months-years but not days-weeks. A hemisphere-scale eruption might produce a temporary silicate atmosphere that dramatically changes the albedo and/or apparent radius of the planets causing erratic variations in transit depth and triggering rejection by the \Kepler \emph{Robovetter}. These volcanic eruptions would also occur on rocky Earths, but their lower gravity allows the silicate atmosphere to rapidly escape so the albedo/radius variations are rarely seen. 

Rocky planets with extremely small orbits may also be disintegrating leading to erratic variations in transit depths \citep{vanLieshout18}. This phenomenon may have been observed in K2-22 b, a rocky exoplanet with an extreme UltraShort Period of 0.4 day.  It exhibits erratic variations in transit depths from $\sim$0\% to $\sim$1.3\% on timescales of days \citep{SanchisOjeda15}. 

\item \citet{2Schlichting11} indicate that silicate or rocky rings can be stable around short-period planets in protoplanetary disks inside the ice line.  If these rings endure into the main sequence phase, they may be present around some hot Neptune candidates.  These planets may exhibit either anomalously large transit depths due to the presence of rings \citep{Zuluaga15} or varying transit depths due to the precession of rings \citep{Heising15}.  Zuluaga et al. and \citet{Santos15} discuss the possibility that exoplanets with rings may have been automatically vetted out of \Kepler due to the inflated transit depth. \citet{Piro20} found that some super-puff planets may be best explained by a denser planet with a ring system.

\item The star may be variable on timescales that would produce apparent variations in transit depth that would lead to rejection by the \Kepler \emph{Robovetter}.  Stellar variations would lead to erroneous normalization that is based on \TESS and Gaia magnitudes.  However, it seems unlikely that stars with Neptune-sized hot planets would preferentially exhibit such variations compared to other \Kepler host stars. 

\end{enumerate}

The explanations listed above are speculative and not direct interpretations of our findings.  While these possibilities may not be correct in detail, they illuminate how unusual physical properties of hot Neptunes, or their host stars, involving changes on months-to-years timescales could explain their relative absence in \Kepler KOI lists but presence in \TESS TOI and DTARPS-S lists.  Planetary candidates in the hot Neptune desert region not only need to be confirmed with mass estimates from radial velocity studies, but need to be monitored in order to determine whether their properties vary on months-to-years timescales.

\section{Ultra Short Period Candidate Planets \label{sec:usp}}

Ultra short period planets (USPs) have periods less than 1 day, and define here DTARPS-S candidates with $0.2 < P < 0.5$ days to be {\it extreme UltraShort Period (xUSP)} candidates.  Our TCF periodogram covers periods down to 0.2 days (Paper I, \S2.3) whereas many other planet detection procedures do not cover periods shorter than 0.5 days\footnote{The DTARPS-S analysis does not include effects such as beaming, ellipsoidal star shapes, and mutual reflection that can affect light curves of very close binaries \citep{Faigler11}}.  As the \Kepler periodicity search stopped at 0.5~days, most USPs from the \Kepler sample were discovered by independent pipelines such as \citep{SanchisOjeda14}. 

Most of the USPs in the Confirmed Planet list on the NASA Exoplanet Archive have radii $<$ 1.8 $R_{\oplus}$. However, there are no DTARPS-S Candidates with radii $<$ 1.8 $R_{\oplus}$ due to the short \TESS FFI sectors (Figure~17 in Paper~I).  Therefore we will use the term USP more generally to include all planetary radii including Neptunian- and Jovian-size planets.  Almost all of the DTARPS-S USP Candidates lie in the Neptune desert addressed in \S \ref{sec:Neptune_desert}.\footnote{
Seven DTARPS-S USPs are missing from Table~\ref{tab:neptunedesert.members} as they lie outside the Neptune desert boundaries: three hot Jupiters in the DTARPS-S Candidates catalog (DTARPS-S 167, 358, and 368) and four stars in the Galactic Plane list (TIC 4616346 and 385267507 are hot sub-Neptunes, TIC 340889095 and  468958331 are hot Jupiters).}  They are noted in footnotes $g$ and $h$ in  Table~\ref{tab:neptunedesert.members}.

\begin{figure}[tb!]
\begin{minipage}[b]{0.49\textwidth}
    \centering
    \includegraphics[width=\textwidth]{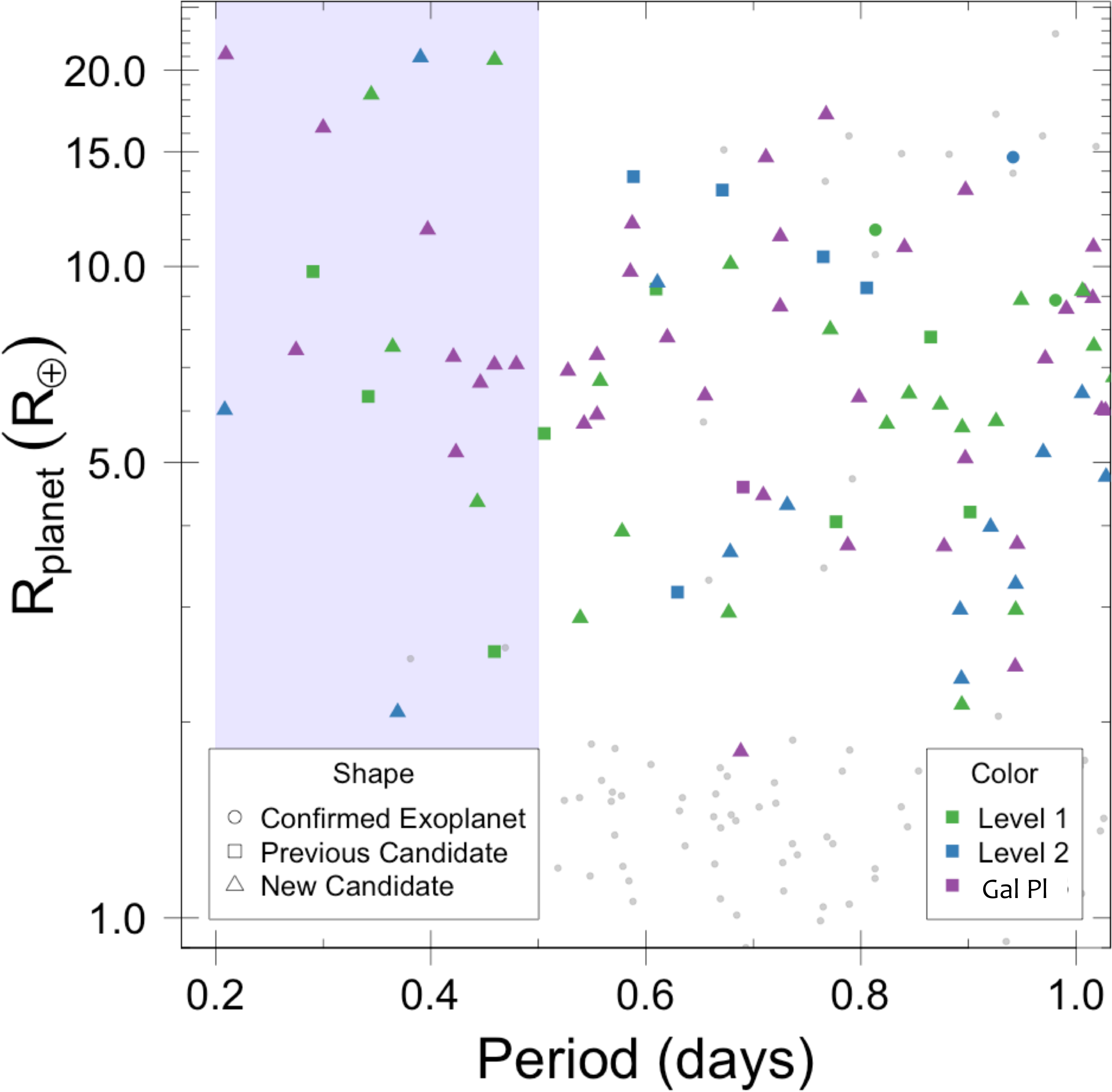}
    \caption{Period-radius plot of DTARPS-S ultra short period (USP) Candidates. The extreme ultra short period (xUSP) region corresponding to periods of less than 12 hours is highlighted in purple. The light gray points are some of the Confirmed Planets that transit from the NASA Exoplanet Archive. The color of the DTARPS-S Candidate points indicate their DTARPS-S disposition level and the shape of the point indicates any previous identification.  \label{fig:cand_only_usp_per_rad_aR_dist1}}
\end{minipage}
\hspace{0.2in}
\begin{minipage}[b]{0.49\textwidth}
    \centering
    \includegraphics[width=\textwidth]{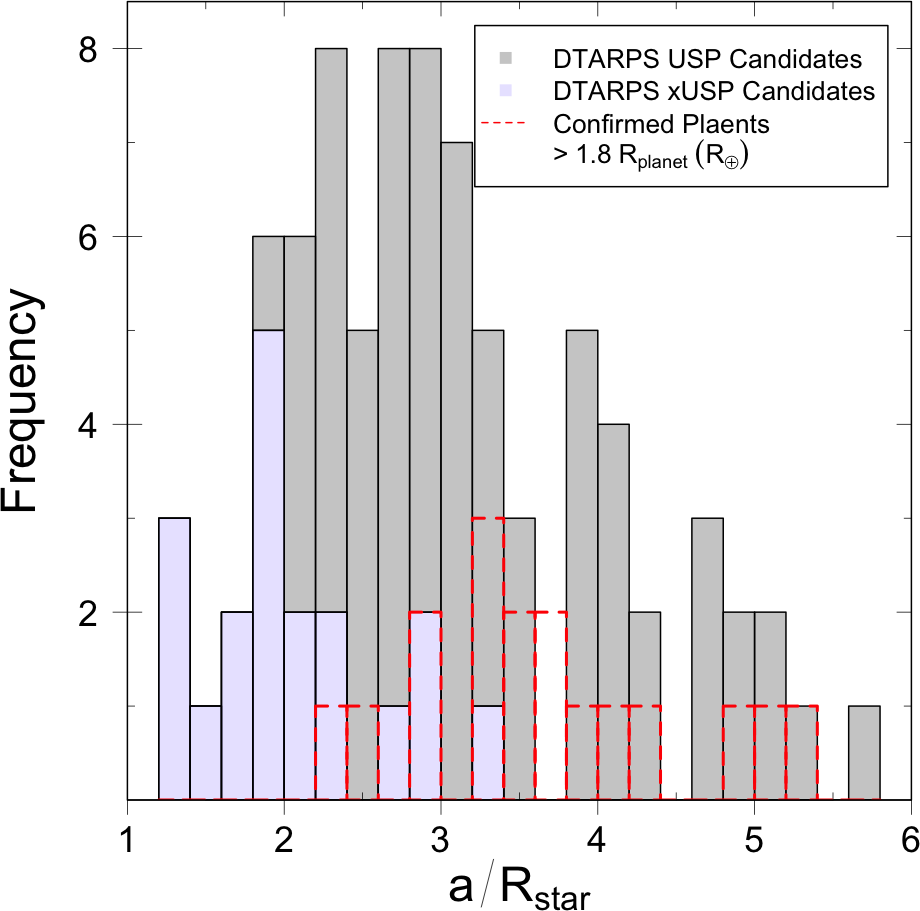}
    \caption{Distribution of the ratio of the semi-major axis to the stellar host radius for DTARPS-S candidate USPs (in gray, omitting Confirmed Planets) and our DTARPS-S candidate xUSPs (in purple, omitting Confirmed Planets). For comparison the $a$/$R_{\star}$ distribution for USP exoplanets in the NASA Exoplanet Archive with radii $>$ 1.8 $R_{\oplus}$ is shown by the red dashed histogram. \\ \label{fig:cand_only_usp_per_rad_aR_dist2}}
\end{minipage}
\end{figure}

The DTARPS-S analysis identifies 82 USP candidates $-$ 48 in the DTARPS-S Candidate Catalog and 34 in the DTARPS-S Galactic Plane list $-$ of which 19 are xUSP candidates. This USP sample is shown in the period-radius diagram in  Figure~\ref{fig:cand_only_usp_per_rad_aR_dist1} with the xUSP region highlighted in purple.  They include three Confirmed Planets and 14 previously identified Planetary Candidates.

Generally,  USPs are thought to be rocky planets, possibly with lava oceans on star-facing surfaces, whose orbital energies are subject to tidal dissipation \citep{Winn18, Dai21}.  Tidal dissipation effects depend on the ratio of orbital semi-major axis and stellar radius as shown in Figure \ref{fig:cand_only_usp_per_rad_aR_dist2}.  Many DTARPS-S Candidate USPs and xUSPs have smaller $a$/$R_{\star}$ than the Confirmed Planets with radii $>$ 2 $R_{\oplus}$.  This is due to a combination of the slightly hotter (and therefore larger) stellar host population for DTARPS-S Candidates (Figure~\ref{fig:st_temp_hist}) and the ability of the DTARPS-S planet detection methodology to detect very short periods giving smaller semi-major axes. The DTARPS-S xUSP sample extends the $a$/$R_{\star}$ distribution to 1.25  where tidal effects can be very strong for candidates with sub-Neptune to Jovian radii.

An important question is whether the DTARPS-S USPs suffer a higher rate of False Positives than DTARPS-S candidates at longer periods.  This is true for the 0.9~million DIAmante sample that serves at the test dataset for DTARPS: more than 50\% of the objects in external surveys with periods $<$ 1 day are labeled as False Positives compared to about 30\% objects with periods 1$-$10 days. However, the Random Forest classifier was trained against injected sinusoidal variables for short periods to help mitigate False Positives masquerading as USPs (\S5.2 in Paper I). This appears to be very effective: the classifier confusion matrix misclassifies only 2 out of 9,095 injected False Positives in the DTARPS-S Analysis List (Figure~11 in Paper I). The DTARPS-S Candidate catalog objects with $P<1$~day contains only 2 out of 94 astronomically known False Positives. The classifier 'specificity' is therefore close to perfect (\S9.3 in Paper~I). The evidence therefore indicates that the USP and xUSP lists have high reliability.  However, this can be easily checked with reconnaissance spectroscopy as radial velocity variations for stellar companions at short periods will have high amplitudes. 

If the DTARPS-S candidate USP and xUSP samples are confirmed as valid transiting planets, they will provide an excellent laboratory for study of tidal effects on large radii planets.  While small, rocky USPs have been shown to be stable against tidal orbital decay, hot Jupiters experience tidal inspiral and can be destroyed on rapid timescales \citep{Hamer19, Hamer20, Yee20}. The tidal forces experience by a planet and the orbital decay of a USP depend on factor tidal quality parameter $Q_{\star}'$ that is often difficult to estimate \citep{Winn18}.  

It seems likely that some of the DTARPS-S USPs, and especially xUSPs, will show orbital decay due to tidal dissipation.  The clearest case for orbital inspiral of a hot Jupiter, WASP-12, has  mass 1.5~M$_J$, radius 1.9~R$_J$, and $a/R_\star = 3.0$ \citep{Yee20}.  It exhibits transit timing deviations of $\sim 4$ minutes over a decade and a timescale of several million years for orbital decay.  The DTARPS-S sample has over 40 candidates with  $a/R_\star \lesssim 3.0$ involving host stars ranging from K to F types (Figure~\ref{fig:cand_only_usp_per_rad_aR_dist2}; spectral types are available in Tables~1 and 4 of Paper II).  The DTARPS-S sample is thus likely to be important for understanding tidal dissipation.  

USPs also experience very high levels of insolation flux \citep{Dai21}. Planets experiencing insolation $S \gtrsim$ 990 $S_{\oplus}$ could be disintegrating rocky planets despite their large radii \citep{Jones20}.  Dozens of DTARPS-S candidates with radii $< 5$~R$_\oplus$ exhibit insolation fluxes above this threshold and a few exceed 10,000 S$_\oplus$ (Figure~\ref{fig:insol_flux}). 

The three disintegrating rocky planets discovered with \Kepler have transit depths that correspond to 1-18 $R_{\oplus}$ \citep{vanLieshout18}. The dust from the disintegrating planet creates a dust cloud enveloping the planet with a dust tail behind that causes deeper transit depths.  They state that disintegrating planets are characterized by short periods, varying transit depths asymmetric transit shapes, and positive bumps in the light curve before the transit. DTARPS-S and follow-up light curves could be examined for these effects.

\section{Candidate DTARPS-S Planets for Atmospheric Transmission Spectroscopy\label{sec:atmos}}

A primary goal of the \TESS mission is the identification of planets smaller than Neptune that transit stars sufficiently bright for transmission spectroscopy to characterize the planetary  atmosphere.\footnote{\url{https://heasarc.gsfc.nasa.gov/docs/tess/objectives.html}} Atmospheric characterization gives insight into planetary mass, composition, formation, and evolution. While many Jovian planets are known to be promising targets for atmospheric follow-up, fewer are available for smaller planet sizes.  The DTARPS-S Candidates catalog has a considerable number of candidate planets that may satisfy criteria for high priority transmission spectroscopy with radii down to $\sim 2$~R$_\oplus$ (Figure \ref{fig:nep_desert_per_rad}).

The Near-Infrared Imager and Slitless Spectrograph (NIRISS) instrument on NASA's James Webb Space Telescope is a primary instrument for transmission spectroscopy.  
\citet{Kempton18} created a metric for identifying \TESS candidates for follow-up using the signal-to-noise ratio for the detection of spectral features in a 10 hour observing campaign with NIRISS based on work by \citet{Louie018}.  Their transmission spectroscopy metric is
\begin{equation}
    \textrm{TSM} = A \times \frac{R^3_{pl} \, T_{eq}}{M_{pl} \, R^2_{\star}} \times 10^{-m_J/5}
\end{equation}
where A is a scale factor given the planet radius from Table 1 in \citet{Kempton18}, $R_{pl}$ is the planet radius, $T_{eq}$ is the equilibrium temperature of the planet, $M_{pl}$ is the mass of the planet \citep[estimated from the \texttt{forecaster} model in][]{Chen17}, $R_{\star}$ is the radius of the star, and $m_J$ is the apparent magnitude of the star in the $J$ Band.  

The planet equilibrium temperature, assuming an albedo of zero for the planet and full night-day heat redistribution is given by
\begin{equation}
    T_{eq} = T_{\star} \sqrt{\frac{R_{\star}}{a}}\left(\frac{1}{4}\right)^{1/4}
\end{equation}
where $T_{\star}$ is the effective temperature of the star in Kelvin and $a$ is the semi-major axis of the planet.  The scale factor includes an assumed planetary atmosphere composition, the fixed mass-radius relationship for planets, and a cloud-free atmosphere.  The planetary atmospheres are assumed to be dominated by water/steam for planets with radii $<$ 2 $R_{\oplus}$ and otherwise are assumed to have a solar composition, H$_2$ dominated, atmosphere.

\begin{figure}[h]
    \centering
    \includegraphics[width=0.95\textwidth]{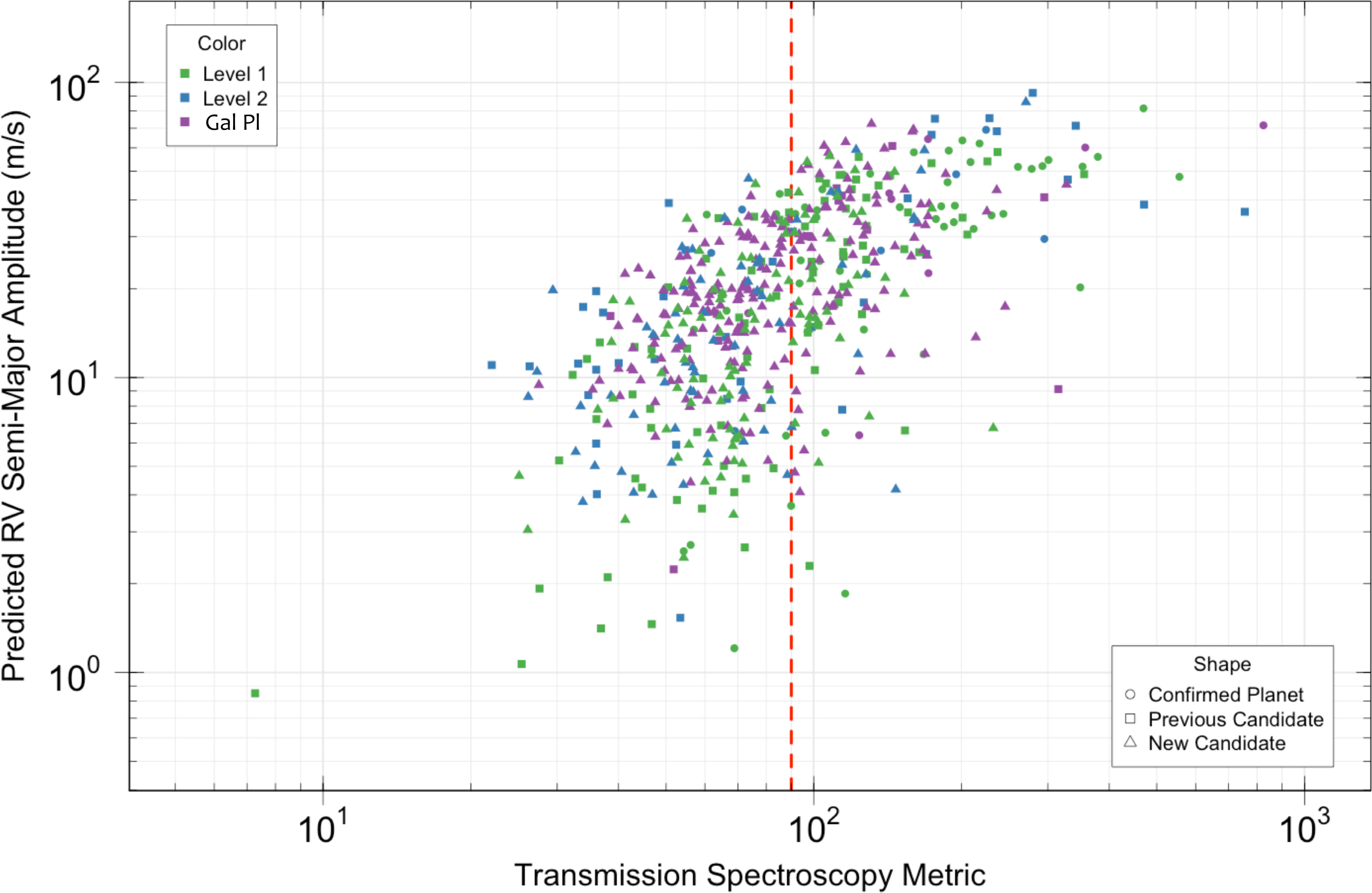}
    \caption{Predicted radial velocity semi-major amplitude as a function of the transmission signal-to-noise metric for the JWST NIRISS instrument. The red line shows the suggested limit for high priority atmospheric transmission spectroscopy \citep{Kempton18}.  The color of the points indicates the disposition and the shape indicates the category of the DTARPS-S candidates. \label{fig:cand_tsm}}
\end{figure}

Figure \ref{fig:cand_tsm} shows the TSM distribution of DTARPS-S Candidates as a function of predicted amplitude of their radial velocity curves.  \citet{Kempton18} recommend TSM $>$ 90 for planets with radii $>$ 1.5 $R_{\oplus}$ as a guideline for selecting exoplanets for atmospheric study. The radial velocities are based on planet mass predictions from the \texttt{forecaster} model assuming a circular orbit.  

The figure shows 238 DTARPS-S Candidates above the TSM $>$ 90 boundary: 52 Confirmed Planets, 61 previously identified planet candidates, and 125 new candidates from DTARPS-S. All but one of the 15 DTARPS-S Candidates with TSM $>$ 300 are previously known cases (either Confirmed Planets or planet candidates), but six newly identified DTARPS-S Candidates with 200 $<$ TSM $<$ 300 are found with a wide range of predicted masses. Eighteen of the DTARPS-S Candidates above the suggested boundary are around very bright stars ($m_J$ $<$ 9) that may be strong targets for ground-based atmospheric transmission spectroscopy. 

\begin{figure}[h]
    \centering
    \includegraphics[width=0.45\textwidth]{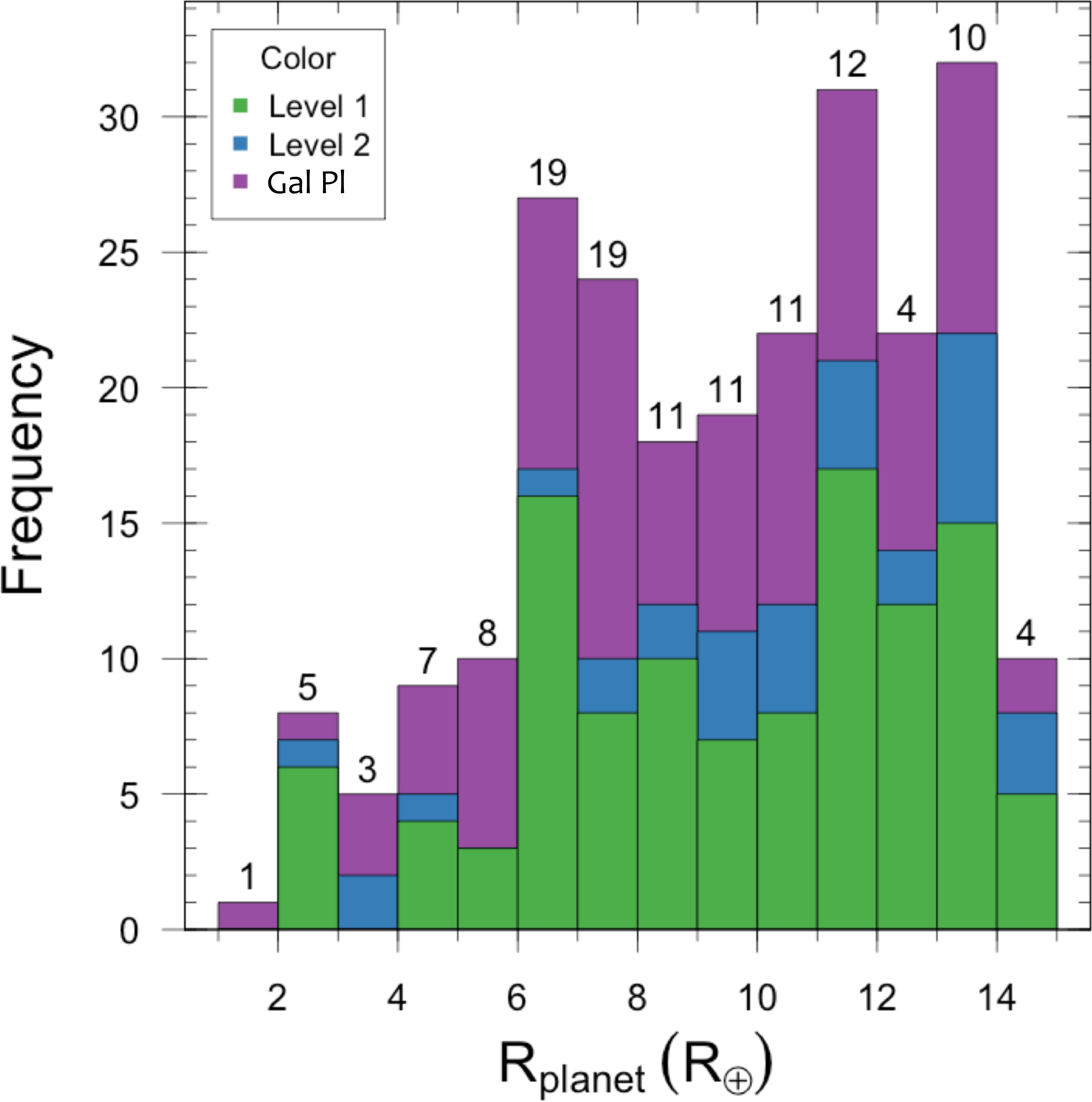}
    \caption{Distribution of planet radii for high priority transmission spectroscopy DTARPS-S Candidates.  Each bar is colored by the DTARPS-S disposition of the candidates.  The number of newly identified DTARPS-S Candidates is given for each bar.  \label{fig:tsm_hist}}
\end{figure}

The distribution of planet radii for the DTARPS-S sample having $TSM > 90$ is presented in Figure \ref{fig:tsm_hist}.  Here highly inflated Jovian candidates with $R_{pl}$ $<$ 14 $R_{\oplus}$ are omitted due to degeneracy in the mass estimates 
\citep{Chen17, Louie018}.  Most of the high-TSM DTARPS-S candidates have Neptune or Jupiter sizes, and thus do not satisfy the primary goal of the \TESS mission.  However, a handful have sub-Neptune or super-Earth radii.

\section{M Dwarf Hosts of DTARPS-S Candidates}\label{sec:m_dwarf}

M dwarfs ($T_{eff}$ $<$ 3,870 K) are the most populous stars in the Galaxy \citep{Henry06} and have advantageous qualities for detection of  planets with Earth-like radii and Habitable Zone surface temperatures.  M dwarf planets produce deep transits and strong radial velocity signals although spectroscopic followup is hampered by their faintness.  Despite these advantages,  only a handful of confirmed hot Jupiters orbiting M dwarfs are known, although Earths and super-Earths are common \citep{Dressing15}. Planet formation theory attributes the paucity of giant planets to a failure of massive cores to reach the threshold for runaway growth before the planetary disk dissipates \citep{Liu20} although rapid formation by disk gravitational instability may be possible \citep{Boss06}. Larger samples of M dwarf gaseous planets can help constrain these formation processes and can probe planetary migration theory by measuring planetary obliquities. 

\begin{figure}[h]
\begin{minipage}{0.49\textwidth}
    \centering
    \includegraphics[width=\textwidth]{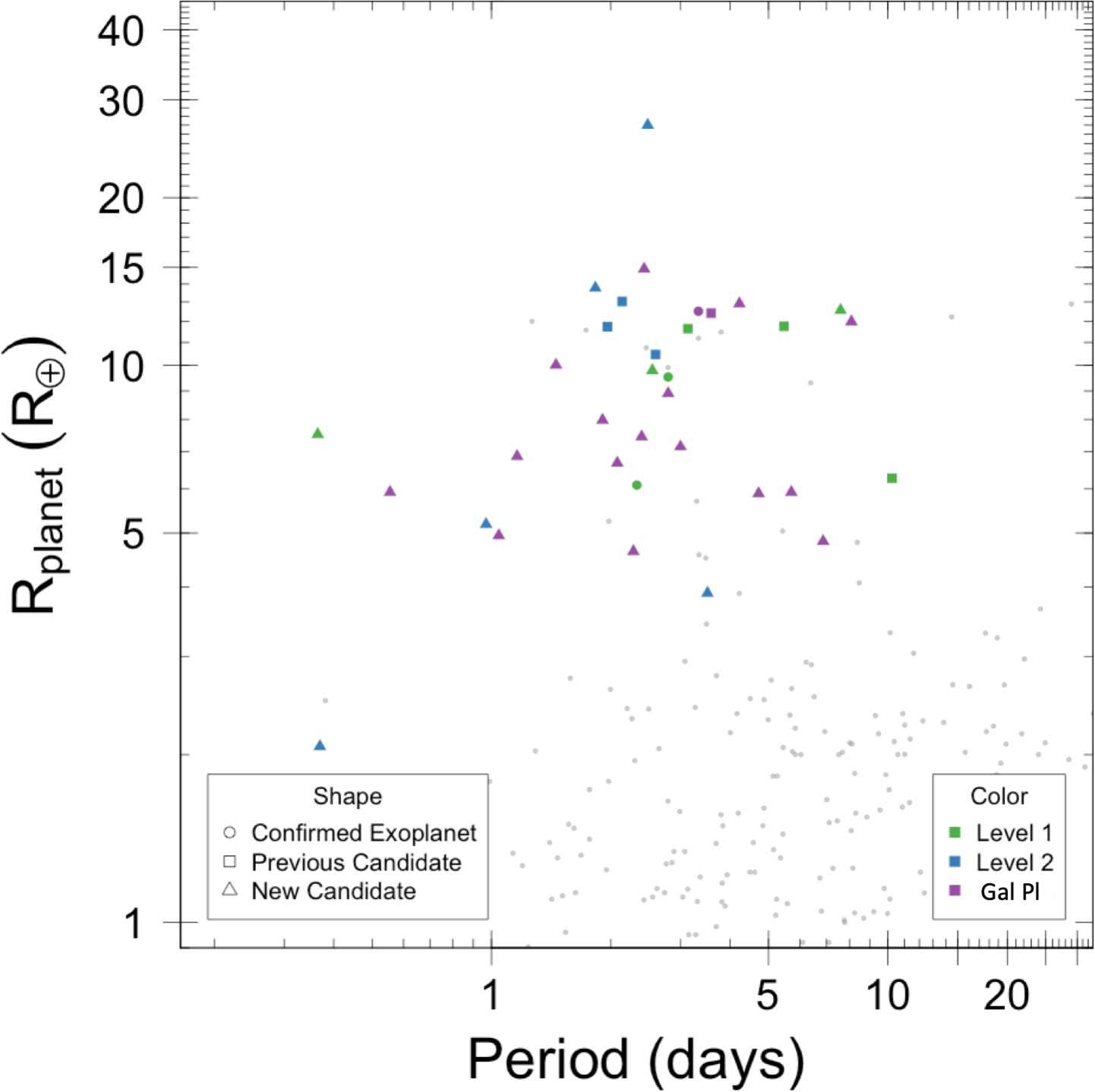}
    \caption{Period-radius plot for DTARPS-S Candidates around M dwarfs ($T_{eff}$ $<$ 3,870 K). The light gray points are the Confirmed Planets from the NASA Exoplanet Archive. DTARPS-S Candidates are colored by their disposition level with symbols representing their novelty. \label{fig:cool_dwarfs1}}
\end{minipage}
\begin{minipage}{0.49\textwidth}
    \centering
    \includegraphics[width=0.8\textwidth]{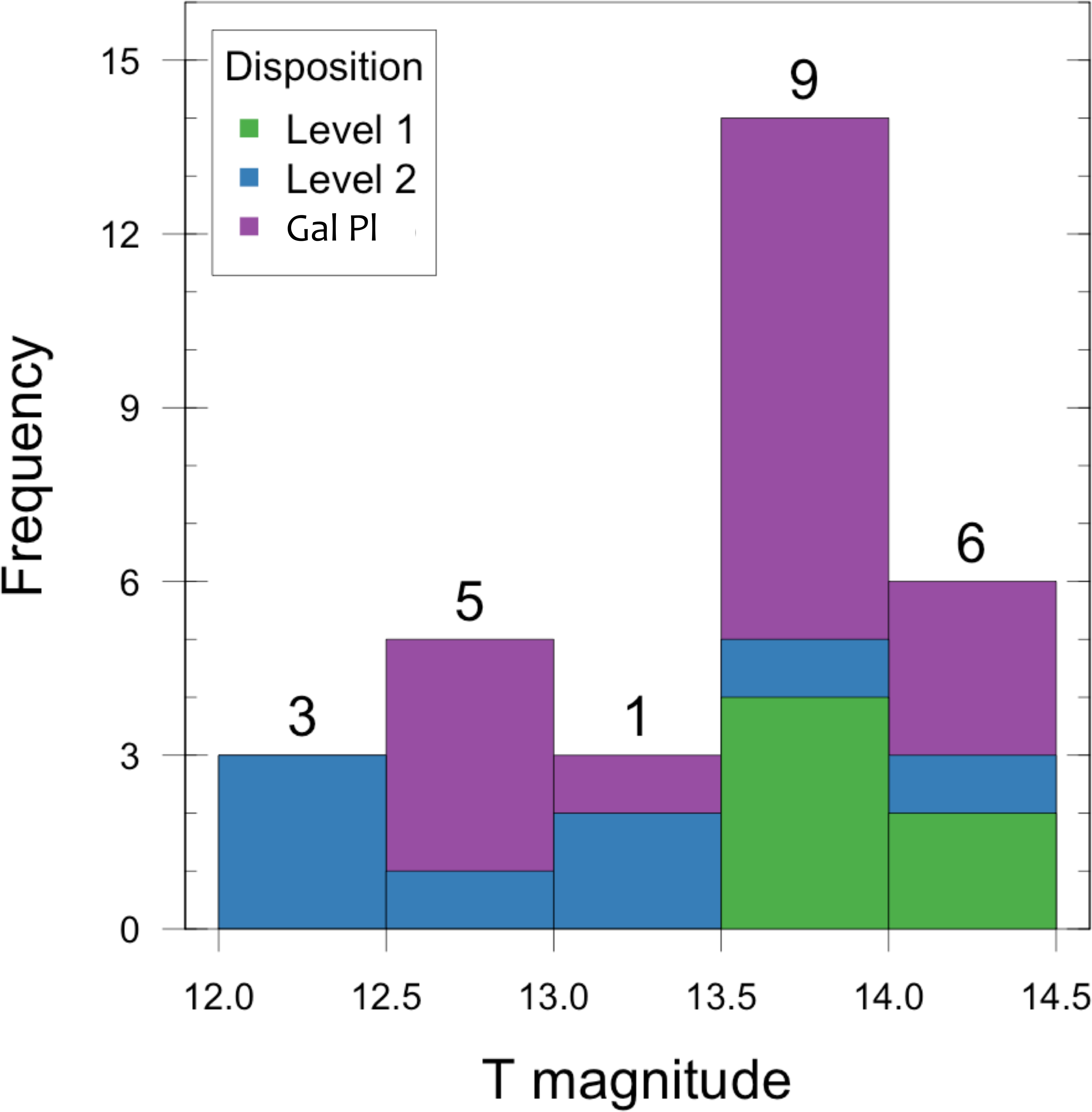}
    \caption{Histogram of the \TESS T magnitude of the candidates with hosts around M dwarfs. Only newly identified DTARPS-S Candidates and previously identified planetary candidates are included in the histogram. The number of newly identified DTARPS-S Candidates in each column is labeled above it.\label{fig:cool_dwarfs2}}
\end{minipage}
\end{figure}

DTARPS-S identifies 34 candidate planets orbiting M dwarfs, roughly half from the DTARPS-S Candidates catalog and half from the Galactic Plane list.  They are listed in Table~\ref{tab:Mdwarf.members} and shown in a period-radius diagram in Figure \ref{fig:cool_dwarfs1}.   Six have TOI designations and five are previously identified in the DIAmante project \citep{Montalto20}.  The sample includes rediscoveries of three well-studied planets orbiting M dwarfs: HATS-6b \citep{Hartman15}, TOI-532b \citep{Kanodia21}, and HATS-75b \citep{Jordan22}.  Most of the M dwarf hosts are fainter than 13th magnitude, but the DTARPS-S Candidate list does include 8 new planet candidates around M dwarfs brighter than a \TESS $T < 13$ (Figure \ref{fig:cool_dwarfs2}).

About half of the DTARPS-S M dwarf sample reside in the Neptune desert; most of the others have radii $\sim 10-15$~R$_\oplus$ above the Neptune desert boundary.  One member is unusual:  DTARPS-S 101 = TIC 83391903 appears to be an inflated Jovian planet with radius 28~R$_\oplus$ orbiting a 12.4~mag M2 V star with orbital period 2.5~days. For details on this and other DTARPS-S Candidates, see the Figure Set and Machine Readable Table of DTARPS-S results, and Appendices of prior published information, in Paper II.

\begin{deluxetable}{rrrrrrrr}[h]
\label{tab:Mdwarf.members}
\tablefontsize{\footnotesize}
\tablecaption{M Dwarf DTARPS-S Systems}
\startdata
& \\
\multicolumn{8}{c}{\bf DTARPS-S Candidates Catalog (N=16)} \\
23\phn{} & 57$^a$ & 76\phn{} & 101\phn{} & 155$^b$ & 162$^b$ & 164$^a$ & 214$^b$ \\ 
224$^b$ & 244\phn{} & 259$^b$ & 273\phn{} & 284\phn{} & 400$^b$ & 417\phn{} & 449\phn{} \\
& \\
\multicolumn{8}{c}{\bf Galactic Plane list (N=18)} \\
  5108864\phn{} &  33521996$^a$ & 71728605\phn{} & 95057860$^b$ & 122798633\phn{} & 146323580\phn{} & 176242777\phn{} & 178120324\phn{} \\ 
231155960\phn{} & 231383819\phn{} & 235548135\phn{} & 237594977\phn{} & 262605715\phn{} & 266657247\phn{} & 389040826\phn{} & 438098149\phn{} \\ 
441131571\phn{} & 452867348\phn{}
\enddata
\tablenotetext{a}{Confirmed planet \quad $^b$ Planetary candidate }
\end{deluxetable}

\section{DTARPS-S Candidate Planetary Occurrence Rates \label{sec:occ_rate}}

The DTARPS-S analysis procedures are sufficiently well-defined in Papers I and II that conversion from $observed$ planet transit rates to $intrinsic$ planet occurrence rates can be estimated in a preliminary fashion.  The  strongest limiting factor to accurate occurrence rates is the uncertain False Positive rate of the DTARPS-S analysis procedure and the possibilities that it varies acriss the Period-Radius diagram).   The injection recall map in Paper I gives a true positive rate of the classifier with respect to the different radius period bins, but unfortunately without widespread follow-up we must estimate the false positive rate in each bin with a global false positive rate for the entire classifier. The lack of a robust measure of the false positive rate for each bin is one of the greatest limiting factors to our occurrence rate estimation.  This is the first such estimate from \TESS survey data for solar-type stars; \citet{Feliz21} gives a occurrence rate for M dwarfs based on \TESS FFI data. Our estimates can be compared to planetary occurrence rates calculated for the 4-year \Kepler survey \citep{Howard12, Fressin13, ForemanMackey14, Hsu19, Neil20, Bryson20}.  Our goals are to examine whether \TESS and \Kepler occurrence rates are compatible, and to identify regimes  of the period-radius diagram where \TESS FFI rates may outperform \Kepler rates due to its access to more stars over the full celestial sphere.

The occurrence rates presented here are preliminary estimates that assume all DTARPS-S Candidates are true planets.  This will overestimate the true occurrence rates because up-to-half of the DTARPS-S Candidates may be False Positives (\S\ref{sec:cand_pure}).  We compare to the \Kepler rates obtained by \citet{Hsu19} who perform a Bayesian analysis based on the \Kepler DR25 and Gaia DR2 surveys. We calculate approximate planet occurrence rates from the DTARPS-S Candidates following the procedure outlined in \citet{Howard12} and \citet{Dressing15} that utilize the \Kepler exoplanet sample, calculating occurrence rates within two dimension bins in the Radius-Period diagram. Errors on the planetary occurrence rates were found by calculating the binomial confidence intervals (Appendix).

Planetary occurrence rates quantify the true distribution of planets around an unbiased sample of stars based on orbital and stellar parameters. The calculation accounts for survey coverage and depth, non-transiting geometries, detection completeness, and reliability. Here we calculate occurrence rates from the 462 DTARPS-S Candidates catalog that represents a sample of uniformly vetted exoplanet candidates from DIAmante-extracted \TESS FFI light curves processed through the DTARPS-S pipeline (Papers I and II).  Most candidates have had no follow-up observations performed to support or reject the planetary nature of the candidate.  

Our  methodology for calculating planetary occurrence rates, largely based on \citet{Howard12}, is presented in the Appendix. This analysis is preliminary and incomplete in various respects.   We omit consideration of radius uncertainties, smoothing between arbitrary radius and period bins, and a likelihood-based statistical analysis \citep{ForemanMackey14, Hsu19, Kunimoto20, Neil20, Bryson21}. The rates derived here are further limited by utilizing an approximate global correction for contamination by False Positives; however, FPs affect less than half of the sample (\S \ref{sec:cand_pure}).   The false positive rate can and likely varies across the entire radius period space, especially in the hot Neptune desert. 
 A global false positive rate is used as an approximation because due to the limitations of trying to characterize a robust false positive rate..  We omit DTARPS-S Galactic Plane objects where FPs may be more abundant.  The Appendix describes the calculation of uncertainties for the occurrence rates and further discusses caveats and limitations of our occurrence rate analysis.

\subsection{Occurrence Rates for DTARPS-S Candidates  \label{sec:occ_results}}

\begin{figure}[tb]
    \centering
    \includegraphics[width=\textwidth]{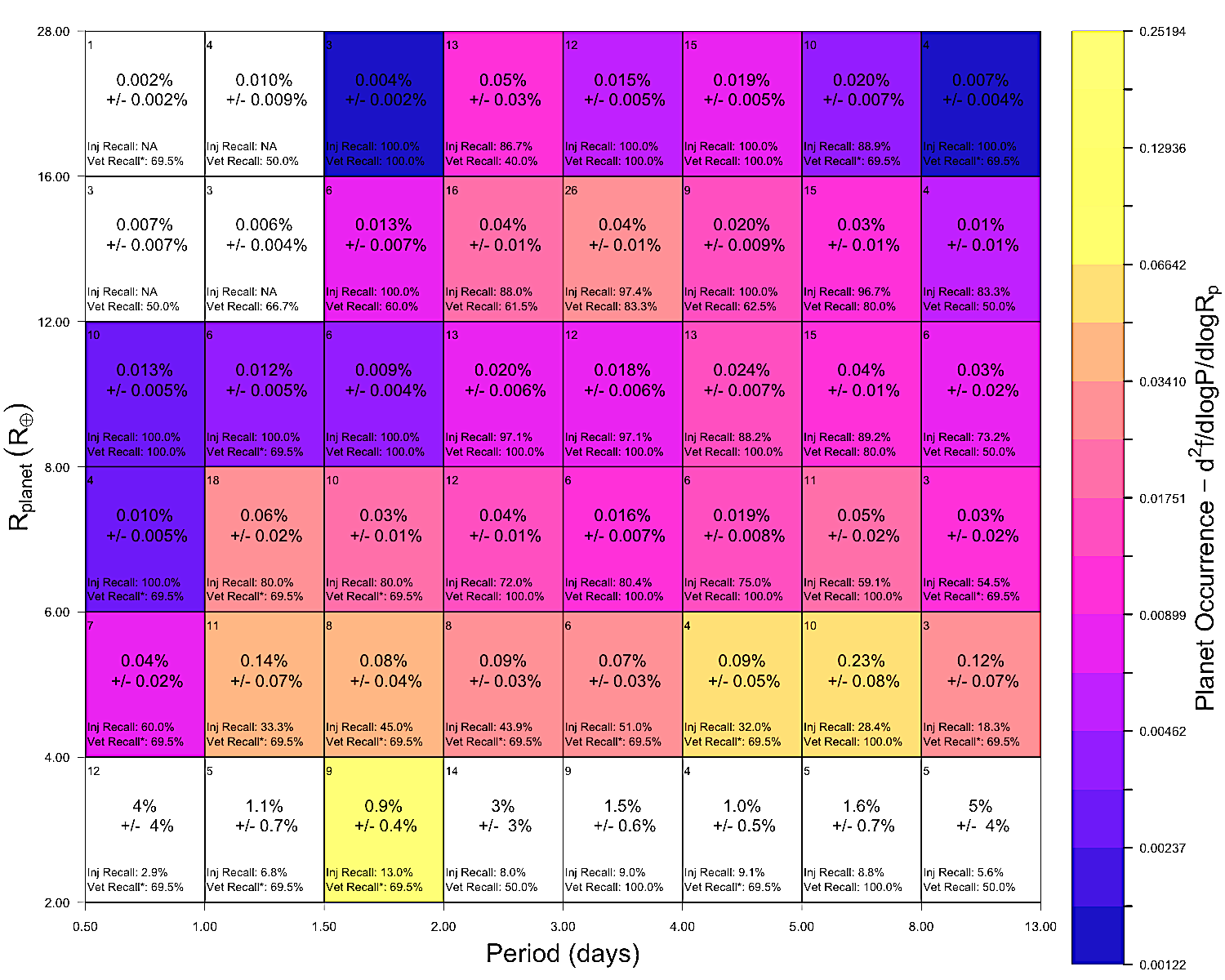}
    \caption{DTARPS-S planet occurrence as a function of planet radius and orbital period. The occurrence rates are given as percentages. Cell color indicates planet occurrence per logarithmic unit area. Note that the bin sizes are not constant. White cells have unreliable DTARPS-S occurrence rates because the injected recall is less than 10\%. Text annotations in each cell list occurrence statistics: upper left—the number of DTARPS-S Candidates, $n_{\textrm{cands, cell}}$, at the bottom of each cell the RF classifier recall rate, $r_{\textrm{RF, cell}}$, and the vetting recall rate, $S_{\textrm{CP, cell}}$. Cells for which the global vetting recall value was used are marked with an asterisk.\label{fig:dtarps_occ}}
\end{figure}

Figure \ref{fig:dtarps_occ} shows the results of this preliminary occurrence rate calculation from the DTARPS-S Candidates catalog over the planetary period-radius space covered by our planetary injections. Each cell has the number of DTARPS-S Candidates in the upper left hand corner, the occurrence rate with uncertainty in the middle, and the injection recall (completeness) and vetting recall (sensitivity to confirmed planets) rates below. The grid squares are colored by the planet occurrence rate as a function of the logarithmic area of each cell. The grid cells with a classifier completeness less than 10\% are not colored due to high uncertainties.

The estimated occurrence rates for DTARPS-S Candidates with $R$ $<$ 4 $R_{\oplus}$ are very uncertain due to the paucity of stars around which these candidates would be able to be detected from \TESS Year 1 FFI light curves.  Figure 13 in Paper I shows that due to the short light curve durations typical of \TESS data, planets with radii less than 5 $R_{\oplus}$ are unlikely to achieve high transit SNR.  The estimated occurrence rates have large errors at long periods ($P$ $>$ 8 days) as well.  This effect is a result of the sparsity of photometric observations during transits for longer period transit signals in typical \TESS FFI light curves with the DTARPS-S method.  This sparsity affects both classifier completeness and vetting completeness.  

\citet{Shabram20} showed that the planet occurrence rates are sensitive to the size of the stellar population examined: smaller star samples lead to larger errors in the planetary occurrence rates. This effect likely biases the occurrence rates for planets with radii $<$ 4 $R_{\oplus}$ where the detection sensitivity of the DTARPS-S method on the \TESS FFI images is low.  The number of stars around which a planet with a radii $<$ 4 $R_{\oplus}$ could be detected  is much smaller compared to other cells in Figure \ref{fig:dtarps_occ}, biasing the occurrence rates towards larger values.

\subsection{Comparison to Kepler Planetary Occurrence Rates \label{sec:occ_comp}}

\begin{figure}[t]
    \centering
    \includegraphics[width=\textwidth]{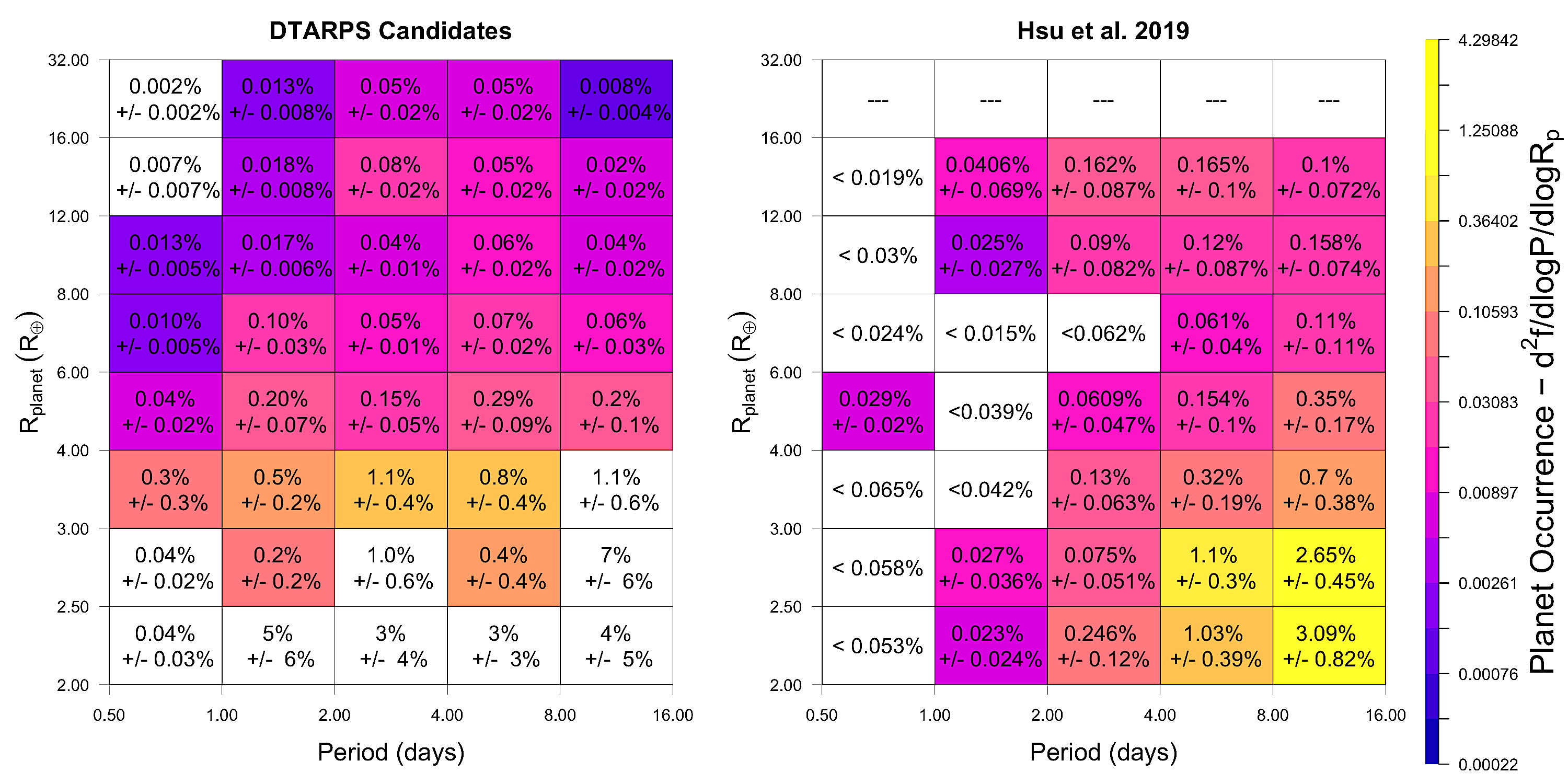}
    \caption{Occurrence rates from DTARPS-S Candidates (left panel) and from \citet{Hsu19} (right panel) using the same bins and color scheme for easy comparison.}
    \label{fig:occ_comp}
\end{figure}

\begin{figure}[tb!]
    \centering
    \includegraphics[width=0.49\textwidth]{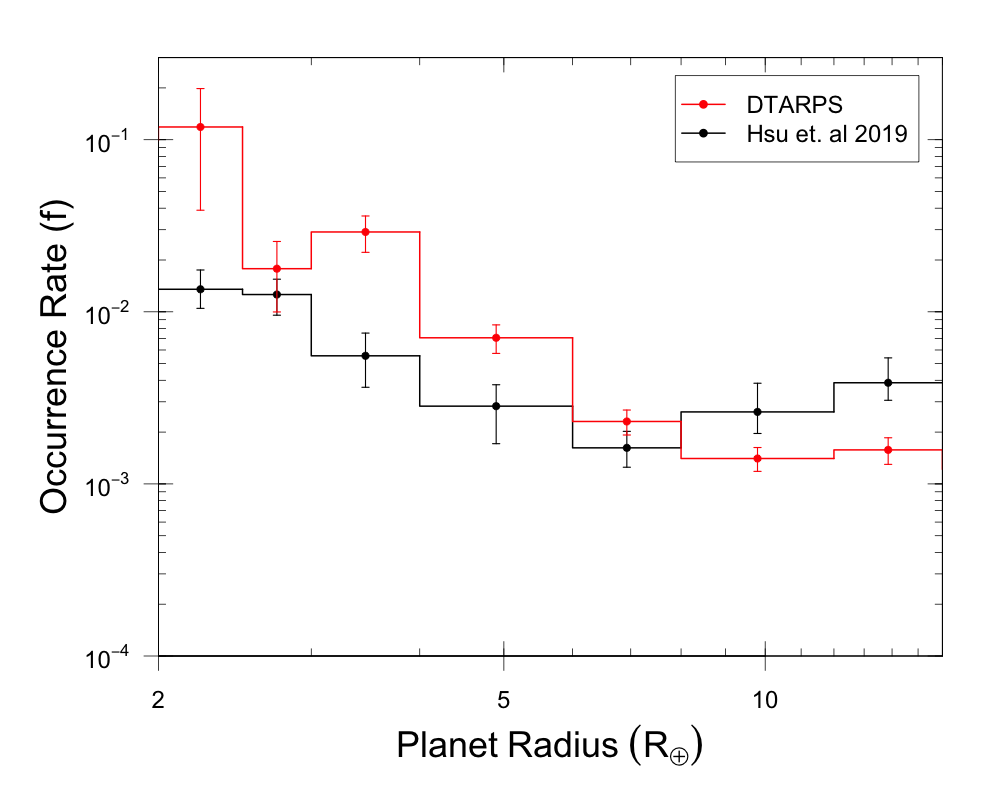} 
    \includegraphics[width=0.49\textwidth]{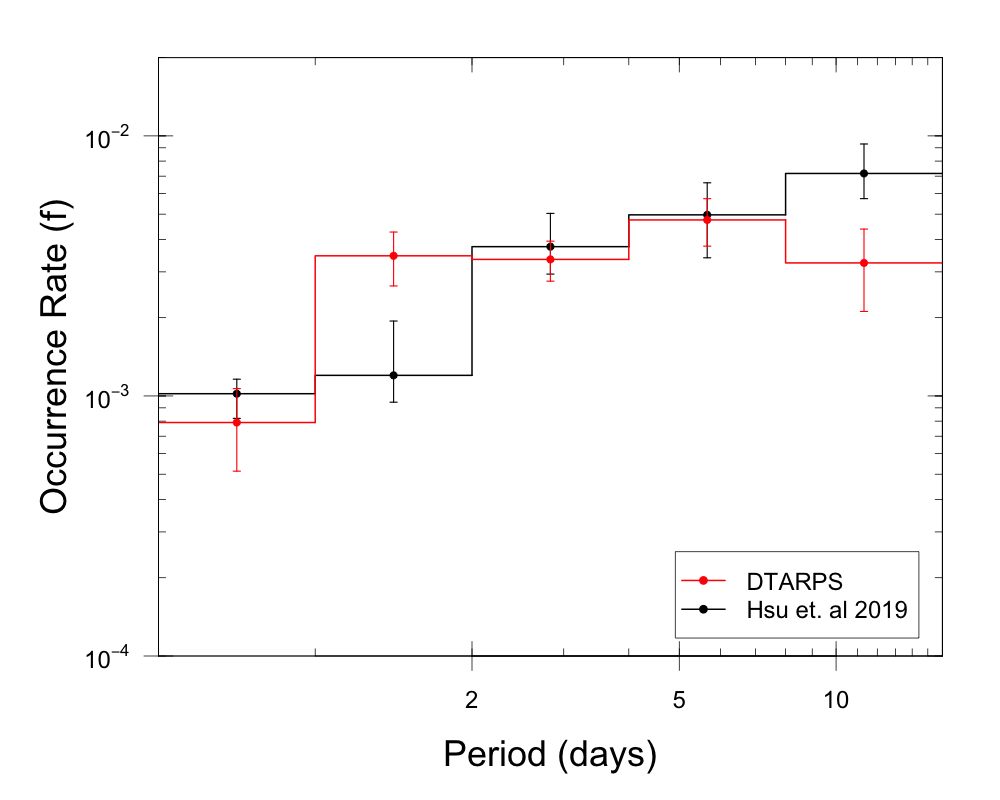}
    \caption{Occurrence rates from DTARPS-S Candidates (red) and from \citet{Hsu19} (black) marginalized over planetary radii for orbital periods less than 8 days (left) and marginalized over orbital periods for radii between 4 and 16 $R_{\oplus}$ (right).}
    \label{fig:occ_marg}
\end{figure}

We compare our occurrence rate calculations to those obtained from \Kepler DR25 by \citet{Hsu19}. They used an Approximate Bayesian Computation procedure to estimate occurrence rates treating measurement uncertainties and other effects that we neglect. Figure \ref{fig:occ_comp} shows the occurrence rates for the DTARPS-S Candidates catalog recalculated using the procedure from Appendix \ref{sec:occ_method} onto the same grid as the occurrence rates from \citeauthor{Hsu19} were presented. If the cell's occurrence rate is well constrained, the cell is colored by the occurrence rate per logarithmic unit area. If the cell's classifier completeness is less than 10\% or gives an upper limit to the occurrence rate, the cell is not colored. Marginalized occurrence rates by planet radius for orbital periods between 0.5 and 8 days and marginalized occurrence rates by orbital period for planets with radii 4$-$16 $R_{\oplus}$ are presented in Figure \ref{fig:occ_marg} for both DTARPS-S Candidates and \citeauthor{Hsu19} \Kepler occurrence rates.

Except for the Neptune desert seen as the white bins around $1<P<4$~days and $3<R<8$~R$_\oplus$ in the right panel of Figure~\ref{fig:occ_comp}, occurrence rate values for the two missions are generally consistent with each other, although uncertainties allow factor-of-two differences in some cells. The occurrence rates estimated from DTARPS-S agree within error bars with the results from the \Kepler DR25 sample for every cell with a period less than 8 days and radii greater than 4 $R_{\oplus}$ except in the Neptune desert. In particular, the DTARPS-S occurrence rates for the ultra short period planets (USPs) with periods less than 1 day are consistent with \citet{Hsu19}, although with considerable uncertainties in both surveys. This overall agreement  supports other evidence (\S\ref{sec:cand_pure}) that the False Positive rate of the DTARPS-S Candidates catalog is not high and that the analysis in Appendix \ref{sec:occ_method}, despite various approximations, is basically reliable. 

However, as expected from the discussion in \S\ref{sec:Neptune_desert}, a major difference is seen in the Neptune desert regime. The DTARPS-S analysis finds occurrence rates at least several times higher than found in the Kepler analysis where occurrence rates in the hot Neptune regime. Possible causes for this strong discrepancy between \TESS and Kepler occurrence rates are discussed in \S\ref{sec:nep_explain}. 

Our results are consistent with a study of \TESS planets orbiting pre-main sequence and young-ZAMS stars by \citet{Fernandes22}.  They find $49\% \pm 20\%$ for sub-Neptunes and Neptunes $1.8 < R < 6$~ R$_\oplus$ with periods $P < 12.5$ days, a regime where the \Kepler's occurrence rates around older stars is $6.8\% \pm 0.3\%$.  Our corresponding occurrence rate is around 20\%; but this is very uncertain because it is dominated by super-Earths with radii $\simeq 2$~R$_\oplus$ where our completeness is poor.  

Some other differences are apparent.  The DTARPS-S occurrence rates for periods $>$ 8 days and radii $>$10 R$_\oplus$ appear lower the occurrence rates reported in \citet{Hsu19} (Figure \ref{fig:occ_marg}, right panel).  Weak recovery of long periods may arise from the \TESS FFI cadence giving sparse light curves with (in most cases) a  27 day duration with a 13.7 day gap from the satellite orbit, making it difficult to find periodic signals with periods longer than $\sim 13$ days. The classifier may be biased against large radius objects because the injected False Positives outnumber the True Positives in this regime (\S5.2 in Paper I). 

We can evaluate the sensitivities of the two surveys in different regimes by examining the uncertainties to the estimated rates in the cells of Figure~\ref{fig:occ_comp}. With 4 years of data compared to a typical 1 month of data, the \Kepler survey is more sensitive than \TESS to common smaller planets with radii $< 4$~R$_\oplus$. But the \Kepler sample covered a smaller area of the sky and retrieved fewer rare species than \TESS; \TESS thus gives  more accurate occurrence rates for (inflated) Jovian planets.  The DTARPS-S occurrence rates have smaller error bars than \citet{Hsu19} due to larger samples in the period-radius cell for planets with radii $>$ 6 $R_{\oplus}$ and have larger error bars for radii $<$ 4 $R_{\oplus}$. \TESS occurrence rates should be more accurate when \TESS full mission surveys are treated. 

Overall, the occurrence rates from the DTARPS-S Candidate sample are in general agreement with the \Kepler DR25 occurrence rates presented by \citet{Hsu19} for candidates with radii $>$ 4 $R_{\oplus}$ and periods $<$ 8 days, with the important exception of the Neptune desert. Extension of DTARPS-S analysis to the full celestial sphere and \TESS Extended Mission raises good prospects for improved accuracy of occurrence rates in several parts of the Radius-Period diagram.

\section{Final remarks  \label{sec:final_remarks}}

\subsection{Why is the DTARPS-S Approach Successful?} \label{sec:why_it_works}

In Paper I, we emphasized that our statistical methodology can be unusually effective. First developed by \citet{Caceres19a}, the AutoRegressive Planet Search procedure starts with ARIMA modeling that reduces uninteresting flux variations including short-memory autocorrelation missed by other detrending methods. ARIMA has been enormously effective in econometrics and signal processing, so its success for modeling stellar variability is not surprising. The Transit Comb Filter is unusually effective in extracting faint transit-like periodicities from the ARIMA residuals.   \citet{Gondhalekar23} demonstrate that the statistical capabilities of ARIMA followed by TCF procedures are favorably compared to more common procedures involving kernel- or polynomial-based detrending followed by Box-Least Squares (BLS) periodogram.  In particular, the TCF periodogram has better noise properties with fewer false spectral peaks than the BLS periodogram in many situations so it can detect smaller planets. 

We then apply a sophisticated machine learning classifier.  Random Forest classification is now two decades old, but the specific code we use $-$ {\it randomForestSRC} by \citet{Ishwaran22} $-$ incorporates many improvements to the method originally developed by \citet{Breiman01}.  With $\sim 70,000$ annual downloads, {\it randomForestSRC} is used by researchers in many fields. For imbalanced astronomical classification problems like transiting planet detection, design of classifier training sets is critically important. We adopt from \citet{Montalto20} the strategy of adding injected EBs to the $negative$ training set for the classifier, complementing the injected planetary transits in the $positive$ training set. This helps remove the astronomical False Positives injurious to transit detection efforts at an early stage of analysis. 

This procedure leads to the DTARPS-S Analysis List (DAL) of 7,743 objects in Paper I, a list with high recall of true planets but dominated by False Alarms and False Positives.  It was followed by aggressive application of standard vetting tools applied to the DAL to give the much smaller, high-confidence DTARPS-S Candidates catalog with 462 objects, supplemented by the somewhat less secure 310 objects in the DTARPS-S Galactic Plane list in Paper II.  

\begin{figure}[h]
    \centering
    \includegraphics[width=0.85\textwidth]{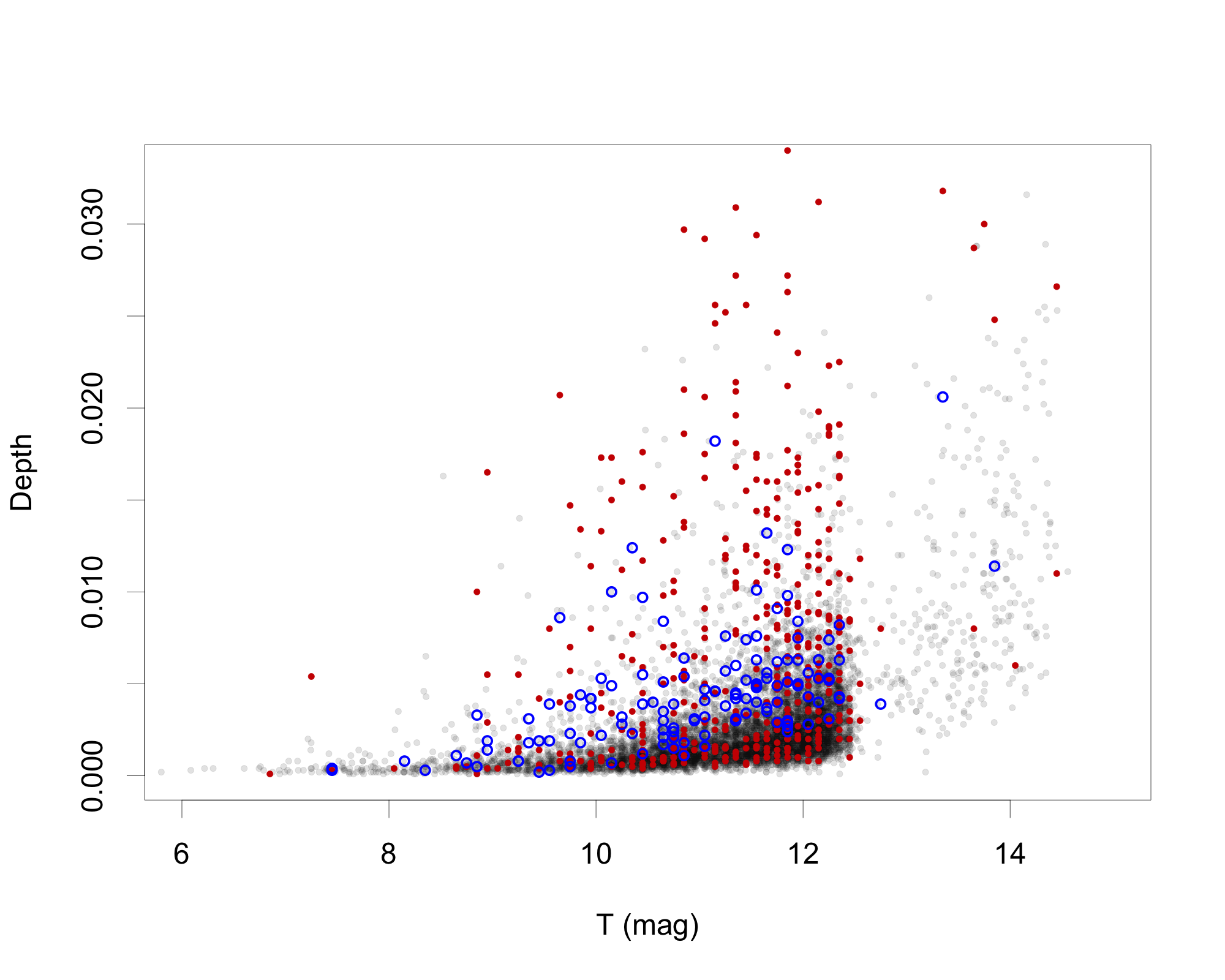} 
    \caption{Plot of fractional transit depth and host star brightness for DTARPS-S stars.  The gray points are 7,743 objects from the DTARPS-S Analysis List of Paper I and the red points are 462 objects from the DTARPS-S Candidates catalog of Paper II.  Blue circles are the subset of astronomical Confirmed Planets from the NASA Exoplanet Archive.}      \label{fig:purity}
\end{figure}

Figure~\ref{fig:purity} shows the result of these efforts from an astronomical, rather than statistical, viewpoint. The pale gray points show the host star magnitude and transit depth distribution of the DAL.  Most objects have depths around $0.001-0.003$ corresponding to planets in the super-Earth and sub-Neptune regime when they represent true planets.  The red points show the objects surviving vetting procedures in the DTARPS-S Candidates catalog and DTARPS-S Galactic Plane list.  Here we see that some of the larger planets are confirmed with similar depth-magnitude distributions as Confirmed Planets (blue circles).  

But a majority of the DTARPS-S Candidates have smaller depths than Confirmed Planets derived mostly from TOI lists.  This is expected from the analysis of \citet{Gondhalekar23}: DTARPS-S should recover smaller planets due to its use of ARIMA and TCF rather than other detrenders and the BLS periodogram.  Unless these candidates are found to be very badly contaminated with False Positives (which we argue in \S\ref{sec:summary_validation} and \S\ref{sec:Neptune_contam} is not the case), then DTARPS-S has a significant discovery space for Neptune-size and smaller planets orbiting stars with $T < 12.5$ in a single year of \TESS survey.  

\subsection{Summary of DTARPS-S Year 1 Findings}
\label{sec:summary3}

In this Paper III, we show that the combination of strong ARPS analysis and strict vetting procedures produce catalogs with relatively high purity: at least half of the sample are likely to be true planets (\S\ref{sec:cand_pure}). This conclusion is supported by several measures of high classifier precision and specificity; reconnaissance spectroscopy of a limited subsample; application of the \texttt{TRICERATOPS} validation tool; inflated Jupiter properties; and more (\S\ref{sec:summary_validation}). Surprisingly, the Galactic Plane sample without crowding or centroid vetting appears to have similar False Positive rates as the more carefully vetted DTARPS-S Candidates catalog. DTARPS-S thus provides one of the largest and most reliable catalog of \TESS exoplanet candidates independently of the official \TESS Objects of Interest developed by the \TESS Science Center.  

The DTARPS-S Candidate catalog and Galactic Plane list give large samples for important exoplanetary populations. Some are independent rediscoveries of Confirmed or candidate planets while many are newly reported here. 
\begin{enumerate}

\item Half (387 of 772) of the DTARPS-S candidates are hot Neptunes, lying in the `Neptune desert’ regime of the \Kepler exoplanet period-radius diagram (\S\ref{sec:Neptune_desert}).   Many of these candidates lie in the central hot Neptune region which is sparsely populated in Confirmed Planets.  The continuing identification of hot Neptunes in \TESS data by various research groups occurs mainly near the borders of the hot Neptune desert, but new planets have been recently discovered in the central region (\S\ref{sec:nep_desert_Kep_TESS}). It is possible that the central desert region is especially prone to False Positive contamination in DTARPS-S catalogs, but we find no internal evidence for this despite considerable opportunities where the effect would appear (e.g. Paper I, section 10.5; Paper II section C.4; sections 3 \& 4.1 here). Using the best available resources for exploring False Positives, we find that some candidates are likely contaminant EBs with cool companions (\S\ref{sec:Neptune_contam}), but the DTARPS-S procedures presented in Papers I and II are effective in eliminating many classes of False Positives (\S\ref{sec:Neptune_TOI}). Overall we estimate the purity of the DTARPS-S catalog to be around 50\% (\S\ref{sec:summary_validation}). Possible astrophysical explanations for the discrepancy between \TESS and \Kepler hot Neptune populations include differences in stellar ages and instabilities in apparent hot Neptune radii or albedos (\S\ref{sec:nep_explain}).  

\item We report a large sample of 82 candidate Ultra Short Period ($P < 1$~day) planets (\S\ref{sec:usp}). Nineteen  of these lie in the poorly explored $0.2 < P < 0.5$~day regime of extreme USPs; several these systems have semi-major axes that are less than twice the stellar radius where orbital decay will be rapid.  Insolation fluxes can exceed 10,000 time current Earth levels and cases of planetary disintegration may be present.  These Ultra Short Period planets lie almost exclusively in the hot Neptune desert region. False positive rates can vary widely in the period-radius space and could be higher for Neptune-size USPs, but our vetting procedures are designed to removed most of these contaminants (\S4.1).  Any USP contaminants will be easily identified through reconnaissance spectroscopy follow-up observations. 

\item DTARPS-S also gives promising new targets for atmospheric transmission follow-up spectroscopy (\S\ref{sec:atmos}) and a handful of new candidates orbiting M stars (\S\ref{sec:m_dwarf}).  As above, the purity of these objects is estimated to be around 50\%, but this is not firmly established.
\end{enumerate}

DTARPS-S methodology is sufficiently well-characterized at each step that preliminary planet occurrence rates can be estimated (\S\ref{sec:occ_rate}). The DTARPS-S Candidates have a overabundance of candidates in the Neptune desert region, but otherwise agree within errors with the \Kepler calculations by \citet{Hsu19}.  This may be the first occurrence rate calculation for FGK stars estimated from the \TESS survey, based on completeness calculations (Paper I) similar to those developed for the \Kepler survey by \citet{Christiansen20}.  

Detailed graphics, statistical and published properties of individual The DTARPS-S Candidate catalog and Galactic Plane list from the \TESS Year 1 survey are provided in Paper II to assist researchers with spectroscopy and other follow-up studies.  Our research group is proceeding with extension to the Year 2 northern sky survey and with a more comprehensive planet detection effort using the \TESS Extended Mission surveys. Our group is beginning a limited program of radial velocity spectroscopic followup, but observations by other groups are strongly encouraged.  We are continuing methodological developments to improve recall (completeness) and precision (False Alarm and False Positive rejection) in \TESS surveys.  

\begin{acknowledgements}

We are enormously grateful to David Latham (Smithsonian Astrophysical Observatory), Lars Bucchave (Technical University of Denmark) and the TRES team for obtaining and analyzing reconnaissance spectroscopic observations of DTARPS-S candidates. TRES resides at the Fred L. Whipple Observatory operated by the Smithsonian Astrophysical Observatory.  We thank \TESS scientists Jon Jenkins (NASA-Ames) and Andrew Vanderburg (MIT) for valuable discussion of methodology and science.  MIT \TESS Science Office scientists (N. Guerrero, K. Hess, M. Kunimoto, A. Rudat) provided useful correspondence on technical matters. 

The AutoRegressive Planet Search project is supported at Penn State by NASA grant 80NSSC17K0122 and NSF grant AST-1614690.   The DTARPS-S project benefits from the vibrant community of Penn State’s Center for Exoplanets and Habitable Worlds that is supported by the Pennsylvania State University, the Eberly College of Science, and the Pennsylvania Space Grant Consortium. This study is also a product of the Center for Astrostatistics supported by the Eberly College of Science. We appreciate comments on the manuscript by members of these Centers: Ian Czekala, Rebekah Dawson, Hyungsuk Tak, Jason Wright, as well as Joel Hartman (Princeton).  

This paper includes data collected with the TESS mission. Funding for the TESS mission is provided by NASA’s Science Mission Directorate. This research has made use of the NASA Exoplanet Archive, which is operated by the California Institute of Technology, under contract with the National Aeronautics and Space Administration under the Exoplanet Exploration Program.  This work has made use of data from the European Space Agency (ESA) mission Gaia processed by the Gaia Data Processing and Analysis Consortium. Funding for the DPAC has been provided by national institutions, in particular the institutions participating in the Gaia Multilateral Agreement. 

\end{acknowledgements}

\facility{\TESS, NASA Exoplanet Archive, \emph{Gaia}}

\software{R: \emph{binom} \citet{binom22}; Python: \texttt{TRICERATOPS} \citet{triceratops21}}

\bibliography{DTARPS_main}{}
\bibliographystyle{aasjournal}

\appendix
\section{Occurrence Rate Methodology \label{sec:occ_method}}

Our star sample starts with 823,099 DIAmante stars processed by the Random Forest classifier, reduced to 725.933 with the deletion of M stars. We further exclude regions of the period-radius diagram where the DTARPS-S Analysis List is badly incomplete which reduces the DTARPS-S Candidate catalog from 462 to 385 objects\footnote{We note that \citet{Howard12} restricted their occurrence rate calculation to regions where the \Kepler completeness was near unity.}.  Restrictions in the DIAmante sample or DTARPS-S analysis are summarized in Table \ref{tab:occ_param}.    

\begin{table}[tbh!]
    \centering
    \fontsize{9}{11}\selectfont
    \caption{Stellar and Planetary Parameters for Occurrence Rate Calculation \label{tab:occ_param}}
    \begin{tabular}{ll}
        \\\hline \hline
         Parameter & Value \\
         Stellar effective Temperature, $T_{eff}$ & 3,870 $-$ 7,000 K \\
        Stellar gravity, log $g$ (cgs) & 3.8 $-$ 5.1 \\
        V  & $<$ 13 mag\\
        Number of stars & 725,933 \\
        Orbital period & 0.5 $-$ 13 days \\
        Planet radius & 2 – 28 $R_{\oplus}$ \\
        Number of planet candidates & 385 \\\hline
    \end{tabular}
\end{table}

The 385 candidates are grouped into period-radius cells in Figure \ref{fig:dtarps_occ}.  The number of DTARPS-S Candidates, $n_\textrm{cands,cell}$,  is given at the upper left corner of each cell. The number of DTARPS-S Candidates in each period-radius cell is corrected for the completeness of DTARPS-S based on the completeness of transit identification by the Random Forest (RF) classifier (Figure~13 in Paper I) and the vetting process based on the planet injections and Confirmed Planets.  The classifier and vetting completeness measures are assumed to be independent and are multiplied together to calculate the completeness measurement for the full DTARPS-S method. The number of candidates in each cell is divided by the completeness fraction to estimate the occurrence rate. Errors are assumed to be entirely due to counting uncertainties and likely underestimate the true uncertainties.

The completeness of the RF classifier, designated $r_{\textrm{RF, cell}}$, is measured from the recall rate of the injections of synthetic planets drawn from the \Kepler sample into random DIAmante light curves in each cell (\S9 in Paper I).  Cells without measured RF completeness had the RF completeness set to 1. The completeness from the vetting process, $S_{\textrm{CP, cell}}$, is measured from the fraction of Confirmed Planets from the NASA Exoplanet Archive and the \TESS Object of Interest list in the DTARPS-S Analysis list (Paper I) that passed the vetting process to be listed in the DTARPS-S Candidate catalog (Paper II). Cells without measured vetting completeness (due ot a lack of Confirmed Planets in the DTARPS-S Analysis list) had the vetting completeness set to the global vetting completeness for the entire DTARPS-S Analysis list.

The number of candidates in each cell must be corrected for the fraction of astrophysical False Positives labeled as DTARPS-S Candidates. The purity of the candidates is discussed in \S \ref{sec:cand_pure}; there is insufficient information to estimate FL rates in each cell independently, so a global value is used here. We use the 35 labeled False Positives in the DTARPS-S Candidate list to estimate the vetting reliability.  We calculate the True Positive rate for the DTARPS-S Candidates utilizing only candidates with at least some follow-up observations.  The True Positive rate, designated $\textrm{TPR}_{\textrm{vet}}$, is 70\%; the 82 Confirmed Planets in the DTARPS-S Candidate list divided by the 82 Confirmed Planets, and 35 False Positives in the DTARPS-S Candidate catalog.

Only a small fraction of exoplanets transit their host star due to the inclination of their orbit with respect to line of sight and therefore a large geometric correction must be applied to the number of candidates in each cell. Assuming a uniform distribution of the cosine of orbital inclination angles, the probability that an exoplanet on a circular orbit will transit is $\sim$ $R_{\star}$/$a$, where $R_{\star}$ is the radius of the host, and $a$ is the semi-major axis of the planet. Following \citet{Howard12}, the geometric transit probability correction is applied to each cell by transforming the number of candidates in each cell to 
\begin{equation}
    n_{\textrm{cands,geom.,cell}} =  \sum_{j=1}^{n_\textrm{cands,cell}} \frac{a_{j}}{R_{\star,j}} 
\end{equation}
where $n_\textrm{cands,geom.,cell}$ is the geometrically-corrected number of DTARPS-S Candidates in the cell. This method corrects each individual candidate for the geometric transit probability rather than correcting the entire cell with the geometric probability of transit for a planet at the midpoint of the cell.  

The number of stars around which the planet candidate would have been identified by the DTARPS-S method if such a transiting planet candidate existed must be estimated for each DTARPS-S candidate. For each of the 725,933 DIAmante stars in the sample, the orbital period for each DTARPS-S candidate and the scaled transit depth is used to estimate an effective signal-to-noise ratio (SNR$^{eff}$) of the transit signal the planet candidate would have in the photometric light curve for that host star. The effective SNR of the transit associated with DTARPS-S Candidate, $j$, around a star, $k$, is the calculated using equation 7 in Paper I. The IQR of the ARIMA residuals (our substitute for the more commonly used standard deviation of the light curve during the transit) and the number of points in the ARIMA residuals are properties of each star in the stellar sample that do not change based on the theoretical planet signal.  The period of the DTARPS-S candidate is used directly, but the depth and duration of the theoretical candidate transit on a star, $k$ depends on the stellar radius. The duration of the theoretical transit was calculated from Equation 19 in \citet{Winn10}, assuming a circular orbit. We used the radius of star from the TIC catalog \citep{Stassun19}.

The transit depth for the theoretical transit signal is calculated from the DTARPS-S candidate planetary radius and stellar radius. We found that the depths from the best peak of the TCF periodogram tended underestimate the injected transit depth (especially for large transit depths).  This effect was corrected for in each of the DTARPS-S Candidates (\S4 in Paper II).  However, the DTARPS-S method detectability for a theoretical transit signal depends on the effective SNR of the transit as identified by the TCF periodogram since the RF classifier uses the transit depth as a feature. Therefore the transit depth for the theoretical transit signal is corrected to reflect the theoretical transit depth reported by the TCF periodogram for the candidate around the star.  This correction was performed by fitting a line to the log depths from the TCF periodogram as a function of the log depth of the injected for the injected planets that were recovered by the TCF periodogram.

Instead of setting a hard SNR$^{eff}$ cutoff to determine whether a theoretical candidate orbiting a star in the stellar sample would have been detected \citep{Howard12}, we loosely follow the procedure in \citet{Fressin13} for determining detectability. Figure~18 in Paper~I shows the recall rate of injected planet signals as a function of the SNR$^{eff}$ of the transit. The theoretical SNR$^{eff}$ for a DTARPS-S candidate around a star is used to find the recall rate of injected planets for the SNR$^{eff}$ of the transit. For each DTARPS-S candidate $j$, the number of stars in the DIAmante stellar sample around which a transiting planet with the same orbital period and planet radius as $j$ is the sum of the recall rates for the injected planets given the theoretical SNR$^{eff}$,
\begin{equation}
    n_{\star,j} = \sum_{k = 1}^{n_{\star}} r_{inj}\left(\textrm{SNR}^{eff}_{k,j}\right)
\end{equation}
where $n_{\star}$ is the number of stars in the DIAmante stellar sample from Table \ref{tab:occ_param} and $r_{inj}$ is the recall rate from Figure~14 in Paper I given the SNR$^{eff}$ of the theoretical transit of candidate $j$ around star $k$.

The estimated occurrence rate for each cell in Figure \ref{fig:dtarps_occ} is then  
\begin{equation}
    f_{cell} = \sum_{j=1}^{n_\textrm{\small cands,cell}} \frac{n_{\textrm{\small cands,geom.,cell}}}{n_{\star,j}}\times \frac{1}{r_\textrm{\small RF,cell}} \times \frac{1}{S_\textrm{\small CP,cell}} \times \textrm{TPR}_{\textrm{\small vet}} 
\end{equation}
where $r_\textrm{RF,cell}$ is the completeness of the planetary injections from the Random Forest classifier (Figure~13 in Paper I), $S_\textrm{CP,cell}$ is the sensitivity (completeness) of the vetting procedure from the Confirmed Planets in each cell, and $\textrm{TPR}_{\textrm{vet}}$ is the False Positive reliability correction from the vetting True Positive rate for objects with follow-up observations. 

Error estimates for $f_{cell}$ were estimated using binomial statistics following \citet{Howard12}. The error for each term of the occurrence rate calculation is assumed to be a binomial probability distribution. The first term is treated as the binomial probability of drawing of $n_{pl,cell}$ candidates from $n_{pl,cell}$/$f_{cell}$ ``effective'' stars for the entire cell. The second term is treated as the binomial probability of drawing the number of recovered injected planets from the set of injected planets in the cell. The third term is treated as the binomial probability of drawing the number of Confirmed Planets in the candidate list from the number of Confirmed Planets in the cell. The fourth term is treated as the binomial probability of drawing 82 Confirmed Planets from a population of 117 objects with follow-up observations (Confirmed Planets and False Positives). The $\pm 1 \sigma$ errors for each term of $f_{cell}$ are calculated using the asymptotic confidence intervals for a binomial distribution. The final error for $f_{cell}$ was calculated through propagation of errors.  The error bars can be quite large for some of the cells in Figure \ref{fig:dtarps_occ}. 

While the DTARPS-S method and its resulting Candidate catalog has been sufficiently well characterized to allow the calculation of preliminary occurrence rates, there are a number of caveats and simplifications that were applied to the sample and the process:

\begin{itemize}
    \item[] \textit{Injections on processed light curves}: The Random Forest classifier completeness was characterized by the performance of the classifier on synthetic transit signals injected into random DIAmante light curves. The injection process (\S5 in Paper I) added a very basic trapezoidal transit shape to DIAmante light curves after trend removal and data cleaning. The injections used to characterized the \Kepler detection were performed with injections based on more complicated models into the pixel signals before any trend removal or processing \citep{Christiansen20}. Therefore our analysis of the classifier performance does not take into account potential systematics from the DIAmante extraction pipeline and trend removal on the detectability of exoplanets. 
    
    \item[] \textit{Injection light curves had no pre-vetting}: The light curves that the injections were performed on were not pre-vetted to determine if a periodic photometric signal was present.  If a periodic signal (planetary or non-planetary) is present in the light curve before injecting a new planet signal and the new transit signal is weaker than the inherent signal, then, the injection would not have been recovered skewing the injection recall rates. 
    
    \item[] \textit{Assumed circular orbits}: All candidates are assumed to be on circular orbits for simplicity. This affected the geometric transit probabilities for each candidate as well as  he transit duration calculation when estimating the number of stars in the sample around which a candidate would be detected. \citet{Burke15} found that while eccentricity did affect occurrence rate calculation, it was not significant and on the order of the statistical errors of the occurrence rates. 
    
    \item[] \textit{Multi-planet systems ignored}: The DTARPS-S Candidate occurrence rates do not address multi-planet systems. The DTARPS-S method only examines the most likely periodic transit signal for each light curve. Each object in the DTARPS-S Candidates is around a unique star. 
    
    \item[] \textit{Assumed universal False Positive Rate}: We assumed that the False Positive rate after the DTARPS-S vetting process was the same across all radius-period bins when correcting the occurrence rates for reliability. The distribution of the previously identified False Positives in the DTARPS-S sample is not uniform across the period radius bins (\S10.5Paper I) so it is unlikely that the distribution of any False Positives that contaminate the sample are evenly distributed. However, given that we only have 35 previously identified False Positives in the DTARPS-S Catalog, we cannot quantify a distribution of False Positives for our methodology.
    
\end{itemize}  

\end{document}